\documentclass[aps,prb,twocolumn,nofootinbib,groupedaddress,superscriptaddress,longbibliography]{revtex4-2}

\usepackage{float}
\usepackage{comment}
\usepackage{booktabs}
\usepackage[percent]{overpic}
\usepackage{microtype}
\usepackage{dsfont}
\usepackage{algorithm}
\usepackage{algorithmic}
\usepackage{array}
\usepackage{amsmath}
\usepackage{braket}
\usepackage{amsthm}
\usepackage{amssymb}
\usepackage{graphicx}
\usepackage{bbold}
\usepackage{hyperref}
\usepackage[usenames,dvipsnames]{xcolor}
\usepackage{tikz}
\usetikzlibrary{arrows.meta, automata, positioning, quotes, shapes}
\usepackage{multirow} 
\usepackage{bm}
\usepackage{mathtools}
\usepackage{dsfont}
\usepackage{amsfonts}
\usepackage[normalem]{ulem}
\usepackage{url}
\usepackage{wasysym}

\makeatletter

\newcommand{\bra@nostar}[1]{\langle #1\rvert}
\newcommand{\bra@star}[1]{\left\langle #1\right\rvert}
\newcommand{\expval}{\@ifstar{\expval@star}{\expval@nostar}}
\newcommand{\expval@nostar}[1]{\langle #1\rangle}
\newcommand{\expval@star}[1]{\left\langle #1\right\rangle}
\newcommand{\mel}{\@ifstar{\mel@star}{\mel@nostar}}
\newcommand{\mel@nostar}[3]{\langle #1\rvert #2 \lvert #3\rangle}
\newcommand{\mel@star}[3]{\left\langle #1\middle\rvert #2 \middle\lvert #3\right\rangle}
\newcommand{\abs}{\@ifstar{\abs@star}{\abs@nostar}}
\newcommand{\abs@nostar}[1]{\lvert #1\rvert}
\newcommand{\abs@star}[1]{\left\lvert #1\right\rvert}
\makeatother

\allowdisplaybreaks

\bibpunct{[}{]}{,}{n}{}{}
\usepackage{hyperref} 
\hypersetup{colorlinks=true,citecolor=Red,linkcolor=Blue,urlcolor=Blue}
\newcommand{\honey}[4]{
	\def \w {white}
	\def \g {lightgray}
	\begin{tikzpicture}[baseline={([yshift=-.5ex]current bounding box.center)}, scale=0.5]
	  \begin{scope}[%
	every node/.style={anchor=west,
	regular polygon, 
	regular polygon sides=6,
	draw,
	minimum width=1cm,
	outer sep=0,
	},
	      transform shape]
	    \node (A)[fill=#1] 	    	        {};
	    \node (B)[fill=#2]  at (A.corner 5) {};
	    \node (C)[fill=#3]  at (A.corner 1) {};
	    \node (D)[fill=#4]  at (C.corner 5) {};
	  \end{scope}
	\end{tikzpicture}
}

\begin{document}
% \title{\ourtitle}
%%%%%%%%%%%%

%%%%%%%%%%%%
\title{Magnetic Field Induced Phenomena in Kitaev Spin Liquids}
\author{Shi Feng}
\email{shi.feng@tum.de}
\affiliation{Department of Physics, The Ohio State University, Columbus, Ohio 43210, United States}
\affiliation{Technical University of Munich, TUM School of Natural Sciences, Physics Department, 85748 Garching, Germany}
\affiliation{Munich Center for Quantum Science and Technology (MCQST), Schellingstr. 4, 80799 M{\"u}nchen, Germany}
\author{Nandini Trivedi}
\email{trivedi.15@osu.edu}
\affiliation{Department of Physics, The Ohio State University, Columbus, Ohio 43210, United States}

\date{\today}
%%%%%%%%%%%%

%%%%%%%%%%%%
\begin{abstract}
Quantum spin liquids (QSLs) host a variety of fractionalized particles. In Kitaev's paradigmatic honeycomb model a spin-$\tfrac{1}{2}$ fractionalizes into $Z_2$ flux due to emergent $Z_2$ gauge field and matter Majorana fermions. Although these excitations have well-defined dynamics in the integrable limit, their direct experimental identification is notoriously challenging: realistic materials inevitably host additional symmetry-allowed interactions that break integrability and hybridize gauge and matter sectors, while magnetic fields, which are often required to suppress competing order and stabilize a putative QSL regime, further entangle the responses of different fractionalized quasiparticles and may even drive the system into field-induced spin-liquid phases that are not adiabatically connected to the integrable limit. A prominent example is the quantum Majorana metal, in which the distinct dynamics of fractionalized Majorana fermions can become directly visible in scattering.
This review highlights recent progress on these related questions: in which field-stabilized QSL regimes and nearby emergent phases, and under what conditions, can the response of a specific fractionalized quasiparticle be isolated and positively understood, thereby clarifying the existence and the experimental scope of putative spin liquids? 
We review the progress on these questions across Abelian, non-Abelian, and emergent quantum phases under magnetic field that are not perturbatively connected to the integrable limit.
We connect these field-induced dynamical phenomena to concrete experimental observables, relevant for neutron scattering, resonant inelastic X-ray scattering, and pump-probe spectroscopy, techniques that are capable of identifying the nature of different magnetic-field-induced phases and resolving specific types of fractionalized particles, including Majoranas and $Z_2$ fluxes.
\end{abstract}
%%%%%%%%%%%%
\maketitle
\tableofcontents

\section{Introduction}
At the heart of magnetism lies the interplay of quantum mechanics and electron-electron interactions. Starting with general Hamiltonians with electronic structures arising from multi-orbitals and spin-orbit coupling, upon including strong Coulomb interactions one obtains magnetic interactions that are isotropic between spins (Heisenberg interactions) as well as anisotropic bond-dependent interactions (Kitaev and Gamma interactions). In the absence of spin-orbit coupling, only the Heisenberg interaction (and its components such as Ising and XY) survive and have been extensively discussed both theoretically and experimentally.
Besides the sign of the magnetic interaction that determines the nature of the ground state (ferromagnetic vs antiferromagnetic), another key feature that determines the nature of the phase are the competing magnetic interactions or frustration between different bonds \cite{balents2010spin}. For example, for antiferromagnetic interactions on a triangular lattice, the three spins cannot be simultaneously satisfied and that leads to degenerate ground states with diminished long range magnetic order.
The degree of frustration can be modified by changing the lattice structure and by introducing competing interactions across bonds. 

Frustration essentially allows the system to fluctuate or resonate between several energetically favorable configurations rather than picking a specific configuration \cite{Blote_1982,Zeng97,Moessner01,Moessner01a,Moessner01c,Moessner01d,Patrick08}. These quantum fluctuations are enhanced by reducing the spin and lowering the dimensionality.  
The idea of resonance between different configurations perhaps first goes back to Pauling when he proposed that the energy of the benzene ring is reduced by linearly superposing the two quantum states \cite{Pauling1933,pauling1960nature}. Later in the context of high temperature superconductivity in copper-oxide materials it was proposed that the resonating valence bond (RVB) states between a macroscopic number of spins could form the parent state of the Mott insulator from which superconductivity is born upon doping \cite{Rokhsar88,ANDERSON1973153,anderson1987resonating,BASKARAN1987973,Anderson_2004}. While the Mott insulating state of the cuprates undergoes spontaneous symmetry breaking to an antiferromagnetic state, the RVB idea is conceptually important not only for framing the high temperature superconducting problem but also more generally as a possible ground state for frustrated quantum magnets.

The triangle is the first example of frustration due to the non-bipartite nature of the lattice \cite{Wannier50}. Upon including additional interactions that frustrate the system even further, such as nearest neighbor AF $J_1$ and next neighbor AF interactions $J_2$, and its variant on the Shastry-Sutherland model with alternating diagonal interactions \cite{Shastry81}, or ring-exchange interactions \cite{Motrunich05,Sheng09}, or the kagome lattice with honeycomb and triangular motifs \cite{Sachdev92,Simeng11,Ulrich12}, that lead to additional frustration and can drive the local order parameter to zero, yielding a quantum spin liquid ground state.

\begin{figure*}[t]
    \includegraphics[width=0.85\linewidth]{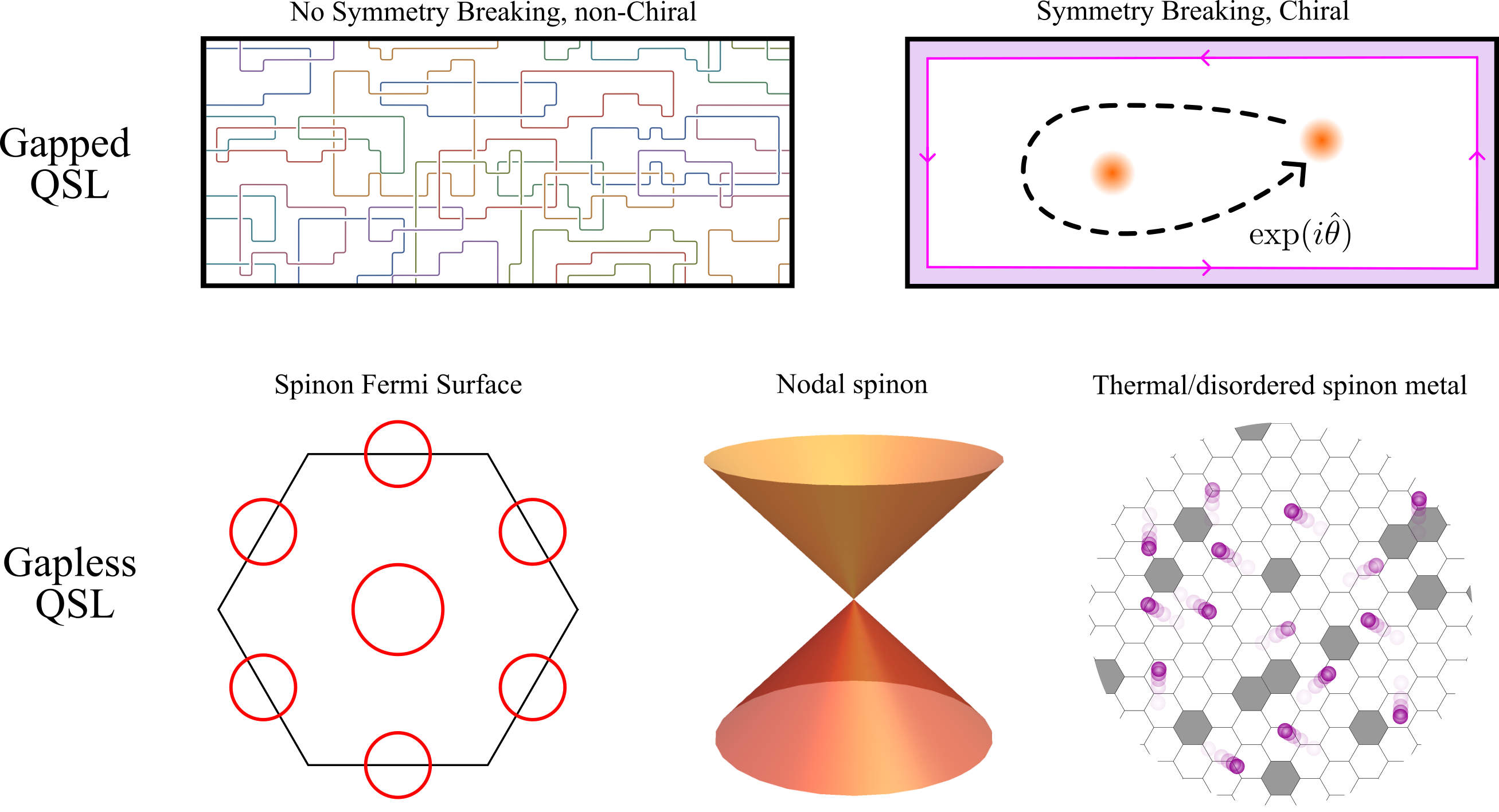}
    \caption{Bifurcation of quantum spin liquids in 2D into \emph{gapped} (top) and \emph{gapless} (bottom) classes.
    Top left: Nonchiral gapped (string-net states or short-range RVB \cite{Wen19}): intrinsic topological order with a finite anyon set, long-range entanglement, and no chiral edge; includes Abelian and non-Abelian examples \cite{Levin05}.
    Top right: Gapped chiral RVB: time-reversal broken with a chiral edge mode (pink) and quantized thermal Hall conductance.
    Bottom row (gapless): spinon Fermi-surface liquid, Dirac/nodal spin liquid, and thermal metal (disorder-enabled spinon metal).
    Other exotic examples and formal classification such as symmetry-enriched topological order are beyond the scope of this article, and are not shown. }
    \label{fig:intro}
\end{figure*}

An alternative route to frustration is driven by spin-orbital interactions rather than by geometrical frustration of the underlying motif. Here, a quantum spin liquid ground state with no magnetic long range order is obtained via bond-dependent interactions between non-commuting operators that generate frustration even in the absence of geometric frustration, as in the materials resembling Kitaev's honeycomb model \cite{kitaev2006anyons,Khaliullin2009,Trebst2022,Chou2025}. The importance of the honeycomb model is that there is an exact solution by Kitaev which explicitly shows how frustration causes the spins to fractionalize into new degrees of freedom, Majorana fermions that form the emergent matter and $Z_2$ gauge degrees of freedom. Given this limit of analytical control it is now possible to investigate the effects of a magnetic field in different directions, anisotropy of exchange couplings, strain and other perturbations on a class of $Z_2$ quantum spin liquids.

These different routes to frustration open the door to Mott insulating phases without conventional spontaneous magnetic symmetry breaking. When order is avoided down to the lowest temperatures, the resulting low-energy physics falls into two phenomenological classes distinguished by their low-energy spectra: gapped versus gapless quantum spin liquids~\cite{Sachdev_2023}. Figure~\ref{fig:intro} organizes this bifurcation.
By a gapped QSL we mean an intrinsically topologically ordered phase with long-range entanglement and deconfined anyons. In two dimensions, this splits naturally into (i) nonchiral topological orders (string-net/short-range RVB type) with no chiral edge and (ii) chiral topological orders (chiral-RVB/Chern–Simons) with broken time-reversal symmetry, chiral edge modes, and quantized thermal Hall conductance \cite{Levin05,Wen19}. In contrast, gapless QSLs host low-energy spinful matter: (i) spinon Fermi-surface liquids with finite zero-energy spinon density of states \cite{Patrick18}, (ii) Dirac/nodal liquids with algebraic correlations \cite{Iqbal16,Shijie19,Drescher23}, and (iii) disorder-enabled gapless delocalized spinon phases \cite{Huse2012,KAO2021168506}. We note that this is a phenomenological classification for the focus of this article; for formal classification by projected symmetry groups, we refer interested readers to \cite{Wen_PRB_2002}. 
In what follows we use this dichotomy as an organizing principle and focus on how gapped and gapless regimes are realized, tuned, and diagnosed in Kitaev honeycomb materials and their extensions, where fields, anisotropies, and disorder reshape the spectra of Majorana matter and $Z_2$ fluxes.

Lately, we have witnessed remarkable progress on both the materials and theory fronts in the search for Kitaev quantum spin liquids. On the experimental side, intense efforts have focused on honeycomb and related lattice magnets in which strong spin-orbit coupling and edge-sharing octahedra generate substantial bond-dependent interactions. Early work concentrated on the 5$d$ iridates A$_2$IrO$_3$ (A = Na, Li, Cu) \cite{Jackeli2010,2010Na2IrO3,2011Na2IrO3,2015LiIrO3,2017CuIrO3,Kitagawa2018}, and on $\alpha$-RuCl$_3$ \cite{Knolle2018,banerjee2017neutron,Zheng17,Stahl2024}, where a combination of inelastic neutron scattering, Raman spectroscopy, thermal transport, and thermodynamics revealed broad continua and field-tuned phases suggestive of proximate Kitaev QSL behavior. More recently, 3$d$ and 4$d$ transition-metal compounds such as Na$_3$Ni$_2$BiO$_6$ \cite{Shangguan2023}, Na$_2$Co$_2$TeO$_6$ \cite{Lin2021,Yao2022,Pilch23,Gaoting24,chen2024planar}, YbOCl \cite{YbOCl2022,YbOCl2024}, Na$_3$Co$_2$SbO$_6$ \cite{Vavilova23,Poldi2025}, as well as Rydberg-atom array simulations for spin-orbital frustrated systems \cite{Ruben22,Willian25},  have emerged as promising platforms with predominantly Kitaev exchange, although the proximity of solid-state systems to the Kitaev QSL is not without controversy in different thermodynamic measurements \cite{Fang25}, thereby significantly expanding the landscape of candidate materials and synthetic systems, and highlighting the diversity of potential microscopic routes to Kitaev-like frustrated systems. 

In parallel, recent unbiased numerical advances such as density matrix renormalization group (DMRG) \cite{white1992density,xiang2023density}, Projected Entangled Pair States (PEPS) \cite{cirac2021matrix,xie2014tensor,verstraete2004renormalization}, corner transfer matrix renormalization group (CTMRG) \cite{CTMRG96,Orus_CTM09,Corboz_tJ14}, optimization techniques like automatic differentiation \cite{Liao19,Corboz22}, neural quantum states and emerging efficient quantum Monte-Carlo algorithms \cite{chen2025,gu2025,kong2025fft}, have provided new tools to greatly sharpen our understanding of frustrated magnetism, including field-induced phenomena in the paradigmatic Kitaev honeycomb model and its extensions. Despite the exact solvability of the zero-field model, the phases under other symmetry-allowed exchanges, e.g. Heisenberg, $\Gamma$, $\Gamma’$, longer-range exchange, and by external magnetic fields, have been discussed extensively in recent literature in terms of statics, but whose dynamical nature regarding fractionalized quasi-particles remains far from settled. Even for the pure Kitaev honeycomb Hamiltonian, the character of the putative field-induced gapless phase at intermediate fields is under active debate, with proposals ranging from U(1) spinon Fermi-surface states and gapped parton Chern insulators to gapless Majorana metals stabilized by random glassy $Z_2$ flux backgrounds \cite{David2019,hickey2019emergence,Patel12199,Jiang2020,Han21,Baskaran2023,ZhangNatComm2022,wang2024,penghao24}. Related questions arise in more realistic models relevant for candidate materials, where additional interactions and lattice distortions compete with the Kitaev term and where disorder, stacking faults, and phonons further complicate the interpretation of experimental data \cite{KAO2021168506,Mengxing24,Knolle21,Natalia24}.

A common thread running through these developments is the crucial role of the magnetic field as a clean and versatile tuning parameter \cite{Trebst2022}. A field can partially or fully suppress magnetic order, close or open gaps, polarize or proliferate gauge fluxes, and thereby reshape the spectrum and mobility of fractionalized excitations \cite{Kasahara2018}. In Kitaev and Kitaev-like systems, weak magnetic fields along [111] direction can generate chiral spin liquids with non-Abelian anyons \cite{kitaev2006anyons}, tune the Chern number of fractionalized particles \cite{Yip22,Kyusung22,Kyusung25}, induce intermediate gapless regimes with enhanced low-energy density of states, or realize regimes where fluxes and Majorana fermions acquire distinct dynamics \cite{Roser19,Knolle2014,Aprem21,wang2024,penghao24}, leading to emergent glassiness \cite{Baskaran2023}, fractonic constraints, or tune the quasi-particle mobility to enhance or suppress thermal transport \cite{Nasu17,Zekun,Brenig21,Feng2025}. Importantly, these field-driven crossovers and phase transitions imprint themselves directly on experimentally accessible observables, such as the dynamical spin structure factor, dimer correlations, and thermal transport, offering rare opportunities to probe specific species of fractionalized quasiparticles in a controlled manner.

This review is devoted to recent numerical and theoretical developments on magnetic-field-induced phenomena in $Z_2$ quantum spin liquids, with a particular emphasis on Kitaev honeycomb systems and closely related models, both in the pure Kitaev limit and in extended models with symmetry-allowed non-Kitaev interactions relevant to real materials. Compared with existing broad reviews of quantum spin liquids, Kitaev materials, and the zero-field Kitaev model \cite{Knolle2018,Trebst2022,matsuda2025}, the distinctive focus of this review is the behavior of Kitaev spin liquids across the full range of magnetic fields that is relevant both theoretically and experimentally. We review recent progress on the weak-field perturbative regime, where the Kitaev spin liquid evolves into a non-Abelian chiral spin liquid; the intermediate-field regime, where numerical studies of both pure and extended Kitaev models have revealed possible gapless spin-liquid phases and where substantial debate remains about their microscopic origin; and the strong-field regime, where the system approaches a polarized state while retaining sharp remnants of the anisotropic Kitaev exchange interaction. This field-oriented organization is especially important for solid-state candidate materials, where magnetic fields are often required to suppress low-temperature magnetic order and access a non-magnetic regime, and where additional Heisenberg, $\Gamma$, $\Gamma^\prime$, longer-range interactions, disorder, and lattice effects are unavoidable. For each field regime, we discuss complementary theoretical pictures and unbiased numerical results, emphasizing the emergent fractionalized degrees of freedom, Majorana fermions, $Z_2$ fluxes/visons, and their possible hybridized composites, that control thermodynamic and dynamical responses.
For broader overviews, we refer the reader to comprehensive reviews of topological order \cite{Wen19}, of quantum spin liquids \cite{ZhouRMP,broholm2020quantum,Blents1}, of the (extended) Kitaev honeycomb model \cite{Knolle2018,Knolle_ARCMP_2019,Nasu2020}, and of their solid-state realization under the banner of Kitaev materials \cite{Winter_2017,takagi2019concept,Trebst2022,Chou2025,ojedaaristizabal2025,matsuda2025}.

The central guiding question of this article is therefore: under what conditions (magnetic fields or potentially non-Kitaev terms), and to what extent, can thermodynamic and response functions be interpreted as signatures of specific fractionalized excitations in field-induced phases of QSL models? More concretely, can one identify the fingerprints of Majorana fermions, fluxes/visons, spinons, or their hybridized composites directly in unbiased numerical computations and experimentally relevant response functions, instead of merely inferring fractionalization from the absence of sharp conventional quasiparticles? This question has become one of the most important and puzzling issues in current QSL research. It lies at the heart of ongoing debates over field-induced QSL candidates and spinon-Fermi-surface-like states, where apparently conflicting observations, such as gapless thermodynamic or spectroscopic signals coexisting with weak or vanishing thermal transport, are currently under intense investigation~\cite{matsuda2025,James25,Fang25}.

The structure of this review is organized as follows. In Sec.~\ref{sec:review} we briefly review the exact solution of the Kitaev model, the structure of its gauge and matter sectors, and the implications for spin and dimer spectroscopy that are under-presented in existing reviews. In Sec.~\ref{sec:diagram} we survey the field-induced phase diagrams of the pure and extended Kitaev models, highlighting the theoretical proposals and numerical evidence for intermediate gapless phase(s), whose nature concerning fractionalized excitations has been recently under debate. Sec.~\ref{sec:mm} focuses on the interplay between vison (flux) fluctuations and Majorana fermions, including the recent theory on the emergence of quantum Majorana metal and the consequences for dynamical response and heat transport. In Sec.~\ref{sec:dim} and Sec.~\ref{sec:dim2} we discuss how weak and strong magnetic fields can generate sub-dimensional and fractonic dynamics of emergent excitations in Kitaev honeycomb model, providing sharp, symmetry-enforced constraints on their mobility and experimental signatures. We conclude in Sec.~\ref{sec:conclude} with an outlook, emphasizing open questions—such as the universality versus non-universality of finite-energy signatures, the limitations of translation-invariant mean-field approaches, and the prospects for unambiguous detection of fractionalized particles in current and future Kitaev material candidates.

\begin{figure*}[t]
    \centering
    \includegraphics[width=0.9\linewidth]{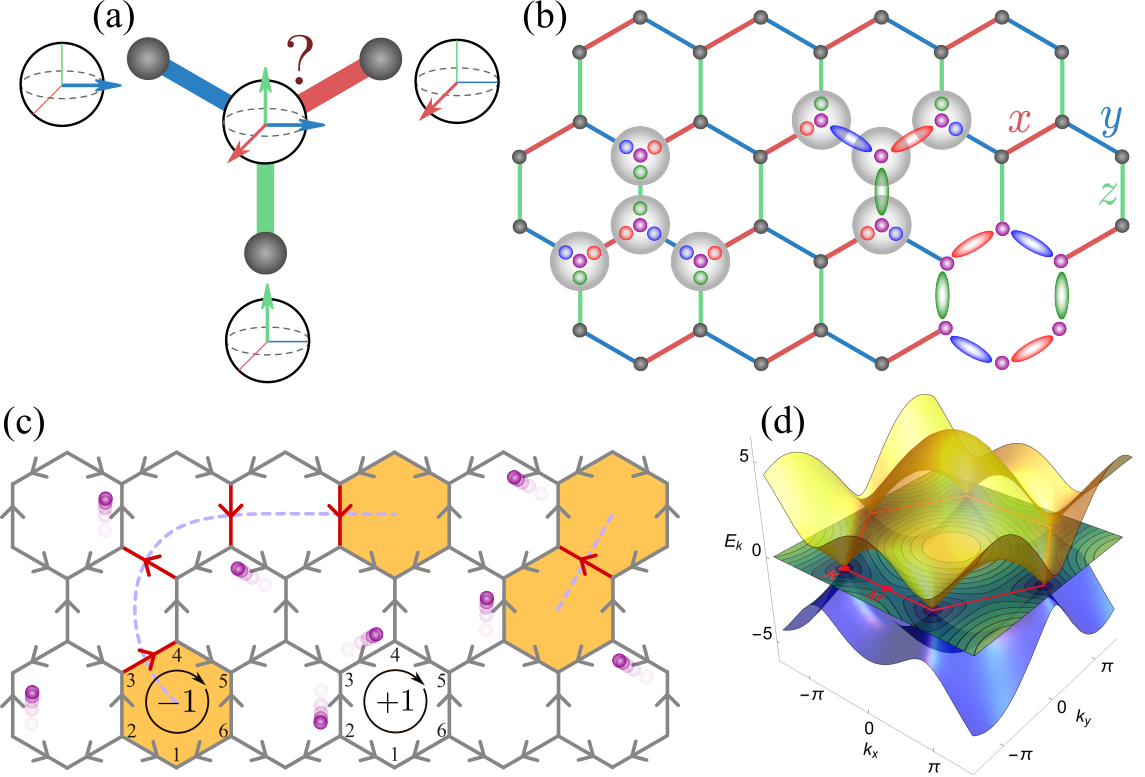}
    \caption{(a) Illustration of spin-orbital frustration. (b) Spin-orbital frustrated honeycomb model with anisotropic compass interactions along the $x$, $y$, and $z$ axes, depicted by distinct colors. Each local spin degree of freedom (gray dots or circles) fractionalizes into four Majorana fermions, indicated by four colors inside larger gray circles. Three Majoranas combine into conserved localized bond fermions (colored ovals), determining the local $Z_2$ gauge field on each bond. (c) The remaining itinerant Majorana fermion (purple dot) propagates in a static background of $Z_2$ gauge fields, with positive directions indicated by arrows. Flipping an odd number of bonds around a hexagonal plaquette (red arrows) excites a $Z_2$ flux within the plaquette (orange). These fluxes sit at the endpoints of dashed purple string operators, and all link variables intersected by the string operator change sign. (d) In the ground-state zero-flux sector, itinerant Majorana fermions form Dirac cones in the absence of time-reversal symmetry-breaking perturbations.   }
    \label{fig:kitaev}
\end{figure*}

\section{A brief review of the honeycomb model} \label{sec:review}
\subsection{Exact solution}
The Kitaev honeycomb model, introduced by Kitaev in 2006~\cite{kitaev2006anyons}, represents a paradigmatic example of an exactly solvable spin model exhibiting quantum spin-liquid phases with fractionalized excitations, and has been widely used as an important reference point from which various controlled perturbative expansions can be launched.
It describes $S=1/2$ spins arranged on a two-dimensional honeycomb lattice with strongly anisotropic, bond-dependent Ising interactions:
\begin{equation} \label{eq:kitaev_o}
    H_K = \sum_{\langle ij \rangle_\alpha} K_{\alpha}\sigma_i^\alpha\sigma_j^\alpha, 
\end{equation}
where the coupling strengths $K_\alpha$ ($\alpha=x,y,z$) differ along the three distinct bond directions of the honeycomb lattice, leading to spin-orbital frustration as illustrated in Fig.~\ref{fig:kitaev}(a).
A central result of Kitaev's seminal work \cite{kitaev2006anyons} is that this model can be solved exactly by fractionalizing the spin degrees of freedom into Majorana fermions coupled to a static $Z_2$ gauge field. 
Specifically, as illustrated in Fig.~\ref{fig:kitaev}(b), each spin operator at site $i$ is expressed in terms of four Majorana fermions $(b_i^x,b_i^y,b_i^z,c_i)$, satisfying the anti-commutation relations: $\{b_i^\alpha, b_j^\beta\} = 2\delta_{ij}\delta_{\alpha\beta}$, $\{c_i, c_j\} = 2\delta_{ij}$, $\{b_i^\alpha,c_j\}=0$, and $(b_i^\alpha)^2 = c_i^2 = 1$. 
In this representation, the spin operators become $\sigma_i^\alpha = ib_i^\alpha c_i$, $\alpha = x,y,z$.

The four-Majorana representation enlarges the local Hilbert space, and the physical spin Hilbert space is recovered by imposing a local gauge constraint. A single Majorana mode carries formal Hilbert-space dimension $\sqrt{2}$; equivalently, two Majoranas form one complex fermion with a two-dimensional Fock space. Thus the four Majoranas $(b_i^x,b_i^y,b_i^z,c_i)$ span $(\sqrt{2})^4 = 4$ local states, or equivalently two complex fermionic modes, whereas a physical spin-$1/2$ has only two states. The redundant factor of two is removed by imposing the local constraint
\begin{equation} \label{eq:proj}
    D_i \equiv b_i^x b_i^y b_i^z c_i = +1,
    \qquad 
    P_i=\frac{1+D_i}{2},
\end{equation}
where $D_i^2=1$ and $P_i$ projects onto the physical two-dimensional subspace. The full projection is $P=\prod_i P_i$.
Within this physical subspace, the Majorana representation reproduces the Pauli algebra. Indeed, $(\sigma_i^\alpha)^2=(i b_i^\alpha c_i)^2=1$.
For $\alpha\neq\beta$, let $\gamma$ be the remaining spin component, with orientation fixed by $\epsilon_{\alpha\beta\gamma}$. Then, the operator identity $\sigma_i^\alpha\sigma_i^\beta=(i b_i^\alpha c_i)(i b_i^\beta c_i)=b_i^\alpha b_i^\beta$ can be recast using, 
\begin{equation} \label{eq:check}
    b_i^\alpha b_i^\beta
    =
    \epsilon_{\alpha\beta\gamma}
    D_i c_i b_i^\gamma
    =
    \epsilon_{\alpha\beta\gamma}
    c_i b_i^\gamma
    =
    i\epsilon_{\alpha\beta\gamma}\sigma_i^\gamma ,
\end{equation}
where in the second equality we used the physical constraint $D_i=+1$, and in the last equality we used $\sigma_i^\gamma=i b_i^\gamma c_i$.
This gives exactly the Pauli algebra $\sigma_i^\alpha\sigma_i^\beta =\delta_{\alpha\beta}+i\epsilon_{\alpha\beta\gamma}\sigma_i^\gamma$.
Therefore the four-Majorana construction is an exact rewriting of a spin-$1/2$ degree of freedom after the projection in Eq.~\eqref{eq:proj}.

The Kitaev Hamiltonian is thus recast into the form
\begin{equation}\label{eq:Kitaev_Majorana}
    H_K = -i\sum_{\langle ij \rangle_\alpha} K_{\alpha}\hat{u}_{\langle ij \rangle_\alpha}c_i c_j,
\end{equation}
for the convention $\sigma_i^\alpha=ib_i^\alpha c_i$ and $\hat{u}_{ij}=ib_i^\alpha b_j^\alpha$; the sign is often absorbed by reversing the orientation convention for the directed link variable.
Here the operators $\hat{u}_{\langle ij\rangle_\alpha} \equiv ib_i^\alpha b_j^\alpha$ [ovals in Fig.~\ref{fig:kitaev}(b)] commute with each other and with the Hamiltonian, thus forming a set of conserved $Z_2$ gauge fields [arrows in Fig.~\ref{fig:kitaev}(c)].    
The eigenvalues $u_{\langle ij\rangle_\alpha} = \pm 1$ correspond to static gauge configurations, whose gauge-invariant observables are given by hexagonal plaquette operators:
\begin{equation} \label{eq:wpu}
    W_p = u_{21} u_{32} u_{43} u_{45} u_{65} u_{61} \equiv \prod_{\langle ij\rangle_\alpha \in \partial p} u_{\langle ij\rangle_\alpha}.
\end{equation}
In the spin basis it is equivalent to
\begin{equation}
    W_p = \sigma^z_1 \sigma^x_2 \sigma^y_3 \sigma^z_4 \sigma^x_5 \sigma^y_6,
\end{equation}
with indices denoted in Fig.~\ref{fig:kitaev}(c) within a hexagon, whose conservation is readily seen in $[W_p, H_K] = 0$.
Equations~\eqref{eq:Kitaev_Majorana} and \eqref{eq:wpu} thus set up a simple tight-binding problem for the itinerant Majorana fermions in each fixed $Z_2$ gauge sector. 
The Majorana representation should therefore be understood as a free-fermion problem only after choosing a conserved gauge sector and imposing the physical projection discussed above.
Nevertheless, much of the thermodynamics and spectral information can still be retrieved from the unprojected Hamiltonian, with a deviation of order $O(1/N)$ \cite{Pedrocchi11}; for more technical details we refer readers to Refs.~\cite{Pedrocchi11,Vojta15,Mandal_2025}.

Non-trivial physics arises from the sector-dependent interplay between the itinerant $c$ Majorana fermions and the static link variables $\hat{u}_{jk}$ in Eq.~\eqref{eq:Kitaev_Majorana}.
Since $u_{ij}=-u_{ji}$, the Majorana hopping amplitude is naturally represented by a directed binary graph, as shown in Fig.~\ref{fig:kitaev}(c), where a choice of positive direction corresponds to the orientation indicated by gray arrows.  
The gauge-invariant sign accumulated around a hexagonal plaquette is encoded in $W_p$; depending on the directed-bond convention this may differ from the literal product of hopping phases by an overall fixed sign. We follow the standard convention in which
an eigenvalue $W_p = +1$, with an even number of flipped red arrows within a hexagon, represents a $0$-flux or flux-free plaquette (white plaquettes), while $W_p = -1$, with an odd number of flipped red arrows, represents a $\pi$-flux plaquette (orange plaquettes). 
Fluxes are created in pairs by open string operators on a closed lattice. The purple dashed line in Fig.~\ref{fig:kitaev}(c) is a convenient gauge representation of such a two-flux state: changing the signs of the bonds crossed by the string moves the branch cut, whereas only the endpoint plaquette fluxes are gauge invariant.

\begin{figure}[b]
    \centering
    \includegraphics[width=\linewidth]{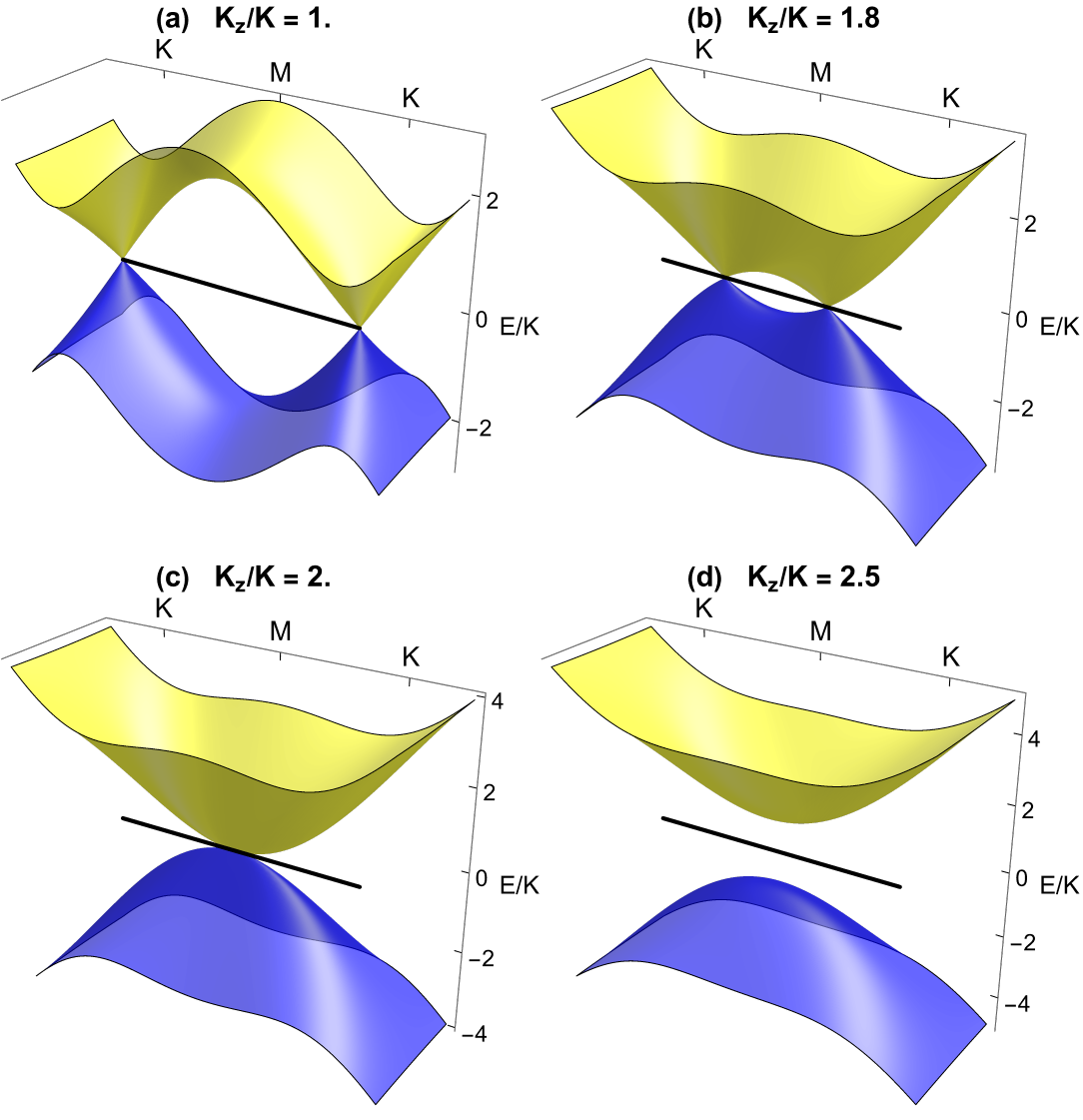}
    \caption{
    Majorana band evolution in the flux-free sector of the Kitaev honeycomb model along the $\rm K$-$\rm M$-$\rm K$ line, with $K_x=K_y=K$ and varying $K_z/K$. 
    (a) At $K_z/K=1$, the isotropic model hosts Dirac nodes at the $\rm K$ points. 
    (b) For the larger anisotropy $K_z/K=1.8$, the two nodes move toward the $\rm M$ point. 
    (c) At $K_z/K=2$, the nodes merge at $\rm M$, marking the weak- to strong-pairing transition of the effective $p$-wave superconductor. 
    (d) For $K_z/K=2.5$, the spectrum is fully gapped in the strong-pairing regime. 
    In the time-reversal-symmetric Kitaev model, the weak-pairing phase (a,b) is nodal. A TR-breaking perturbation, e.g. an effective next-nearest-neighbor Majorana hopping, gaps the weak-pairing side into a chiral topological superconductor with odd Chern number $C=\pm1$, supporting Majorana zero modes bound to $Z_2$ fluxes, see Fig.~\ref{fig:chern}.
    }
    \label{fig:solvableTransition}
\end{figure}

The ground-state sector in the honeycomb model, according to Lieb's seminal theorem~\cite{Lieb1994}, is the flux-free sector with $W_p=+1~\forall p$.
Choosing a uniform gauge in this sector reduces the problem to the band structure of itinerant Majorana fermions, or equivalently to a BdG problem for canonical fermions. 
The flux-free sector takes the form of a spinless $p$-wave superconductor on the triangular lattice. Introducing a canonical fermion on each $z$-bond dimer, for example
\begin{equation}
    f_j=\frac{1}{2}\left(c_{j,A}+i c_{j,B}\right),
\end{equation}
the flux-free Hamiltonian $H_{\rm ff}$ can be written, up to an additive constant, as
\begin{equation} \label{eq:fham}
    H_{\rm ff} = -K \sum_{i,j} \left( f_i^\dagger f_{i+\mathbf{n}_j}
    + f_i^\dagger f_{i+\mathbf{n}_j}^\dagger + {\rm H.c.} \right) 
     - 2 K_z \sum_i f_i^\dagger f_i ,
\end{equation}
where we assumed $K_x=K_y=K$, and $\mathbf{n}_{j}\in\{\mathbf{n}_{1},\mathbf{n}_{2}\}$ are primitive vectors of the triangular Bravais lattice. Varying $K_z/K$ then tunes the system between weak- and strong-pairing regimes. 
At the isotropic point $K_z=K$, $H_{\rm ff}$ describes a nodal weak-pairing $p$-wave spinon superconductor with Dirac nodes at the $\rm K$ points, see Fig.~\ref{fig:solvableTransition}(a) or Fig.~\ref{fig:kitaev}(d). As $K_z/K$ increases, the two nodes move toward each other along the $\rm K$-$\rm M$-$\rm K$ line [Fig.~\ref{fig:solvableTransition}(b)] and merge at the $\rm M$ point when $K_z/K=2$ [Fig.~\ref{fig:solvableTransition}(c)]. 
The motion and annihilation of the two Dirac nodes at the $\rm M$ point is directly analogous to the well-known merging transition of Dirac cones in anisotropic graphene-like honeycomb lattices~\cite{Montambaux2009PRB,Montambaux2009,DEGAIL20121948}.
For positive couplings with $K_x=K_y=K$, the spectrum is gapless for $K_z/K\leq 2$ and becomes gapped for $K_z/K>2$, corresponding to the strong-pairing phase shown in Fig.~\ref{fig:solvableTransition}(d).
This transition at $K_z/K=2$ is the weak- to strong-pairing transition of the effective $p$-wave superconductor. In the time-reversal-symmetric Kitaev model the weak-pairing side is nodal, so its Chern number is not defined until the Dirac nodes are gapped. A perturbative time-reversal-breaking field generates an effective next-nearest-neighbor Majorana hopping, which gaps the nodal weak-pairing regime and turns it into a chiral topological superconductor with odd Chern number $C=\pm 1$, with the sign determined by the sign of the induced chiral term \cite{Read2000}. This will be reviewed in more detail in Sec.~\ref{sec:weakfield}. In this gapped weak-pairing phase, $Z_2$ fluxes/visons bind Majorana zero modes, while the strong-pairing phase for $K_z/K>2$ is topologically trivial. The $p$-wave nature of the fermion model therefore directly connects the band topology of the Majorana sector to the non-Abelian chiral spin liquid obtained once a TR-breaking perturbation gaps out the Dirac points~\cite{kitaev2006anyons}.

It is thus convenient to discern two types of excitations and at low energy scales: itinerant nodal Majorana fermions moving in a fixed uniform gauge background, and local flux-pair excitations generated by flipping a link variable $u_{jk}$, which changes the Majorana Hamiltonian locally. 
Such a flux-pair excitation created by flipping one link variable has a finite gap, which evaluates to $\sim 0.26 K$ in the limit of $K_z =K$ \cite{Coleman23}. 
For separated fluxes, the interaction energy depend on their separation, i.e. the linear length scale of the gauge string in Fig.~\ref{fig:kitaev}(c) \cite{Lahtinen12}.
Indeed, particularly relevant to the topic of this review is the seminal result introduced in Ref.~\cite{kitaev2006anyons}: applying a perturbative TR-breaking Zeeman magnetic field gaps out the Dirac fermions in the weak-pairing phase, leading to a chiral spin liquid (CSL) characterized by chiral edge modes and flux-trapped Majorana zero modes.

This exact solution also clarifies what can and cannot be measured directly. Since an isolated Majorana fermion or a single link variable $u_{ij}$ is not itself a gauge-invariant spin observable, experimental probes access fractionalized excitations only through gauge-invariant combinations. A local spin flip is the simplest example: $\sigma_i^\alpha=i b_i^\alpha c_i$ not only creates or annihilates an itinerant $c$ Majorana, but also flips the adjacent $u_{ij}$ variable on the $\alpha$ bond (See also Sec.~\ref{sec:bondf}), thereby changing the fluxes on the two neighboring plaquettes. Thus the spin structure factor measured by neutron scattering or THz spectroscopy is not a pure Majorana response; in the Kitaev limit it probes a composite process involving both flux excitations and itinerant Majorana fermions~\cite{Knolle2014}. This selection rule is one reason why the spin response can look broad and continuum-like even in an exactly solvable model, rather than exhibiting a conventional magnon pole.

To access the itinerant Majorana sector directly, one should instead consider operators that are gauge invariant and do not change the flux sector. For example, bond-energy or dimer operators on an $\alpha$ bond, $\sigma_i^\alpha\sigma_j^\alpha = -i u_{ij} c_i c_j$ are of this type: in a fixed flux background they reduce to Majorana bilinears and hence couple directly to the Majorana density of states and band structure. This is the theoretical basis for using Raman, RIXS-like, or dimer-response channels as more selective probes of the Majorana sector. Conversely, the $Z_2$ fluxes are strictly static in the integrable Kitaev limit because all $W_p$ commute with $H_K$. Once generic perturbations such as magnetic field, Heisenberg, $\Gamma$, or $\Gamma^\prime$ interactions are included, the fluxes are no longer conserved objects; they acquire dynamics and should be viewed as mobile vison-like excitations with a finite bandwidth and lifetime \cite{Aprem21}. Therefore, direct identification of fractionalization in the model requires comparing flux-changing spin probes with more nearly flux-neutral Majorana bilinear probes, while keeping in mind that perturbations turn the static gauge background of the solvable model into a dynamical vison sector. We shall elaborate on dynamical spin observables in relation to the two types of fractionalized sectors in Sec.~\ref{sec:spec}. Nevertheless, as will be discussed later in Sec.~\ref{sec:mm}, when the system is subjected to an intermediate magnetic field, the aforementioned distinction may be softened in a possible Majorana-metal phase: as fluxes proliferate but remain slow, the spin structure factor may predominantly reflect a disorder-averaged Majorana spectral function at low but finite energies.

\subsection{Flux orthogonality and short-range correlation} \label{sec:bondf}
Although Eq.~\eqref{eq:fham} hosts gapless Dirac nodes, the spin correlations in the ground state remain extremely short-ranged due to the extensive number of conserved fluxes. This constraint significantly influences both the static and dynamical spin structure factors. 
Due to the fractionalization of spins, the spin-spin correlations in the pure Kitaev quantum spin liquid is dependent on both the $Z_2$ gauge sector and the matter Majorana fermion sector \cite{kitaev2006anyons,Baskaran2007}, with the gauge sector contributing to the $S_{\rm topo} = -\ln 2$ topological entanglement \cite{levin2006detecting,kitaev2006topological,xiaoliangqi2010,Roderich18}; and the matter Majorana fermion sector equivalent to a $p$-wave spinon superconductor conditioned on the $Z_2$ gauge fields, as described in the previous section. 
To analyze spin correlations in the joint basis of both sectors, it is convenient to introduce bond fermions $\eta$ \cite{Baskaran2007}: 
\begin{equation}
    \eta_{\langle ij \rangle_a} = \frac{1}{2}(b_i^a + ib_j^a),
\end{equation}
where $i \in A$ and $j \in B$ denote sublattice sites. 
As the original quantum spin is a composite of itinerant matter fermions and localized $Z_2$ gauge fields, the application of a local spin operator involves exciting the composite excitation of both flux and Majorana fermion. This is made simple by 
rewriting Pauli matrices in terms of 
\begin{equation} \label{eq:bondf}
    \sigma_i^a = ic_i(\eta_{\langle ij \rangle_a} + \eta_{\langle ij \rangle_a}^\dagger), \quad
    \sigma_j^a = c_j (\eta_{\langle ij \rangle_a}- \eta_{\langle ij \rangle_a}^\dagger).
\end{equation}
which involves both $c$ for matter Majorana fermions and $\eta$ for $Z_2$ gauge field.
Since local spin operators involve only two-point Pauli matrices sharing a link, each spin operator can be rewritten in terms of bond fermions as  
\begin{equation}
    \sigma_j^a \propto c_j \hat{\pi}_{1,\langle jk \rangle_a} \hat{\pi}_{2,\langle jk \rangle_a},
\end{equation}
where $\hat{\pi}_{1,\langle jk \rangle_a}$ and $\hat{\pi}_{2,\langle jk \rangle_a}$ flip a pair of adjacent fluxes, denoted by subscripts $1$ and $2$, that share the same link $\langle jk \rangle_a$. 
Note that depending on the sublattice index of the Pauli matrix, there can be a factor of $i$, which we left out by using $\propto$ instead of equality. 
It follows that Pauli spin operators transform an eigenstate $\ket{n}$ by altering both the matter and gauge sectors as 
\begin{align}  \label{eq:orth}
    \sigma_{j}^x \ket{\mathcal{M}_{0};\honey{\w}{\w}{\w}{\w}} &\propto c_j \ket{\mathcal{M}_{0};\honey{\g}{\w}{\g}{\w}}, \\
    \sigma_{j}^y \ket{\mathcal{M}_{0};\honey{\w}{\w}{\w}{\w}} &\propto  c_j \ket{\mathcal{M}_{0};\honey{\g}{\g}{\w}{\w}}, \\
    \sigma_{j}^z \ket{\mathcal{M}_{0};\honey{\w}{\w}{\w}{\w}} &\propto  c_j \ket{\mathcal{M}_{0}; \honey{\w}{\g}{\g}{\w}}. \label{eq:sz}
\end{align}
In the graphical notation, we represent $x$ bonds by $(\backslash)$, $y$ bonds by $(/)$, and $z$ bonds by the horizontal link $(-)$, with site $j$ positioned at the left end of the middle horizontal bond. Again, we use proportionality instead of equality, since an additional factor of $i$ arising from the sublattice-dependent definitions of bond fermions, as given in Eq.~\eqref{eq:bondf}, may appear depending on the convention.
We have separated the Majorana fermion $c_j$ from the $Z_2$ gauge field. The state $\ket{\mathcal{M}_0}$ represents the free Majorana sector conditioned on the zero-flux gauge sector. In the gauge sector, flipped fluxes with $\Delta \pi = -1$ are represented by light gray hexagons, whereas the original flux configuration of $\ket{n}$ is depicted by white hexagons. The Bravais lattice index $j$ labels the primitive cell centered on the horizontal link.  
By exploiting the orthogonality between different flux configurations, it follows that the two-point spin correlation in the ground state of the isotropic Kitaev model is given by \cite{Baskaran2007,Lin10,Feng2022}  
\begin{equation} \label{eq:corr}
    \langle \sigma_j^\alpha \sigma_{k}^\beta\rangle  = \frac{\sqrt{3}}{16\pi^2} \int_{\text{BZ}} \frac{A_\mathbf{k}}{\sqrt{A_\mathbf{k}^2+B_\mathbf{k}^2}} d^2\mathbf{k} \,\delta_{\alpha,\beta}  \delta_{\expval{jk},\alpha}
\end{equation}
for $\alpha \in \{x,y,z\}$, where $\delta_{\expval{jk},\alpha}$ selects the nearest-neighbor pair on the $\alpha$ bond; and $A_\mathbf{k}$ and $B_\mathbf{k}$ on the right-hand side are dependent on the compass interaction strength: $A_\mathbf{k} = 2[K_x \cos(\mathbf{k}\cdot \mathbf{n}_1) + K_y \cos(\mathbf{k}\cdot \mathbf{n}_2)+K_z]$, $B_\mathbf{k} = 2[K_x \sin(\mathbf{k}\cdot \mathbf{n}_1) + K_y \sin (\mathbf{k}\cdot \mathbf{n}_2)]$, 
which evaluates to $\pm 0.52$ for the isotropic Kitaev interaction, whose sign depends on the sign of $K$. In the anisotropic limit e.g. $K_z/K_x = K_z/K_y \rightarrow \infty$, $\langle \sigma_j^z \sigma_{j+z}^z\rangle$ evaluates trivially to $\pm 1$, i.e. a dimerized state.

\subsection{Weak field perturbation} \label{sec:weakfield}

Near the isotropic point of the Kitaev compass interactions, the lowest-energy matter sector harbors gapless Dirac modes. Because these Majorana modes are gapless and spatially extended, flux excitations do not bind exponentially localized zero modes in the time-reversal-symmetric model, and their non-Abelian exchange statistics are not sharply separated from the gapless continuum. A finite matter gap is therefore required before the non-Abelian topological character of the weak-pairing phase becomes sharply defined. As discussed in Ref.~\cite{kitaev2006anyons}, a small time-reversal-breaking perturbation generated by a magnetic field can gap out the Dirac cones. A bare Zeeman term,
$H_h=-\sum_{i,a}h_a\sigma_i^a$,
breaks the integrability of the Hamiltonian. However, in leading nonvanishing perturbation theory one can construct an effective time-reversal-breaking term that preserves the conserved-flux structure while gapping out the Dirac modes. At third order, the effect of the Zeeman field is captured by the three-spin interaction
\begin{equation}
    H_{\rm eff}
    =
    H_K
    -
    \lambda
    \sum_{\langle jkl\rangle}
    \sigma_j^x\sigma_k^y\sigma_l^z ,
\end{equation}
with $\lambda\sim h_xh_yh_z/K^2$ for the isotropic model. Here $\langle jkl\rangle$ denotes an ordered two-bond path, so that $j$ and $l$ are next-nearest neighbors connected through the intermediate site $k$. The precise overall sign and numerical prefactor of $\lambda$ depend on the convention used for the spin axes and bond orientations; the essential point is that this term is odd under time reversal and is nonzero only when the field has components along all three spin axes.

\begin{figure}
    \centering
    \includegraphics[width=\linewidth]{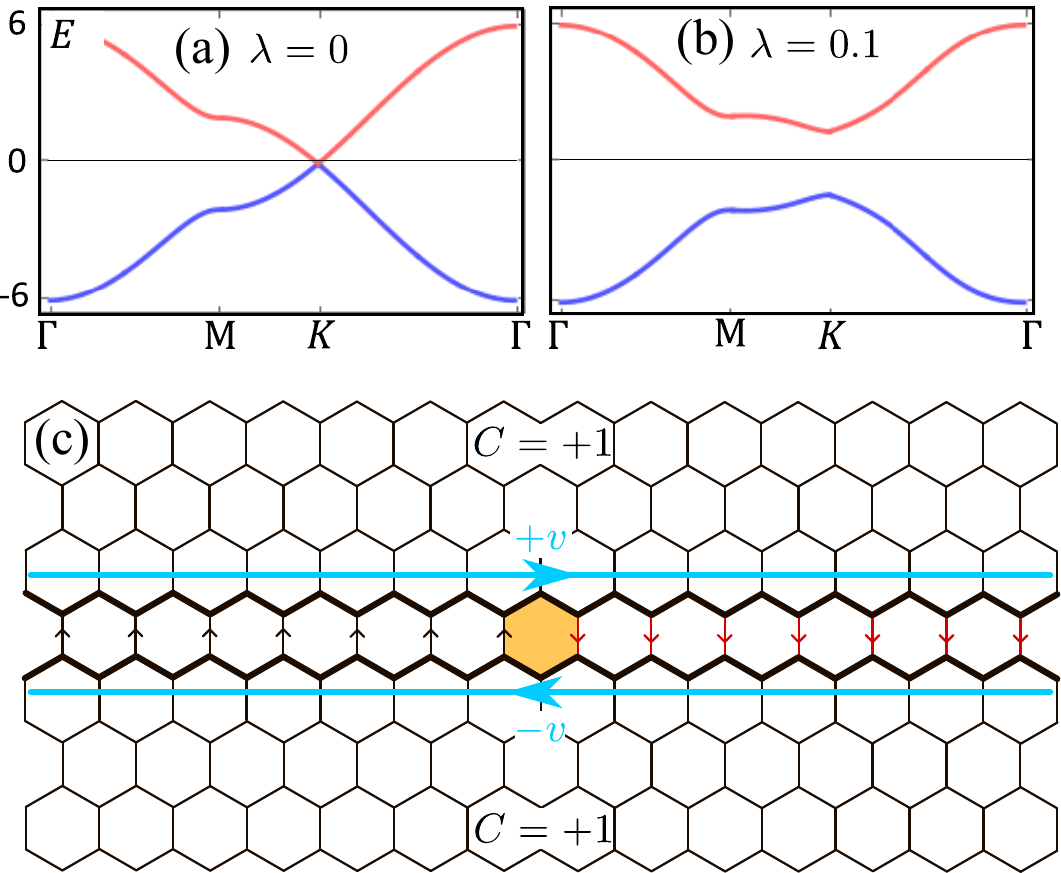}
    \caption{(a) Dirac cone in the absence of perturbation. (b) With $\lambda \sim h_x h_y h_z$ perturbation, the Dirac point is gapped out by $Q(\mathbf{k})$, resulting in a chiral spin liquid (CSL). (c) A single flux (shown in orange) is created in an infinite CSL by
    flipping a line of bond operators starting at the flux and extending horizontally to infinity.
    The system can be rewritten as a pair of Chern insulators which are then
    coupled together using the remaining horizontal line of bonds. Each of the two systems
    contributes an edge mode traveling with velocity $\pm v$ at the boundary, shown in blue. The thick black lines mark out the boundaries of two patches of Chern insulators, whose chiral modes are coupled via $t(x)$ in between the two edges. }
    \label{fig:chern}
\end{figure}

The operator $\sigma_j^x\sigma_k^y\sigma_l^z$ does not change the plaquette fluxes. Hence, in the flux-free sector, at the isotropic point $K_x=K_y=K_z=K$, the effective Majorana Hamiltonian may be written as
\begin{equation}
    H_{\rm ff}
    =
    \frac{i}{4}\sum_{jk} A_{jk} c_j c_k ,
    \;
    A_{jk}
    =
    2K\,u_{jk}
    +
    2\lambda
    \sum_l
    u_{jl}u_{lk}.
\end{equation}
The first term denotes the nearest-neighbor Majorana hopping of the Kitaev model, while the second term denotes the field-induced next-nearest-neighbor Majorana hopping. The sum over $l$ is restricted to the common nearest-neighbor site connecting the next-nearest-neighbor pair $j,k$. This second term is the Majorana analogue of the complex next-nearest-neighbor hopping in Haldane's model~\cite{Haldane88}.

In momentum space, the flux-free Hamiltonian can be written as
\begin{equation}
    H_{\rm ff}
    =
    \frac{1}{2}
    \sum_{\mathbf{q}}
    \begin{pmatrix}
        a_{-\mathbf{q}} & b_{-\mathbf{q}}
    \end{pmatrix}
    \begin{pmatrix}
        Q(\mathbf{q}) & i f(\mathbf{q}) \\
        -i f^*(\mathbf{q}) & -Q(\mathbf{q})
    \end{pmatrix}
    \begin{pmatrix}
        a_{\mathbf{q}} \\
        b_{\mathbf{q}}
    \end{pmatrix}.
\end{equation}
Here $a_{\mathbf{q}}$ and $b_{\mathbf{q}}$ are Fourier transforms of the Majorana operators on the $A$ and $B$ sublattices,
\begin{equation}
    a_{\mathbf{q}}
    =
    \frac{1}{\sqrt{N_c}}
    \sum_j e^{-i\mathbf{q}\cdot\mathbf{r}_j}c_{j,A},
    \;
    b_{\mathbf{q}}
    =
    \frac{1}{\sqrt{N_c}}
    \sum_j e^{-i\mathbf{q}\cdot\mathbf{r}_j}c_{j,B},
\end{equation}
where $N_c$ is the number of unit cells. Up to a convention-dependent phase choice, $f(\mathbf{q})$ is the nearest-neighbor Majorana hopping form factor, while $Q(\mathbf{q})$ is the time-reversal-breaking mass generated by the third-order field perturbation.

The time-reversal-breaking mass term has the form
\begin{equation} \label{eq:qk}
    Q(\mathbf{k})
    \sim
    \lambda
    \left[
    \sin(\mathbf{k}\cdot\mathbf{n}_2)
    -
    \sin(\mathbf{k}\cdot\mathbf{n}_1)
    -
    \sin(\mathbf{k}\cdot\mathbf{n}_3)
    \right],
\end{equation}
where $\mathbf{n}_3=\mathbf{n}_2-\mathbf{n}_1$. This term opens a gap at the Dirac points,
$\Delta = 2|Q(\mathbf{K})|\sim 6\sqrt{3}|\lambda|$,
and gives the Majorana band a Chern number $C=\pm1$, with the sign determined by the sign of the induced chiral term, as shown in Fig.~\ref{fig:chern}(a,b). With a nonzero Chern number, the system acquires gapless chiral Majorana edge modes. In the weak-pairing $p$-wave spinon superconductor~\cite{Read2000}, namely Kitaev's B phase~\cite{kitaev2006anyons}, a single flux binds a Majorana zero mode (MZM) with an exponentially localized spatial profile. These flux-trapped MZMs are directly related to the non-Abelian anyons that underlie topological quantum computation \cite{Read2000,Pachos_2012}. Furthermore, as discussed in Sec.~\ref{sec:mm}, the proliferation and hybridization of such flux-bound MZMs can lead to a novel gapless quantum phase known as a Majorana metal.

To make the discussion self-contained and to prepare for the subsequent analysis of the Majorana metal, here we briefly outline how an MZM becomes localized around a flux excitation in the honeycomb model. Microscopically, this localization can be demonstrated by introducing a flux excitation into the ground-state flux sector by flipping a string of bonds. Strictly speaking, on a closed manifold fluxes are created in pairs; a single flux can be discussed on the infinite plane by sending the partner flux to infinity, or equivalently by attaching the flux to a branch cut that extends to infinity. We choose a simple $Z_2$ gauge configuration consisting of a string of flipped links that originates from the plaquette where the flux excitation resides and extends outward, as shown in Fig.~\ref{fig:chern}(c). This gauge choice conveniently exposes two class-D Chern-superconductor patches joined along the row containing the flipped links.

The effective Hamiltonian is then a tight-binding model of free Majorana fermions with a nonuniform coupling amplitude $t(x)$ along the row of links. The function $t(x)$ changes sign at the link position $x_0$ associated with the flux excitation. We take $t(x)>0$ for $x>x_0$ and $t(x)<0$ for $x<x_0$. Consider first the two patches in the absence of the coupling bonds $t(x)$. At the boundary where the two patches meet, chiral Majorana modes with opposite velocities are present on the two sides of the row, as illustrated in Fig.~\ref{fig:chern}(c). These low-energy modes can be described by a one-dimensional Dirac Hamiltonian,
\begin{equation}
    H_{\rm patch}
    =
    -i\sigma_z v\partial_x ,
\end{equation}
where $v$ is the chiral-mode velocity. Upon joining the two patches by turning on the position-dependent coupling $t(x)$, the two counterpropagating Majorana modes acquire a mass term,
\begin{equation}
    H_{\rm eff}
    =
    -i\sigma_z v\partial_x
    +
    t(x)\sigma_x .
\end{equation}
The existence of a Majorana zero mode follows from solving $H_{\rm eff}\ket{\psi}=0$. For the sign-changing mass profile described above, the normalizable zero-mode wave function is
\begin{equation}
    \ket{\psi_0(x)}
    =
    \mathcal{N}
    \exp\!\left[
    -\frac{1}{v}
    \int_{x_0}^{x} t(x')\,dx'
    \right]
    \chi_-,
\end{equation}
with $\sigma_y\chi_-=-\chi_-$. 
For example, one may choose
\begin{equation}
    \chi_- =
    \frac{1}{\sqrt{2}}
    \begin{pmatrix}
        1 \\ -i
    \end{pmatrix},
\end{equation}
up to a convention-dependent phase and orientation of the branch cut. Thus an exponentially localized Majorana zero mode appears at $x_0$, the location of the $\pi$ flux. Its localization length is set by $v/|t|$, or equivalently by the inverse bulk gap of the chiral Majorana phase.
At the level of the continuum Dirac/Majorana equation, these flux-bound zero modes are closely analogous to zero modes bound to vortices or mass defects in Jackiw-Rossi-type problems and in Kekul\'e defects of graphene-like Dirac systems~\cite{JackiwRossi1981,Ryu2009Masses,Hou2007}. 
However, in the Kitaev spin liquid the defect is not an externally imposed lattice distortion or superconducting vortex, but a $Z_2$ gauge flux/vison of a fractionalized spin system.

\subsection{Spectrum of fractionalized excitations} \label{sec:spec}
Dynamical response is a very useful tool in reflecting fractionalization: a local spin flip can decay into emergent neutral quasiparticles, producing broad continua whose thresholds and line shapes encode the low-energy spectrum \cite{Frost13,Mourigal2013}. For instance, the dynamical spin structure factor directly tests how a single spin flip fractionalizes, while bond (dimer) correlators can sometimes bypass selection rules and more cleanly track the dispersing matter or gauge sector. In practice, these channels are accessed by neutrons (one-spin) and RIXS/Raman/THz (bond/dimer), providing mutually constraining views of the same physics.

In the Kitaev honeycomb model, the emergent quasiparticles are fluxes (visons) and Majorana fermions. A one-spin flip creates a pair of fluxes plus Majorana fermions, yielding a broad continuum with an onset at the flux gap in the spin spectral function; in the integrable limit, flux-sector orthogonality enforces that only nearest-neighbor spin correlations survive \cite{Baskaran2007}, making this channel characteristic of the gauge structure in the pure Kitaev QSL. By contrast, the dimer operator can be diagonal in the flux sector, so it is able to probe the Majorana dispersion without creating fluxes by avoiding mixing the gauge and fermion degrees of freedom.
Using the integrability of the Kitaev honeycomb model, the following text explicates these channels characteristic for the pure Kitaev QSL.

\begin{figure}
    \centering
    \includegraphics[width=\linewidth]{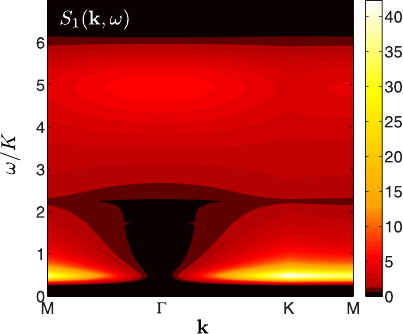}
    \caption{The zero-temperature dynamical spin structure factor $S_1(\mathbf{k}, \omega)$ for the antiferromagnetic isotropic Kitaev honeycomb model ($K_x = K_y = K_z = K > 0$) along the $\rm M\Gamma K M$ path in the Brillouin zone. Figure adapted from Ref.~\cite{Knolle15}. }
    \label{fig:s1}
\end{figure}
\subsubsection{Spin spectrum}
Computing dynamical properties of spin liquids is of particular importance because they contain information about fractionalized quasiparticles, namely the itinerant Majorana fermions and the emergent $Z_2$ gauge fluxes, and are directly relevant to inelastic neutron scattering experiments. 
In the Kitaev model this connection is especially sharp: a neutron spin flip is a local spin operator, and a local spin operator fractionalizes into a matter Majorana operator together with a local change of the $Z_2$ gauge background. Thus the spin spectral function does not simply measure the free Majorana band; rather, it measures a composite dynamical process involving both Majorana fermions and flux excitations.

Given the integrability of the model and the aforementioned orthogonality between different flux configurations, the two-point spin spectral function can nevertheless be solved exactly. After the seminal work~\cite{Baskaran2007} by Baskaran \textit{et al.} on the static spin correlations, Knolle \textit{et al.}~\cite{Knolle2014,Knolle15} analytically solved the dynamical spin-spin correlation function in the integrable limit. The formalism for the time-dependent correlation is greatly simplified by flux-sector orthogonality. Using Eq.~\eqref{eq:orth}, static fluxes guarantee that spin correlations vanish beyond nearest-neighbor distance. More precisely, apart from the on-site autocorrelation, the only nonzero off-site correlators are those between the same spin component on the corresponding $\alpha$ bond. This short-ranged real-space structure is not a consequence of the range of Majorana correlation; instead, it follows from the fact that two spin operators must create and then annihilate the same pair of fluxes in order to return to the original gauge sector.

For a nearest-neighbor pair connected by an $\alpha$ bond, one obtains
\begin{equation} \label{eq:knolle}
\begin{aligned}
C^\alpha_{j,j+\alpha}(t)
&\equiv
\expval*{\sigma_j^\alpha(t)\sigma_{j+\alpha}^\alpha(0)}
\\
&=
-i
\mel{\mathcal{M}_0}
{e^{iH_0t}c_j e^{-i(H_0+V_\alpha)t}c_{j+\alpha}}
{\mathcal{M}_0} .
\end{aligned}
\end{equation}
Here $j+\alpha$ denotes the nearest neighbor of site $j$ along the $\alpha$ bond, equivalently selected by $\delta_{\langle jk\rangle,\alpha}$. The Hamiltonian $H_0$ is the flux-free Majorana Hamiltonian defined in Eq.~\eqref{eq:Kitaev_Majorana}, conditioned on the link variables $u_{ij}$ in the flux-free gauge sector. The operator $V_\alpha$ is a local scattering potential generated by flipping the sign of the $\alpha$-bond variable touched by the spin operator. Equation~\eqref{eq:knolle} therefore has the structure of a local quantum-quench problem: before the spin flip the Majoranas evolve with $H_0$, while after the spin flip they evolve in a background containing an adjacent flux pair. This is the origin of the broad continuum in the spin response, even though the Majorana sector itself is free within each fixed gauge sector.

The total dynamical spin structure factor is then
\begin{equation} \label{eq:s1}
    S_1(\mathbf{k}, \omega)
    =
    \sum_\alpha
    \int dt\, e^{-i\omega t}
    \expval{\sigma_{\mathbf{k}}^\alpha(t)
    \sigma_{-\mathbf{k}}^\alpha(0)} ,
\end{equation}
which can be computed from Eq.~\eqref{eq:knolle}. For the pure isotropic honeycomb model, the resulting zero-temperature spectrum is shown in Fig.~\ref{fig:s1}. The Majorana continuum in the presence of a flux pair extends up to $\omega/K\simeq 6$, set by the bandwidth of the Majorana band. Notably, despite the gapless Dirac points in the flux-free Majorana sector, the spin spectrum is gapped because a spin flip necessarily creates a pair of gapped fluxes. The onset of spectral weight in $S_1(\mathbf{k},\omega)$ therefore gives direct access to the flux-pair gap, while the higher-energy broad continuum reflects the propagation of Majorana fermions in the locally quenched flux background. In this sense, neutron scattering is an excellent probe of fractionalization in the Kitaev model, but it does not directly reveal the gapless Dirac cones of the Majorana matter sector. To observe those more directly, one should instead consider flux-neutral probes such as dimer or Raman-type response functions, as discussed below.

\begin{figure*}
    \centering
    \includegraphics[width=\linewidth]{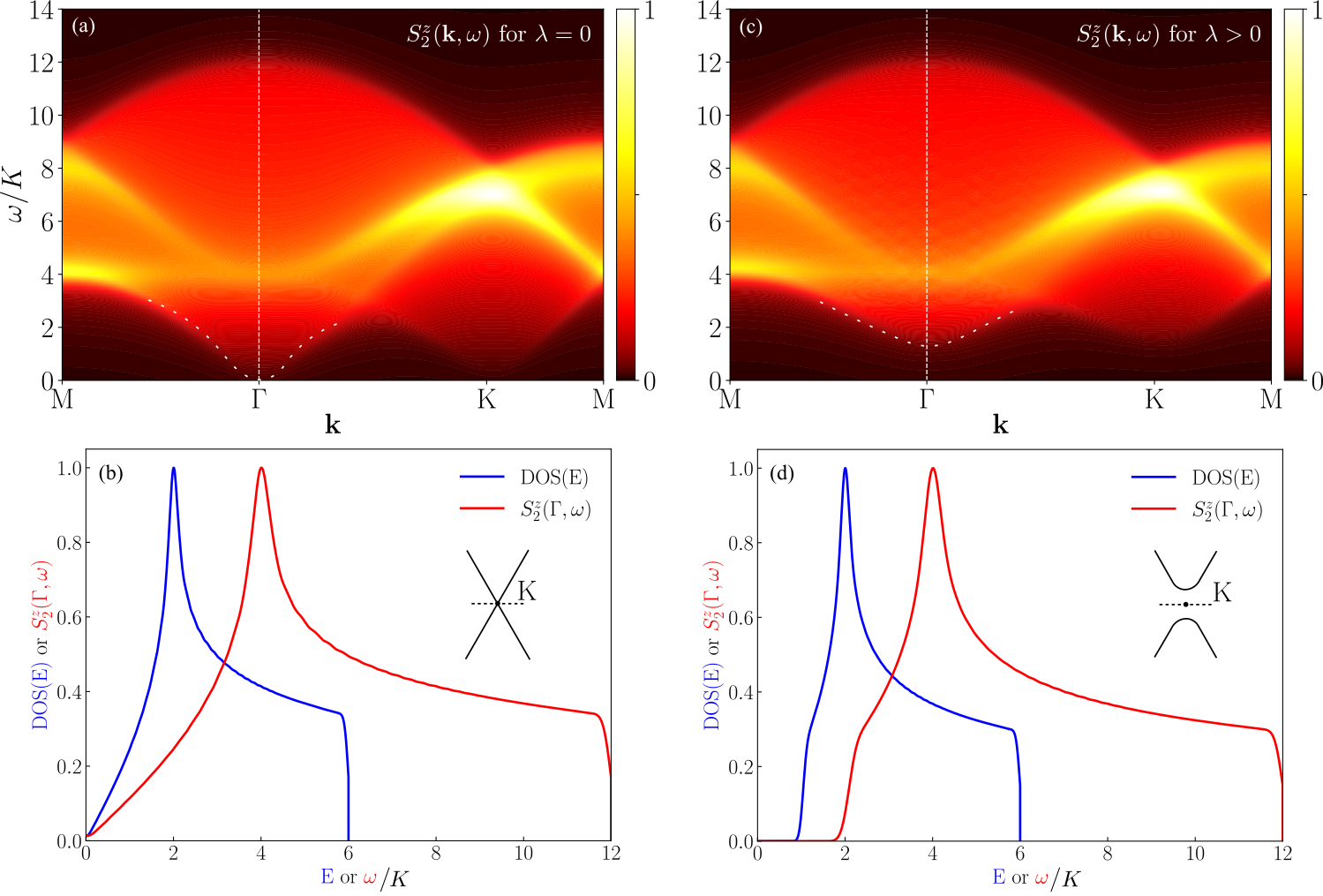}
    \caption{
    Zero-temperature dynamical spin structure factor $S_2(\mathbf{k}, \omega)$ for the isotropic antiferromagnetic Kitaev model. (a) The normalized momentum-resolved spectrum $S_2^z(\mathbf{k}, \omega)$ in the time-reversal symmetric case ($\lambda = 0$). The dotted white line highlights the gapless spectrum at the $\Gamma$ point. (b) The normalized equal-momentum cut at the $\Gamma$ point, $S_2^z(\Gamma, \omega)$, indicated by the vertical dashed white line in panel (a), is plotted as a red solid line. The blue solid line shows the corresponding density of states (DOS) of the Dirac Majorana fermions. Both intensities are rescaled to emphasize the exact correspondence between the spin and Majorana spectra at low energies, that $S_2^z(\Gamma, \omega)$ directly reflects the Majorana DOS up to a scaling factor in energy. (c) $S_2^z(\mathbf{k}, \omega)$ with time-reversal symmetry explicitly broken ($\lambda > 0$). The dotted white line indicates the opening of a finite spectral gap at the $\Gamma$ point due to the TR-breaking perturbation. (d) The rescaled Majorana DOS (blue line) and spin structure factor $S_2^z(\Gamma, \omega)$ (red line) for the TR-broken case.  The similar profiles of DOS and $S_2^z(\Gamma, \omega)$ show that the Majorana DOS can be directly observed from the dimer spectral function, up to a scaling factor of $2$ in energy. 
    }
    \label{fig:s2z}
\end{figure*}

\subsubsection{Dimer spectrum}
In the spin dynamical structure factor discussed above, the gapless Dirac points associated with the Majorana bands in the Kitaev honeycomb model are generally not directly observable, due to the orthogonality between different flux sectors, which introduces finite-energy excitations into the spin spectral function. Nevertheless, a clearer signature of the gapless Majorana fermions may instead be revealed through higher-order correlation functions, such as the two-spin (dimer-dimer) dynamical structure factor. Indeed, this higher-order correlation captures processes where flux sectors remain unchanged or return to their initial state, thus directly probing the matter Majorana spectrum. Such dynamical correlations can be directly measured in experiments like resonant inelastic x-ray scattering (RIXS) and Raman scattering, offering a more accessible window into the fractionalized excitations of the Kitaev quantum spin liquid.

To appreciate this, define the dimer operator $\mathcal{D}_j^\alpha = \sigma_j^\alpha \sigma_{j+\alpha}^\alpha$ for $\alpha \in \{x,y,z\}$. 
The operator $\mathcal{D}_j^\alpha$ is diagonal in all flux patterns; hence, the corresponding dynamics are governed exclusively by the matter Majorana sector. For a dimer operator defined on the $z$-bond, , in contrast to Eq.~\eqref{eq:sz}, we explicitly have:
\begin{align}
\mathcal{D}_j^z \ket{\mathcal{M}_{0};\honey{\w}{\w}{\w}{\w}} = i c_j c_{j+z} \ket{\mathcal{M}_{0};\honey{\w}{\w}{\w}{\w}} \label{eq:dxflux1}
\end{align}
where $j$ labels the site on the $A$ sublattice at one end of the dimer.
The corresponding dynamical dimer structure factor is defined as:
\begin{equation}
    S_2^{\alpha}(\mathbf{k}, \omega) = \int dt\, e^{-i\omega t} \expval{\mathcal{D}_\mathbf{k}^{\alpha}(t)\,\mathcal{D}_{-\mathbf{k}}^{\alpha}},
\end{equation}
which, in contrast to spin-spin correlations, does not suffer from orthogonality between gauge sectors, as it remains entirely within a single flux configuration.

Since the dimer operator defined on the $z$ bonds does not excite fluxes, the spectral function is governed solely by the matter (Majorana) sector. In the flux-free ground state, this quantity can thus be computed exactly from the free-fermion sector, e.g.  for $\alpha = z$ \cite{wang2024}:
\begin{equation} \label{eq:s2z}
\begin{split}
        S_2^{z} (\mathbf{k},\omega) = \frac{\sqrt{3}}{16 \pi^2} \int_{\rm BZ} W(\mathbf{k}-\mathbf{q}) \delta(\omega - \varepsilon_{\mathbf{k},\mathbf{q}}) d^2\mathbf{q}
\end{split} 
\end{equation}
where the energy of a complex-fermion pair is given by $\varepsilon_{\mathbf{k},\mathbf{q}} \equiv E_{\mathbf{k} - \mathbf{q}} + E_{\mathbf{q}}$, and
 the spectral weight factor is:
\begin{equation}
    W(\mathbf{k}-\mathbf{q}) = \frac{E_{\mathbf{k}-\mathbf{q}}^2}{E_{\mathbf{k}-\mathbf{q}}^2 - Q_{\mathbf{k}-\mathbf{q}}^2}.
    \label{eq:weight}
\end{equation}
Here, $E_\mathbf{k}$ denotes the positive-energy Majorana fermion band of the flux-free Hamiltonian, and $Q_\mathbf{k}$ arises from the next-nearest-neighbor hopping terms induced by time-reversal symmetry breaking, as defined in Eq.~\eqref{eq:qk}. We emphasize that unlike the two-spin correlations, which are influenced by gauge-flux excitations and thus mask the low-energy Majorana modes, the dimer correlation function in Eq.~\eqref{eq:s2z} directly reveals the gapless Majorana dispersion without contamination from flux sectors. This makes the dimer dynamical structure factor a particularly effective probe for observing Majorana bands experimentally.

Figure~\ref{fig:s2z}(a,c) presents the momentum- and energy-resolved dimer structure factor $S_2^z(\mathbf{k}, \omega)$ for the cases of TR symmetry-preserving (a) and TR symmetry-breaking (c) perturbations induced by next-nearest-neighbor hopping. In the absence of TR-breaking perturbation (a), the Dirac cones of the gapless Majorana fermions lead to a continuum spectral feature, characterized by a gapless excitation at the $\Gamma$ point. 
Such a gapless two-Majorana continuum has also been numerically demonstrated by PEPS calculation, see \cite{Mei24,wang2024}.
In contrast, introducing TR symmetry-breaking terms opens a gap in the Majorana fermion spectrum, which clearly manifests as a gap at the $\Gamma$ point in the corresponding dimer spectral function, as shown in Fig.~\ref{fig:s2z}(c). 
The Majorana continuum decays rapidly to zero at $\omega/K = 12$, which is twice the band width of Majorana fermions, in contrast to Fig.~\ref{fig:s1}. 
In addition, remarkably, the dimer correlation function directly reflects the density of states (DOS) of the underlying Majorana fermions. Evaluating Eq.~\eqref{eq:s2z} at zero momentum, we obtain:
\begin{equation}
    S_2^z(\Gamma, \omega) \sim \int_{\rm BZ} d^2\mathbf{q}\, \delta(\omega - 2E_{\mathbf{q}}),
\end{equation}
where we have used $Q_{\mathbf{k}=0}=0$. This demonstrates that the dimer spectral function at the $\Gamma$ point exactly reproduces the Majorana fermion DOS, scaled by a factor of two in energy, i.e. $S_2^z(\Gamma, E) \propto {\rm DOS}(E/2)$, as shown in Fig.~\ref{fig:s2z}(b,d). 
Indeed, in the absence of perturbation, the gapless Dirac QSL is manifested by the real-space correlation between dimers, which decays as a power law according to $\sim |\mathbf{r}_i-\mathbf{r}_j|^{-4}$. 
Thus, in theory, the $Z_2$-gauge-invariant dimer correlations, relevant for scattering probes such as Raman or RIXS, could provide a direct route to demonstrate the otherwise mixed signals of Majorana and flux excitations in Kitaev spin liquids.

\section{Field-induced phases and phase diagrams} \label{sec:diagram}
In this section, we present an overview of recent progress concerning the phase diagram associated with the Kitaev honeycomb model under an external magnetic field. We summarize key developments, clarify existing controversies, and discuss updated interpretations that have emerged in light of recent numerical and theoretical insights. A detailed analysis of the distinct phases and their characteristic properties will be provided in subsequent sections.
\subsection{Field induced phases in pure Kitaev model}
\begin{figure*}
    \centering
    \includegraphics[width=0.9\linewidth]{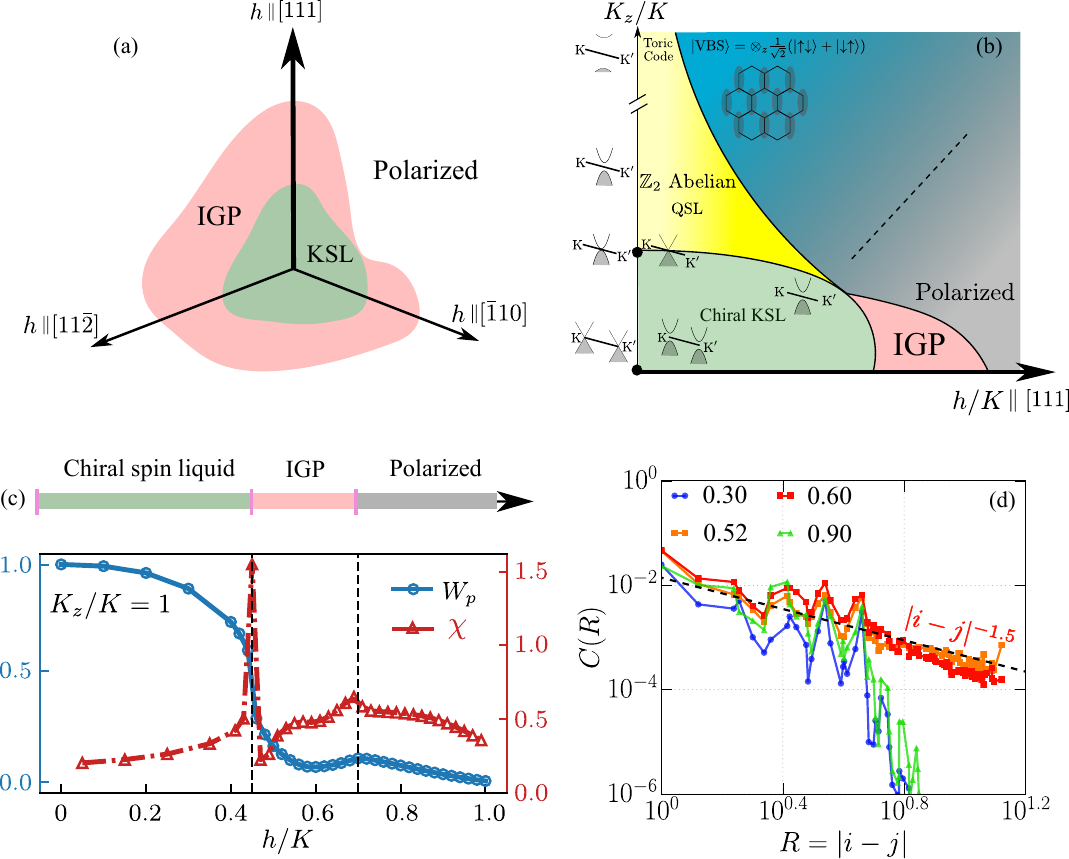}
    \caption{
        Phase diagrams of the antiferromagnetic Kitaev honeycomb model under uniform magnetic fields. (a)~The isotropic limit ($K_x = K_y = K_z = K$) under magnetic fields along various directions. An intermediate gapless phase (IGP, highlighted in pink) emerges between the Kitaev spin liquid (KSL) and the polarized phase. (b) Phase diagram under a [111] magnetic field, parametrized by the anisotropy ratio $K_z/K$ and field strength $h$. In the large $K_z/K$, small-$h$ regime, the system hosts a $Z_2$ Abelian QSL, whose ground state is equivalent to the toric code (TC). The gradient in yellow represents the increasing hybridization between gauge fluxes and matter Majorana fermions. At lower anisotropy, as the magnetic field increases, the system transitions from the chiral QSL (CSL) into the IGP (pink region) before reaching the partially polarized phase (gray). Insets illustrate representative Majorana fermion bands for different anisotropy and magnetic fields in the perturbative limit. Panel (b) is adapted from Ref.~\cite{Feng2023}. (c) Magnetic susceptibility $\chi$ (red line, right axis) and flux expectation value $W_p$ (blue line, left axis) as a function of the [111] magnetic field strength, obtained using iPEPS simulations with bond dimension $D=5$. Two transitions are indicated by sharp changes at $h_{c1}\approx 0.35$ (from CSL to IGP) and $h_{c2}\approx 0.70$ (IGP to the partially polarized phase). Figure adapted from Ref.~\cite{wang2024}. (d) Real-space spin correlation function $C(R)$ plotted versus distance $R$ on a 160-site cylinder obtained via DMRG. While the correlation decays exponentially in both CSL and the polarized phase, it decays algebraically in the IGP, consistent with a gapless Majorana metal to be discussed in Sec.~\ref{sec:mm}. Figure adapted from Ref.~\cite{Patel12199}.
        }
    \label{fig:phasesall}
\end{figure*}

A significant body of recent theoretical and numerical work has explored the intricate behavior of the Kitaev honeycomb model under an external magnetic field. Consider the anisotropic Kitaev Hamiltonian defined on a honeycomb lattice. 
For ferromagnetic Kitaev couplings ($K_\alpha < 0$), the QSL phase is generally susceptible to non-Kitaev terms and transitions directly into a partially polarized phase under moderate external fields, regardless of field orientation \cite{hickey2019emergence}. In contrast, for antiferromagnetic exchange ($K_\alpha > 0$), the situation becomes far richer. For example, a moderate magnetic field can induce an intermediate magnetically disordered phase between the spin liquid and the polarized state, as illustrated in Fig.~\ref{fig:phasesall}(a,b), exhibiting distinct features depending sensitively on the field direction and interaction anisotropies. These intermediate states are frequently proposed as emergent quantum spin liquids with intriguing fractionalized excitations, such as spinon Fermi surface and kinetically constrained excitations, which may survive even under the presence of non-Kitaev interactions \cite{Erik21}, though, their exact nature remains actively debated due to the accessible scale of numerics.

These theoretical considerations are directly relevant to many Kitaev material candidates, most of which require an external magnetic field to suppress their native magnetic orders and reveal potential QSL behavior. Although many putative Kitaev materials are thought to host ferromagnetic Kitaev interactions—an assertion still debated due to the complexity of fitting experimental results—several promising materials exhibiting antiferromagnetic Kitaev interactions have emerged recently. Examples include Na$_2$Co$_2$TeO$_6$, which has been reported to host a field-induced magnetically disordered state~\cite{Gaoting24}; the honeycomb spin-1 magnet Na$_3$Co$_2$SbO$_6$; the spin-1 material Na$_3$Ni$_2$SbO$_6$~\cite{Shangguan2023}; and the spin-$\frac{1}{2}$ candidate YbOCl~\cite{YbOCl2022,YbOCl2024}. 

Furthermore, recent scanning tunneling microscopy (STM) experiments on monolayer $\alpha$-RuCl$_3$ have reported evidence of field-tunable Friedel-type oscillations attributed to an emergent neutral Fermi surface~\cite{Kohsaka24}. Intriguingly, the periodicity of these real-space oscillations varies with the external magnetic field, suggesting a field-tunable Fermi momentum associated with an emergent Majorana or (complex) spinon Fermi surface. These fascinating experimental developments demand a deeper theoretical investigation into the existence, mechanisms, and characterization of QSL phases stabilized by external magnetic fields, and particularly, the fractionalized excitations and their unique experimental signatures in Kitaev systems.

A natural starting point to discuss the intricate phenomena emerging from spin-orbit coupled quantum magnets is the Kitaev honeycomb model under an external magnetic field:
\begin{equation}
    H = \sum_{\langle ij\rangle_\alpha} K_\alpha \sigma^\alpha_i \sigma^\alpha_j - \mathbf{h}\cdot\sum_{i}\boldsymbol{\sigma}_i,
    \label{eq_ham}
\end{equation}
This model is known to host a rich phase diagram, including diverse spin-liquid phases and potentially even gapless states characterized by an emergent Fermi surface. Crucially, the exact integrability of the pure Kitaev model provides an ideal theoretical anchor, enabling controlled perturbative analysis as well as guiding insights into the more exotic non-integrable regimes that emerge under magnetic fields.

Beyond the integrable limit, however, the theoretical and numerical exploration becomes significantly more challenging. In the presence of a magnetic field, the Majorana fermion representation loses integrability, complicating analytical treatments. Numerically, identifying distinct phases using exact diagonalization (ED) or density-matrix renormalization group (DMRG) methods in two dimensions becomes particularly challenging due to finite system sizes, fast entanglement growth, limited momentum resolution, boundary effects, and the convergence issues of iDMRG in long-range correlated states \cite{Erik21}. Specifically, distinguishing gapped from gapless QSL phases can be particularly problematic, as correlation lengths in gapless states often exceed accessible cylinder widths leading to ambiguous or spurious power-law correlations, and the time scale accessible in time evolution confines the scope of numerics to finite energy and thus cannot directly access the late-time information relating to potential neutral Fermi surfaces.

Moreover, various microscopic self-consistent mean-field theories have produced inconsistent results, both among themselves \cite{Jiang2020,ZhangNatComm2022,Will24,Yuanming2018}, due to varying underlying ansatz choices, and with numerical benchmarks \cite{David2019,hickey2019emergence,Patel12199}. The validity of mean-field approaches in Kitaev systems is further complicated by the potential emergence of intrinsic disorder even in translationally symmetric systems with emergent gauge theories \cite{Knolle17,Zhai20}. Such emergent disorder is known to arise naturally in gauge systems, including the $Z_2$ Kitaev honeycomb model, yet poses fundamental difficulties for mean-field treatments~\cite{Fratini21}. 

These theoretical and numerical challenges simultaneously offer fertile ground for further exploration, especially when viewed in the context of rapidly evolving experimental evidence for promising fractionalized physics. As evidence from different perspectives accumulates and numerical techniques advance, such as two-dimensional projected entangled pair states (PEPS) and differentiable programming tensor networks~\cite{Liao19}, the subtleties and controversies in this active research of Eq.~\eqref{eq_ham} and its extensions gradually become clearer.

Figure~\ref{fig:phasesall} summarizes recent results on the phases and their key signatures for the Hamiltonian given in Eq.~\eqref{eq_ham}. Panel (a) shows the schematic phase diagram of the antiferromagnetic Kitaev honeycomb model under external magnetic fields applied in various directions. At weak magnetic fields, perturbation theory predicts distinct field-induced responses of the spin liquid due to the isotropic Kitaev interaction, which are highly dependent on field direction and model anisotropy. However, our primary interest lies in the intermediate gapless phase (IGP) that emerges under moderate fields, outside the perturbative regime; and the phase under large anisotropy. In contrast to the ferromagnetic Kitaev model, which transitions directly into a polarized phase at sufficiently strong fields, the antiferromagnetic model ($K_\alpha>0$) hosts the IGP that is robust across a wide range of field directions. Nevertheless, in presence of other non-Kitaev interactions in realistic materials, it has been reported that a putative gapless QSL phase exists for $K_\alpha<0$ and under strong magnetic field \cite{Han21}. 
Under anisotropic $K_\alpha>0$, as shown in Fig.~\ref{fig:phasesall}(b), and the Abelian QSL also persists for a much larger range of magnetic field than that under the ferromagnetic anisotropic coupling. In the following, we focus primarily on reviewing these intriguing field-induced regimes, discussing its nontrivial nature and diagnostic experimental signatures under moderate field and/or large anisotropy. The detailed properties and characterizations of these phases will be discussed extensively in Sec.~\ref{sec:mm} and Sec.~\ref{sec:dim}. 
Here, we present key diagnostic results obtained via state-of-the-art numerical methods, including DMRG and infinite projected entangled-pair states (iPEPS), that establish the existence of the emergent quantum phases, clearly distinguishing them from the adjacent quantum spin-liquid and the partially polarized phase.

\begin{table*}[t] 
\begingroup
\footnotesize
\setlength{\tabcolsep}{4pt}
\renewcommand{\arraystretch}{1.12}
\begin{tabular}{p{0.22\textwidth}p{0.22\textwidth}p{0.12\textwidth}p{0.38\textwidth}}
\hline
Reference & Method & Gap or Gapless & Remark on the field-induced intermediate phase \\ \hline\hline
Z. Zhu {\it et al.}~\cite{Zhu_PRB_2018} & ED \& DMRG & Gapless & \\ \hline
M. Gohlke, {\it et al.}~\cite{gohlke2018dynamical} & iDMRG & Gapless & \\ \hline
D. C. Ronquillo {\it et al.}~\cite{David2019} & ED & Gapless & \\ \hline 
D. Kaib {\it et al.}~\cite{Roser19} & ED & Gapless & Occurs via proliferation of fluxes \\ \hline
C. Hickey and S. Trebst \cite{hickey2019emergence} & ED & Gapless & U(1) spinon Fermi surface \\ \hline
N. Patel and N. Trivedi \cite{Patel12199} & ED \& DMRG & Gapless & U(1) spinon Fermi surface \\ \hline
Y.-F. Jiang {\it et al.}~\cite{Jiang2019} & DMRG & Gapless & U(1) spinon Fermi surface \\ \hline
H.-C. Jiang {\it et al.}~\cite{Yuanming2018} & ED \& MFT & Gapless & U(1) spinon Fermi surface \\ \hline
M.-H. Jiang, {\it et al.}~\cite{Jiang2020} & VQMC & Gapped & Parton Chern insulator \\ \hline
S.-S. Zhang {\it et al.}~\cite{ZhangNatComm2022} & MFT & Gapped & Parton Chern insulator \\ \hline
W. Holdhusen {\it et al.}~\cite{Will24} & Hierarchical MFT & Gapless & Finite spin scalar chirality \\ \hline
K. B. Yogendra {\it et al.}~\cite{Baskaran2023} & DMRG & Gapless & High density of glassy $Z_2$ fluxes \\ \hline
K. Wang {\it et al.} \cite{wang2024} & iPEPS \& ED & Gapless & $Z_2$ Majorana metal with glassy flux \\ \hline  
P. Zhu {\it et al.} \cite{penghao24} & iPEPS \& Effective Theory & Gapless & $Z_2$ Majorana metal with glassy flux \\ \hline  
S. Feng {\it et al.} \cite{Feng2025} & DMRG \& Effective Theory & Gapless & Transiently localized fermions with slow fluxes \\ \hline
\end{tabular}
\caption{A non-exhaustive summary of recent numerical and theoretical results regarding the intermediate gapless phase (IGP) of the spin-$\frac{1}{2}$ antiferromagnetic Kitaev honeycomb model under moderate magnetic fields. Various numerical methods, including ED, DMRG, different MFTs, iPEPS, and variational quantum Monte Carlo (VQMC), are compared, highlighting their conclusions on the gapped versus gapless nature of the phase and the corresponding effective theories proposed.}
\label{tab:tab1}
\endgroup
\end{table*}

We first describe the field-induced phases of the Kitaev honeycomb model in the strongly anisotropic limit, which we parametrize by $K_z/K >2$, with $K_x = K_y = K$. At small magnetic fields applied along the [111] direction, the Abelian $Z_2$ quantum spin liquid phase emerges, as indicated by the yellow region in Fig.~\ref{fig:phasesall}(b). Upon increasing the magnetic field, a first-order phase transition occurs into a dimerized valence bond solid (VBS) phase, represented by the blue region in Fig.~\ref{fig:phasesall}(b). The critical magnetic field at which this transition occurs scales with the anisotropy parameter as $h_c \propto K^2/K_z$, a result that will be discussed in detail in Sec.~\ref{sec:khmy}.

In the strongly anisotropic limit, the ground state at high magnetic fields is characterized by short-range dimerization on the strong ($z$) bonds, forming a valence bond solid (VBS) or dimer-triplet state:
\begin{equation}
    |\text{VBS}\rangle = \bigotimes_{z-\text{bonds}} \frac{1}{\sqrt{2}}\left(\ket{\uparrow\downarrow} + \ket{\downarrow\uparrow}\right),
    \label{eq:nematic}
\end{equation}
as illustrated schematically in the inset of Fig.~\ref{fig:phasesall}(b). This VBS state, a short-range entangled product state, emerges naturally as a result of gauge confinement induced by the strong magnetic field, and connects smoothly to the partially polarized (PP) phase at lower anisotropy.
The nature of this dimerized valence-bond state can be characterized by the quadrupolar nematic operator, which is sensitive to spin dimerization:
\begin{equation}
    \hat{Q}_{pp'}^{\alpha\beta} = \frac{\sigma_p^\alpha \sigma_{p'}^\beta + \sigma_p^\beta \sigma_{p'}^\alpha}{2} - \frac{\delta_{\alpha\beta}}{3}\boldsymbol{\sigma}_p \cdot \boldsymbol{\sigma}_{p'},
    \label{eq:nem_para}
\end{equation}
where $pp'$ denotes a $z$-bond and $\alpha,\beta \in \{x,y,z\}$. 
For the triplet state defined in Eq.~\eqref{eq:nematic}, this produces a diagonal matrix of $\hat{Q}_{xx} = \hat{Q}_{yy} = \frac{2}{3}$ and $\hat{Q}_{zz} = -\frac{4}{3}$. 
Using exact diagonalization (ED), Ref.~\cite{Feng2023} confirmed numerically that this VBS state indeed emerges clearly as a short-range correlated dimer-triplet state, exhibiting pronounced quadrupolar signatures.
At slightly smaller anisotropy and lower magnetic fields, the model remains in the Abelian QSL regime (yellow region), whose ground state is equivalent to that of the TC. Increasing the magnetic field in this regime eventually induces partial hybridization between gauge fluxes and Majorana fermions, resulting in emergent, partially mobile excitations exhibiting sub-dimensional dynamics. The interplay between gauge fields and matter Majorana fermions at finite fields in this region of the phase diagram will be extensively discussed in Sec.~\ref{sec:dim}.

At lower anisotropy, particularly around $K_z/K < 2$, the system hosts a non-Abelian chiral spin liquid (CSL) phase characterized by bulk Majorana bands whose minima (band edges) occur at the $\rm M$ points of the Brillouin zone. As the anisotropy further decreases toward the isotropic limit ($K_z/K \sim 1$), the Majorana band edges shift to the $\rm K$ and $K'$ points. However, the most intriguing physics emerges near this isotropic limit under a moderate magnetic field, where an intermediate gapless phase (IGP) arises.

To establish the fundamental properties and existence of the IGP, we first examine the magnetic susceptibility, defined as $\chi_h = \partial^2 E_{\text{gs}}(K_z,h)/\partial h^2$, 
as well as the flux expectation value $\expval{W_p}$, which indicates the density of $Z_2$ gauge fluxes. Figure~\ref{fig:phasesall}(c) presents iPEPS results obtained in the thermodynamic limit. The susceptibility $\chi_h$ and the derivative of $\expval{W_p}$ exhibit pronounced peaks at $h_{c1}\simeq 0.45$ and $h_{c2}\simeq 0.70$, clearly marking two distinct phase transitions: one from CSL into IGP and another from IGP into the polarized phase. Specifically, for fields below $h_{c1}$, the system behaves perturbatively, with Majorana fermions becoming gapped and acquiring a nonzero Chern number, while flux fluctuations remain moderate. However, within the intermediate field regime ($h_{c1} < h < h_{c2}$), perturbation theory breaks down due to strong gauge fluctuations, causing a rapid decrease in $\expval{W_p}$ as the system enters the IGP. Finally, at fields exceeding $h_{c2}$, the system becomes fully polarized.
These critical fields ($h_{c1}$ and $h_{c2}$) closely match values reported by earlier finite-size numerical studies~\cite{gohlke2018dynamical,hickey2019emergence,David2019,Patel12199,Jahromi2024,Hickey20}. Further evidence distinguishing the IGP from adjacent phases comes from analyzing real-space spin-spin correlations. Figure~\ref{fig:phasesall}(d) shows the correlation function $C(R)$ computed via DMRG on a finite cylinder geometry. In the intermediate gapless phase, correlations exhibit an approximate power-law decay, $C(R)\propto R^{-m}$, with $m\approx1.5$, as indicated by the dashed black line. Such algebraic decay strongly suggests a gapless spectrum, although finite-size effects in DMRG calculations can complicate a definitive interpretation due to potential correlation lengths exceeding the cylinder circumference, and different theories may be constructed to account for results from different numerical or mean-field methods. In Table~\ref{tab:tab1} we list a few recent results with different methods and claims. This controversy can be addressed due to the limitation in finite-size numerics and self-consistent mean field theories by employing iPEPS calculations, which directly access the thermodynamic limit~\cite{wang2024}.
In contrast, within the neighboring phases, spin correlations clearly decay exponentially. In the CSL phase (e.g., at $h = 0.30$), the exponential decay reflects a finite excitation gap consistent with the $p+ip$ chiral spinon superconductor state emerging from the Majorana sector under third-order perturbation theory~\cite{kitaev2006anyons}. Similarly, in the polarized phase (e.g., at $h=0.90$), exponential decay arises due to the high-energy cost ($\sim h$) associated with flipping individual spins.

As presented in Table~\ref{tab:tab1}, the studies listed there, although non-exhaustive, broadly fall into three major categories regarding the nature of the field-induced IGP near the isotropic limit of Kitaev model: (a) a gapped parton Chern insulator, derived via fermionic parton decomposition and self-consistent mean-field theory; (b) a gapless $U(1)$ spinon metal characterized by the absence of fermion pairing, deduced by numerical evidence including linear-in-temperature specific heat and structure factors; and (c) a recently proposed scenario, validated by tensor network calculations in the thermodynamic limit, describing a gapless quantum phase yet featuring fermion pairing, effectively reconciling elements from both previous interpretations. Before delving into the detailed discussion of scenario (c), we briefly review each of these three approaches in the following section.

\subsection{Field-induced phases in extended models and candidate materials} \label{sec:field-extend}
In various extended Kitaev Hamiltonians \cite{Rau14,Jackeli2010}, putative non-conventional field-induced gapless phases have also been reported, even though they host no conventional gapless quasiparticles. 
A useful way to organize these results is to ask how the intermediate-field regime depends on the sign of the dominant Kitaev exchange together with the accompanying symmetry-allowed non-Kitaev terms. In many spin-orbital coupled van der Waals Mott insulators the sign and relative strength of the Kitaev interaction are not unanimously established and can be non-trivial to determine unambiguously from experiments or microscopic modeling \cite{Banerjee2016,Maksimov20,Martin26,Franz20,Inhee26}. It is therefore important to understand which field-induced phenomena are tied specifically to the antiferromagnetic (AFM) Kitaev limit, and which ones can survive or reappear in extended models with ferromagnetic (FM) Kitaev exchange once additional symmetry-allowed interactions are included.
Indeed, both FM Kitaev and AFM Kitaev couplings in extended Kitaev models can exhibit these intermediate phase(s) under an applied field: FM Kitaev in Refs.~\cite{Leahy17,Zheng17,Gordon2019,Han21}, and AFM Kitaev in Refs.~\cite{Jiang2019,Erik21,Lihan24,Wei2025}. This should be contrasted with the pure ferromagnetic Kitaev honeycomb model, which shows no intermediate gapless phase and is much more susceptible to polarization \cite{Jiang11}. Moreover, extended AFM and FM Kitaev honeycomb models with other symmetry-allowed terms can be related by unitary transformations \cite{dual15,Sanders22}, so thermodynamic and spectral features found in one description often carry over to the other.

\begin{figure*}[t]
    \centering
    \includegraphics[width=0.85\linewidth]{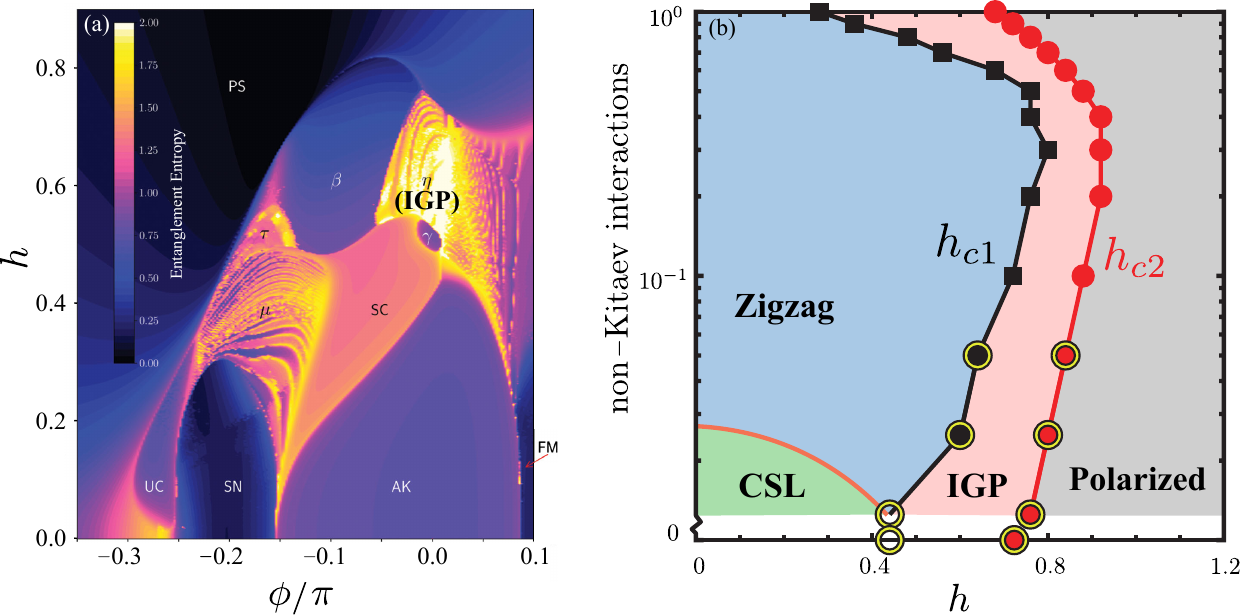}
    \caption{(a) Phase diagram of $\rm K$-$\Gamma$ ladder, with AFM Kitaev exchange, obtained from entanglement entropy, plotted as a function of $\Gamma$ interaction and [111] magnetic field $h$ measured in the magnitude of Kitaev exchange [Eq.~\eqref{eq:kg}]. The colorbar represents the bipartite entanglement entropy. 
    Figure adapted from \cite{Erik21}. The phase labeled by $\eta$ is connected to the IGP defined in Eq.~\eqref{eq_ham} under moderate magnetic field. (b) Phase diagram of $\rm K$-$J$-$\Gamma$-$\Gamma^\prime$ model with AFM Kitaev exchange obtained from spin structure factor, plotted as a function of the magnitude of non-Kitaev interactions and [111] magnetic field $h$ measured in the magnitude of Kitaev exchange. Figure adapted from \cite{Wei2025}.} 
    \label{fig:ladder}
\end{figure*}

We now briefly review the recent progress on the more realistic Hamiltonians beyond the pure Kitaev + Zeeman model that are relevant for candidate solid-state Kitaev QSL materials and that exhibit emergent gapless phases. In these extended Kitaev models, one asks: how robust are the field-induced intermediate phases when additional symmetry-allowed interactions are included? Several such phases have indeed been identified. Strikingly, when additional symmetry-allowed non-Kitaev interactions are included, putative field-induced gapless phases can emerge even for ferromagnetic Kitaev coupling, yet their microscopic nature remains poorly understood, and indeed the very existence of an intermediate gapless phase in the pure Kitaev model is still debated, which will be discussed in the next section. 
Here, the relevant stability of an emergent disordered gapless phase is not that the exactly solvable structure of the pure Kitaev model remains intact. In real candidate materials, additional Heisenberg, $\Gamma$, $\Gamma^\prime$, further-neighbor exchanges, and lattice effects are unavoidable; these perturbations generally render the $Z_2$ fluxes dynamical \cite{Aprem21}, may tune the Chern number in the Majorana fermion sector \cite{Hwang26}, and may also stabilize conventional magnetic orders at low field \cite{Jiang2019}. The more crisp question is whether, after such competing order induced by these non-Kitaev terms is suppressed by magnetic field, the finite intermediate-field phase without conventional order survives over a finite region of extended-coupling parameter space, and whether it can be continuously connected to the intermediate-field QSL of the pure AFM Kitaev model discussed in the previous section. This viewpoint is especially useful for distinguishing a Kitaev-derived intermediate spin liquid from a generic field-induced paramagnet.

One of the simplest extended Kitaev settings is the Kitaev-$\Gamma$ ladder studied in Ref.~\cite{Erik21}, whose Hamiltonian contains the AFM Kitaev exchange together with a symmetry-allowed off-diagonal $\Gamma$ interaction:
\begin{equation} \label{eq:kg}
H_{\rm KG} = H_K + \sin\phi \sum_{\langle ij\rangle_\gamma}
\bigl(\sigma_i^\alpha\sigma_j^\beta+\sigma_i^\beta\sigma_j^\alpha\bigr)
-2h \sum_{i,\alpha}\sigma_i^\alpha\,,
\end{equation}
where $H_K$ denotes the pure AFM Kitaev Hamiltonian [Eq.~\eqref{eq:kitaev_o}] with bond-dependent Kitaev exchange $K=\cos\phi$, the $\Gamma$ term has strength $\sin\phi$, and $(\alpha,\beta)=(y,z),(x,z),(x,y)$ for $\gamma=x,y,z$. At $\phi=0$, Eq.~\eqref{eq:kg} reduces to Eq.~\eqref{eq_ham} on a quasi-one-dimensional ladder. Using iDMRG, Sørensen \textit{et al.} found a rich phase diagram as a function of $\phi$ and field $h$~\cite{Erik21}. In particular, an intermediate gapless phase related to the moderate-field IGP of Eq.~\eqref{eq_ham}, denoted $\eta$ in Fig.~\ref{fig:ladder}(a), persists even for nonzero $\Gamma$. As shown in Fig.~\ref{fig:ladder}(a), the intermediate phase has a large bipartite entanglement entropy, indicating strong long-distance correlations, and occupies a finite region of the $\phi$-$h$ phase diagram.
This ladder result is important because it shows that the field-induced intermediate phase is not immediately destroyed by a realistic non-Kitaev interaction. However, it should also be interpreted with the limitations of the quasi-one-dimensional geometry in mind. A ladder cannot establish the universal ground-state nature of the two-dimensional honeycomb model, and its phases are more naturally described in the language of Luttinger liquids or commensurate-incommensurate/floating phases than by a conventional two-dimensional symmetry-breaking picture~\cite{PokrovskyTalapov79,Giamarchi04,ChitraGiamarchi97}. This caveat is particularly relevant in a [111] magnetic field, where no microscopic symmetries remain; indeed, Ref.~\cite{Erik21} reports incommensurate behavior in the intermediate phase. Thus the precise nature of the corresponding two-dimensional ground state remains an open problem that requires more demanding simulations in the two-dimensional thermodynamic limit.

In addition, Ref.~\cite{Gordon2019} showed that, under a sufficiently strong [111] magnetic field, a similar Kitaev-$\Gamma$ model with FM Kitaev exchange, more relevant to $\alpha$-RuCl$_3$, also hosts an intermediate QSL phase sandwiched between the zigzag-ordered phase and a partially polarized phase.
This demonstrates the stability of the gapless, field-induced phase against realistic symmetry-allowed perturbations. However, due to convergence issues and the confined dimension in the ladder geometry, the precise nature of the $\eta$ phase remains unclear, as does its relevance to the intermediate gapless phase in the two-dimensional thermodynamic limit.

Moreover, both numerical simulation and experiments have reported evidence for field-stabilized gapless QSL phases in realistic materials and in generic extended Kitaev models that include additional symmetry-allowed terms, such as Heisenberg exchanges $H_{J}$,  off-diagonal interactions $H_\Gamma,~H_{\Gamma^\prime}$:
\begin{equation}
    H = H_K + H_{J} + H_\Gamma + H_{\Gamma^\prime} - \mathbf{h}\cdot \sum_i \boldsymbol{\sigma}_i
\end{equation}
with $H_K$ the Kitaev exchange given by Eq.~\eqref{eq:kitaev_o}, and Heisenberg, $\Gamma$ and $\Gamma^\prime$ interactions:
\begin{align}
    H_{J} &= J \sum_{\expval{ij}} \boldsymbol{\sigma}_i \cdot \boldsymbol{\sigma}_j,\;
    H_\Gamma = \Gamma \sum_{\langle ij\rangle_\gamma} \bigl (\sigma_i^\alpha\sigma_j^\beta+\sigma_i^\beta\sigma_j^\alpha\bigr) \\
    H_{\Gamma^\prime} &= \Gamma^\prime \sum_{\langle ij\rangle_\gamma} \bigl( \sigma_i^\gamma\sigma_j^\alpha+\sigma_i^\gamma\sigma_j^\beta+\sigma_i^\beta\sigma_j^\gamma + \sigma_i^\alpha \sigma_j^\gamma\bigr)
\end{align}
A particularly direct test of the stability criterion described above was recently provided by the realistic $\rm K$-$J$-$\Gamma$-$\Gamma^\prime$ model for the cobalt-based honeycomb magnet Na$_2$Co$_2$TeO$_6$~\cite{Wei2025} with AFM Kitaev exchanges ($K>0$). In this case, the dominant interaction is an AFM Kitaev exchange, while the sub-leading non-Kitaev terms are needed to quantitatively account for the material response. DMRG calculations show that the low-field AFM order is suppressed by a [111] magnetic field and is replaced by an intermediate gapless QSL before the system enters the high-field polarized regime. Crucially, this intermediate-field QSL was shown to be adiabatically connected to the intermediate-field QSL of the pure AFM Kitaev model~\cite{Wei2025}, as shown in Fig.~\ref{fig:ladder}(b). This result gives a concrete meaning to stability against realistic perturbations: the intermediate phase need not be identical to the pure-Kitaev phase at the microscopic level, but it can remain a Kitaev-derived spin liquid if it survives over a finite window of field and couplings and can be continuously connected to the pure AFM Kitaev intermediate phase without crossing a phase transition.

\begin{figure}[t]
    \centering
    \includegraphics[width=\linewidth]{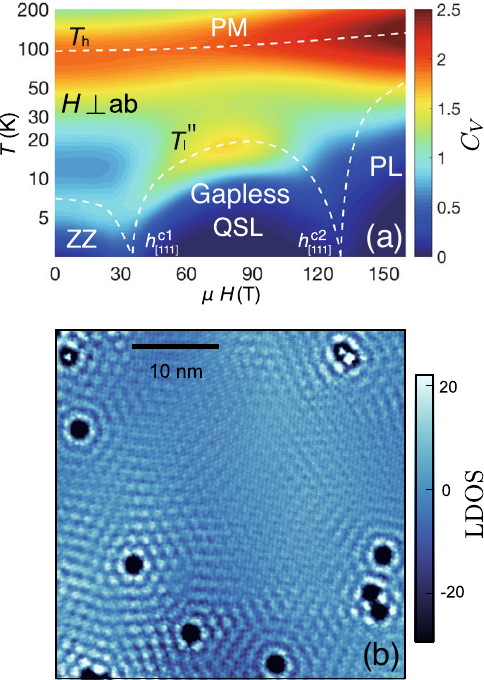}
    \caption{(a) Simulation of specific heat of the extended FM Kitaev model as a function of [111] magnetic field (perpendicular to the $ab$ plane) and temperature. Phase boundaries and characteristic temperature scales are marked using white dashed lines. Figure adapted from \cite{Han21}. 
    (b) Spatial oscillation of the local density of states (LDOS) in monolayer $\alpha$-RuCl$_3$. The LDOS decays
    away from defects, indicative of gapless Majorana fermions akin to spinon Friedel oscillation. 
    Figure adapted from \cite{Kohsaka24}. }
    \label{fig:extended}
\end{figure}

A similar intermediate gapless phase has also been found to be present in extended Kitaev models with FM Kitaev exchange. 
In Ref.~\cite{Han21}, it has also been found that in a similar $\rm K$-$J$-$\Gamma$-$\Gamma^\prime$ model with $K<0$ resides an intermediate-field gapless phase extending a finite parameter range, as evidenced by exact diagonalization calculations of the specific heat and DMRG computations of ground-state observables.
As shown in Fig.~\ref{fig:extended}(a), the specific-heat curve in the gapless QSL regime under [111] field exhibits a characteristic double-peak structure versus temperature, together with algebraic scaling at low temperature. The double peaks are reminiscent of the pure Kitaev model at finite temperature \cite{Han21,Knollethermal}: the lower-temperature peak arises from thermal activation of $Z_2$ flux excitations, while the higher-temperature peak corresponds to activation of the remaining fermionic excitations. The low-$T$ algebraic behavior of the specific heat further signals a gapless QSL.

Here we highlight that although the quantum spin liquid phase in the pure FM Kitaev honeycomb model is not robust against an external magnetic field \cite{hickey2019emergence}, (in contrast to the pure AFM Kitaev model which resists polarization by entering an intermediate phase, it becomes polarized by a Zeeman field almost immediately), FM Kitaev exchange combined with additional symmetry-allowed interactions appears to stabilize a gapless quantum spin liquid over a wide range of field strengths, as reported in the aforementioned example \cite{Han21}. This outcome is not entirely surprising given that a unitary rotation can map an extended Kitaev model with FM Kitaev exchange to one with AFM Kitaev exchange \cite{dual15}. In particular, the parameters for the extended FM Kitaev model used in Ref.~\cite{Han21} in Fig.~\ref{fig:extended}(a) transform into an extended AFM Kitaev model dominated by AFM Kitaev exchange and FM Heisenberg exchange, with sub-leading $\Gamma$ and $\Gamma'$ interactions. The resulting phase diagram and its thermodynamic properties are therefore similar to those of the $J-K$ model studied in Ref.~\cite{Jiang2019}, where a magnetic field stabilized a gapless quantum spin liquid phase.
These examples suggest that the intermediate-field phase in extended models should be viewed as a family of Kitaev-derived phases rather than as a feature restricted to the exactly solvable Hamiltonian. The perturbations determine the location and width of the intermediate-field window and may change detailed spectral features, but the phase can remain the same when its fractionalized low-energy response and its connection to the AFM Kitaev intermediate QSL is identifiable.

We also note that quenched disorder is unlikely to qualitatively destabilize this type of intermediate gapless phase. Recent theories have proposed that the gapless phase arises from strong quantum fluctuations of the fluxes \cite{wang2024,penghao24,Feng2025}, so the fermions effectively propagate in an incoherent ensemble of flux backgrounds (to be extensively discussed in Sec.~\ref{sec:mm}). Vacancies, defects, bond randomness, and local lattice distortions mainly add a static flux disorder component to this pre-existing flux-disorder ensemble \cite{KAO2021168506}; they should not by themselves destroy the intermediate gapless phase as long as they do not induce conventional magnetic order or drive the system through a percolative transition.

This numerical evidence also qualitatively aligns with many experiments and comprehensive simulations on $\alpha$-RuCl$_3$, where thermodynamic measurements such as thermal conductivity, specific heat and spin response point to a gapless QSL stabilized by a magnetic field applied in a plane \cite{Leahy17,Zheng17,Normand18}. Recent experimental progress has allowed access to ultraclean bulk crystals and single-layer candidate materials. For example, experiments on ultraclean $\alpha$-RuCl$_3$ demonstrated an intriguing field stabilized gapless QSL signature in the specific heat data \cite{imamura2025}. Apart from thermodynamic measurements, a recent experiment \cite{Kohsaka24} on monolayer $\alpha$-RuCl$_3$ under a [111] magnetic field observed Friedel-like exotic oscillations around defects \cite{Friedel1952}, as shown in Fig.~\ref{fig:extended}(b). Since $\alpha$-RuCl$_3$ is a Mott insulator, oscillations due to an electronic Fermi surface can be excluded. Therefore, the observed real-space oscillations around defects indicate a neutral Fermi surface of Majorana fermions or complex spinons \cite{Zhang2024,Jahin25}. Their emergent fermionic nature appears as a modulation in real space corresponding to a neutral Fermi wavevector. Although the precise microscopic mechanism behind the emergence of this neutral Fermi surface remains unclear, developing a more comprehensive understanding of these gapless QSL signals in candidate Kitaev materials is an open challenge for future research.

\begin{table*}[t]
\begingroup
\footnotesize
\begin{tabular}{p{0.17\textwidth}p{0.25\textwidth}p{0.25\textwidth}p{0.25\textwidth}}
\hline
Proposal & Low-energy degrees of freedom & Key properties & Main open issue \\ \hline\hline
$U(1)$ spinon metal & Complex spinons coupled to an emergent $U(1)$ gauge field, forming a spinon Fermi surface. & Power-law correlation in 2D;

Linear specific heat;

Friedel oscillation; 

& Requires a route from the low-field $Z_2$ gauge structure to $U(1)$; the universal low-energy transport regime in 2D remains challenging. \\ \hline

Parton Chern insulator & Translation-invariant $Z_2$ parton bands with nonzero Chern number and a finite quasiparticle gap. & Exponential correlations in 2D; 

stable topological character; 

absence of zero-energy spin or dimer spectral weight; 

possible thermal Hall response. & Depends on the assumed mean-field ansatz; self-averaging may miss fluctuating flux sectors and low-energy continua seen in unbiased simulations. \\ \hline

$Z_2$ quantum Majorana metal &  Majorana fermions moving in a dense, slowly fluctuating, 
superposed $Z_2$ flux background. &

Rapid flux proliferation; 

logarithmic divergence of Majorana density of states; 

Possible Friedel oscillation; 

suppressed heat transport from quasi-static flux disorder. & Flux dynamics and the finite-temperature conductivity in the thermodynamic limit remain open; the quasi-static flux description requires further dynamical benchmarks. \\ \hline
\end{tabular}
\caption{Criteria for distinguishing leading interpretations of the intermediate-field phase, and associated open issues. }
\label{tab:igpcriteria}
\endgroup
\end{table*}

\subsection{The debate over the emergent quantum phase under moderate field}
We now turn to the central theoretical debate concerning the intermediate phase under moderate magnetic fields. As summarized in Table~\ref{tab:tab1}, numerical and analytical studies agree on several broad features: the intermediate regime is distinct from the low-field chiral spin liquid and from the high-field partially polarized phase, it shows no obvious conventional magnetic order, and it exhibits unusually strong low-energy responses. However, the nature and the mechanism of this regime remain unsettled. The leading proposals in Table~\ref{tab:tab1} should not be regarded as different names for the same state, but rather as distinct physical scenarios with different gauge structures, quasiparticles, and experimental implications.

The three main interpretations are as follows. First, the intermediate phase may be a gapless $U(1)$ spinon metal, in which the field suppresses pairing of complex fermionic spinons and produces a spinon Fermi surface. Second, it may be a gapped parton Chern insulator obtained in translation-invariant microscopic mean-field theories, where the magnetic field drives topological transitions between different parton bands. Third, it may be a gapless $Z_2$ quantum Majorana metal, in which real Majorana fermions propagate in a dense, slowly fluctuating, quantum-superposed background of $Z_2$ fluxes generated by the magnetic field. These scenarios make rather different assumptions about the fate of the low-field $Z_2$ gauge structure, the role of flux fluctuations, and the origin of the low-energy spectral weight.

For this reason, a reliable assignment of the intermediate-field phase requires a hierarchy of diagnostics. The first requirement is to establish that the low-energy continuum is not simply a remnant of conventional magnetic order, magnon decay, or finite-cylinder artifacts, but is compatible with fractionalized matter excitations such as spinons or Majorana fermions. This requires more than observing a small gap or a broad continuum: one should ask whether thermodynamics, equal-time correlations, and dynamical responses can be consistently described in terms of an emergent low-energy excitation structure, such as a spinon or Majorana Fermi surface, a nodal spectrum, or a gapped Chern band. This question is tied to the gauge structure of the putative phase. A $U(1)$ spinon metal, a $Z_2$ parton Chern insulator, and a $Z_2$ quantum Majorana metal may share some unconventional responses, but they differ intrinsically in whether the emergent gauge structure is $U(1)$ or $Z_2$, whether the fluxes remain well-defined, whether they are dilute or proliferate, and whether the response is controlled by a clean band topology or by a fluctuating flux ensemble that cannot be captured by a translation-invariant self-consistent mean-field theory. The final test is whether the same interpretation survives across different gauge-invariant probes, including thermodynamics, spin and dimer correlations, Raman/RIXS-like responses, and thermal transport. Table~\ref{tab:igpcriteria} summarizes these criteria, with details to be discussed in Sec.~\ref{sec:mm}.
In the following three subsubsections, we briefly review these three competing interpretations.

\subsubsection{Gapless $U(1)$ spinon metal}
It has been proposed that the intermediate phase—sandwiched between the low-field chiral spin liquid and the high-field polarized magnet—may be a gapless U(1) QSL hosting a neutral spinon Fermi surface (FS) \cite{Patel12199,David2019,hickey2019emergence}. Evidence for this scenario arises from exact diagonalization (ED) and density matrix renormalization group (DMRG) studies, which observe a dramatic increase in the low-energy density of states under moderate fields before polarization. The putative spinon FS was shown to reproduce the spin structure factor via the convolution of fractionalized spinon bands near zero energy \cite{Patel12199}.
Conceptually, this theory builds on viewing the low-field chiral spin liquid as a topological p-wave superconductor of itinerant matter fermions. Increasing the magnetic field then suppresses fermion pairing, destroying the $Z_2$ topological order and giving way to a U(1) metallic spinon FS \cite{Yuanming2018,hickey2019emergence}. However, there remains some key unresolved issues, for example, the mechanism for the drastic change in the gauge structure from U(1) to $Z_2$ has yet to be found; assuming a U(1) gauge structure, the construction of such a canonical spinon FS is not numerically founded and may not be unique; and DMRG results on cylindrical geometries may be limited by finite-size effects with restricted bond dimensions, and by the potential underestimation of long-range correlations, well-known to be an especially acute problem in gapless phases.

\subsubsection{Gapped parton chern insulator from microscopic mean-field theories}
The microscopic self-consistent parton mean field theories (MFTs) suggest that the intermediate phase can be a gapped parton Chern insulator. The magnetic field under the mean field picture effectively hybridizes the localized fermions (fluxes in the integrable limit) and itinerate Majorana fermions which leads to topological phases transitions. 
The quantum phase under moderate magnetic field was identified as a gapped topological order with Chern number $\pm 4$ and an emergent $Z_2$ gauge structure \cite{Jiang2020,ZhangNatComm2022}. It was hence proposed that the intermediate phase is gapped with low-energy excitations around the $\Gamma$ point in the momentum space, and the transition between the low-field CSL and the intermediate phase is depicted as a topological phase transition between different parton Chern insulators. 
While the MFTs accurately captures the phase transition evidenced in earlier numerical computations. they are still effectively renormalized product state ansatz in the chosen parton basis, thus  
fall short in representing the entanglement between the two fractionalized quantum sectors, and in representing a flux as a physical degree of freedom with many-body entanglement among multiple sites \cite{feng2023statistical}, a crucial element for capturing non-perturbative physics in the intermediate phase and should be considered in the variational ansatz.
Another caveat regarding the parton mean-field theories assuming translation-invariant self-consistent mean fields is that they suppress the nonlocal quantum interference effects, which can be responsible for potential Anderson-type localization and related anomalous transport phenomena at finite energy scales. We will revisit these caveats in more details in Sec.~\ref{sec:interplay} and Sec.~\ref{sec:mftlimits}. 

\subsubsection{Gapless quantum majorana metal from coherent disorder} \label{sec:cdisorder_mm}
This most recent theory supported by iPEPS in the thermodynamic limit states that the intermediate phase can be an emergent Majorana metal at zero temperature, hosting a finite Majorana FS at zero energy \cite{wang2024,penghao24}. It is describable by an ensemble of tight-binding models of Majorana fermions conditioned on random-flux with an average translation symmetry. Here, the low-energy degrees of freedom are real spinons coupled to a $Z_2$ gauge field, rather than complex spinons with U(1) field. 
Two main arguments supporting this viewpoint are:
(i) The magnetic field induces a “vortex glass” of $\pi$-fluxes on the honeycomb plaquettes, giving rise to an effectively static random flux background \cite{Baskaran2023}. Indeed it was found in earlier numerical works that the flux expectations in the intermediate phases drops quickly from $W_p = 1$ to $W_p\sim 0$ \cite{David2019,hickey2019emergence}, indicating a near-half filling of $Z_2$ fluxes, which allegedly supports a large number of near-degenerate flux configurations, as is confirmed by recent DMRG simulation \cite{Baskaran2023}, thus a vortex glass with negligible dynamics compared to Majorana fermions; and (ii) the dynamical properties of the phase are thus primarily governed by Majorana fermions under random flux configurations, giving rise to a Majorana metallic state of class D~\cite{Senthil2000,Huse2012,Heyl21}.  
In this setup, the $Z_2$ gauge structure persists under the transition from a gapped CSL to a gapless quantum Majorana metal. 
This theory is consistent with recent experiments by Y. Matsuda’s group on monolayer $\alpha$-RuCl$_3$, where real-space Friedel-type oscillations were observed \cite{Kohsaka24}, possibly reflecting an emergent Majorana Fermi surface \cite{baskaran2015majorana}. It also implies that the field-induced gapless phase of Kitaev materials \cite{Zheng17,Han21} may lie outside conventional categories of gapless spin liquids (e.g., nodal complex spinons or U(1) spinon Fermi surfaces), instead constituting an emergent quantum Majorana metal due to the field-induced interplay between $Z_2$ fluxes and itinerate majorana fermions. Outstanding questions include the precise nature and experimental signatures of glassy fluxes under moderate magnetic field, as well as the presence of potential localization physics in Majorana fermions under the emergent random flux disorder.
In the following, we report recent progress on the theoretical framework and numerical evidence for the field-induced quantum Majorana metal. We will show in more details how a magnetic field drives the proliferation of $\pi$-fluxes at low energy and its resulting impact on Majorana fermions. Finally, we present numerical results obtained via iPEPS, which support the theory of a gapless quantum Majorana metal.

\subsubsection{Reliability and limitations of current methods}
\label{sec:reliability}
The disagreement summarized above also reflects the different constraints built into the available numerical and analytical methods~\cite{David2019,hickey2019emergence,Patel12199,Jiang2020,ZhangNatComm2022,wang2024,penghao24}. Exact diagonalization is unbiased and gives direct access to spectra and thermodynamics, but the accessible clusters have limited momentum resolution and cannot reliably determine whether a small gap survives in the thermodynamic limit. DMRG and iDMRG reach much larger systems and are powerful for diagnosing the absence of magnetic order, but they are usually performed on cylinders whose circumference may be shorter than the intrinsic correlation length of a gapless state. Boundary conditions, metastable flux sectors, and finite bond dimension can therefore affect the apparent central charge, spin correlations, and gap extrapolations. iPEPS works directly in the two-dimensional thermodynamic limit and is valuable for distinguishing extended phases from cylinder artifacts, but finite bond dimension, environment truncation, metastable variational states, and finite-time evolution still require systematic extrapolation, especially for highly entangled gapless states and broad dynamical continua. As a result, the true ground-state or asymptotic low-frequency universal regime is particularly difficult to settle with present tensor-network methods. At the same time, though, much can still be inferred from low but finite energy scales and intermediate-time/finite-energy dynamics, which are often precisely the regimes accessed by numerical response functions and experiments at finite energy scales \cite{Feng2025}.

Analytical and variational approaches have complementary limitations. Parton mean-field theories and variational wave functions provide transparent quasiparticles, gauge structures, and topological indices, but their conclusions depend on the chosen gauge ansatz and often impose translation invariance on the emergent gauge fields. They can therefore miss spatially fluctuating flux sectors that may be central to the intermediate-field regime \cite{Fratini21,Hadi24}, which we shall later discuss in detail in Sec.~\ref{sec:mftlimits}. Perturbative effective theories are controlled at weak field or large anisotropy, but they cannot by themselves settle the nonperturbative moderate-field phase. Random-flux or quasi-static-flux descriptions, in turn, give a physically motivated effective picture for the Majorana metal, but their validity must be benchmarked against unbiased ground-state and dynamical calculations.

The most dependable conclusions are therefore those over-constrained by independent diagnostics. One should regard an interpretation as robust only when it accounts simultaneously for ground-state observables, real-space correlations, momentum- and energy-resolved spin, dimer spectra, and the qualitative behavior of thermal (Hall) transport. The quantum Majorana-metal scenario is attractive by this standard because it ties flux proliferation to a finite Majorana density of states and to the spin and dimer response, while also offering a natural explanation for why heat transport can remain weak despite gapless spectra. Nevertheless, this scenario still requires further tests, especially through two-dimensional real-time calculations at larger bond dimension, and purification for finite temperatures, and direct comparisons with material-specific extended Hamiltonians.

\section{Gapless Majorana fermions under vison fluctuation} \label{sec:mm}
Understanding how an external magnetic field modifies the Kitaev honeycomb model has garnered significant experimental and theoretical interest. Most candidate Kitaev materials require a magnetic field to suppress the zigzag magnetic order. For instance, the well-known candidate $\alpha$-RuCl$_3$ and the more recently studied Na$_2$Co$_2$TeO$_6$ \cite{Gaoting24} have both been reported to become magnetically disordered under an in-plane field along the $a$ axis or an out-of-plane field along the $c$ axis, as studied numerically in \cite{Han21,Wei2025}. As discussed in the previous section, when the field increases, they exhibit an intermediate field-induced magnetically disordered phase before ultimately entering a trivial partially polarized phase at higher fields. The exact nature of this intermediate phase remains elusive, especially regarding whether it is an emergent QSL.

Since all known candidates of Kitaev QSL in solid state systems are not integrable due to other symmetry allowed interactions that prefer magnetic orders, requiring the external magnetic fields to prevent the system from ordering at low temperatures. This leads to a fundamental question: if there is indeed an emergent Kitaev QSL stabilized by a magnetic field, how would that field modify the QSL’s hallmark signatures? Since 2018, there has been a burgeoning interest in theoretical efforts exploring how magnetic fields influence the behavior of a Kitaev spin liquid \cite{Patel12199,hickey2019emergence,Jiang_arXiv_2018,Zhu_PRB_2018,Gohlke_PRB_2018,gohlke2018dynamical,jhk21,Kyusung22,Baskaran2023,Lihan24,Will24,Feng24,wang2024,penghao24,Han21,Brenig21,Lukas21}; Research in this direction has also reached the extended Kitaev models with both in-plane and out-of-plane magnetic fields \cite{Normand18,Normand19,Han21,Erik21,Qiang26}. Yet, the nature of these field-induced phases remains an open question under intense debate. The lack of integrability, combined with numerical limitations (finite-size effects in exact diagonalization or DMRG) and potential biases in microscopic mean-field theories, makes it challenging to confirm or refute competing explanations.
After this intermediate phase was discovered numerically in the pure and extended Kitaev honeycomb models, many competing theories emerged to explain its nature from different perspectives, each supported by different numerical techniques. 
In this section, we will focus on the recent parsimonious model whereby an emergent intermediate phase in the non-perturbative regime emerges, namely the antiferromagnetic Kitaev honeycomb model under an out-of-plane magnet field in the $c$ axis; and we focus on the recent proposal of the quantum Majorana metal, including the mechanism, thermodynamics and dynamical responses, and the limitation of relevant theories. We remark that, even though this is a simple setup for the convenience of simulation and theoretical demonstration, its validity goes beyond the pure Kitaev model due to the well-known duality map \cite{dual15} whereby a pure Kitaev model can be mapped to an extended Kitaev model by unitary rotation, making it relevant for a series of experimentally observed gapless QSL candidates stabilized by magnetic field, as discussed in Sec.~\ref{sec:field-extend}.

\begin{figure*}[t]
    \centering
    \includegraphics[width=0.9\linewidth]{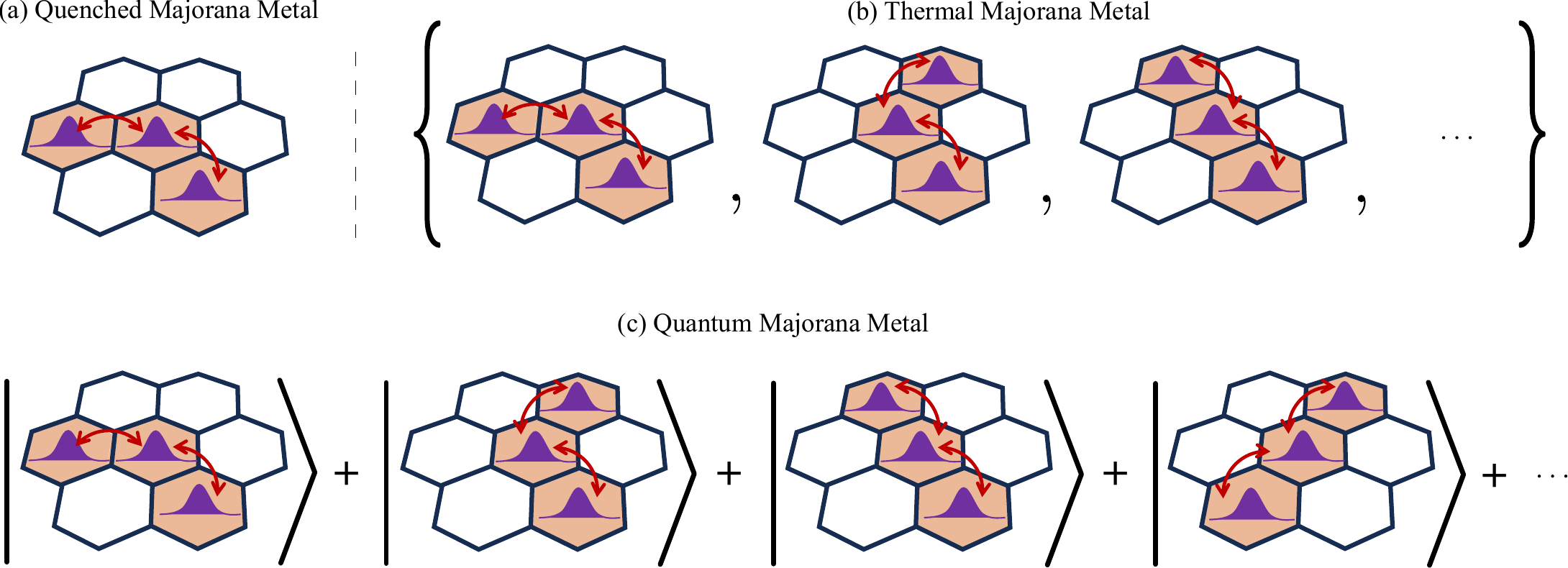}
    \caption{Quenched, thermal Majorana Metal versus the quantum Majorana metal. (a) A quenched Majorana metal due to static flux disorder. The localized wave packets indicate Majorana zero modes trapped at $\pi$ fluxes in orange; and the red arrow denotes the Majorana hopping between two adjacent modes whose separation are comparable to the Majorana localization length. (b) A thermal Majorana metal has thermally agitated flux disorder which forms a set of thermal ensemble, inducing the gapless Majorana fermions.  (c) A quantum Majorana metal consists of an extensive number of disordered flux configurations which are in a coherent superposition at zero temperature, inducing gapless Majorana fermions and gapless spin spectral functions.  }
    \label{fig:quvscl}
\end{figure*}

\subsection{Majorana metals: classical vs quantum}
The theory of field-induced quantum Majorana metal ultimately arises from the topological properties of the Kitaev honeycomb model, specifically the interplay between the topological Chern bands of Majorana fermions and the proliferation of static $Z_2$ fluxes. Therefore, it is useful to first review the simplest setup of this interplay, focusing on Majorana fermions interacting with a single $Z_2$ flux, in the integrable limit of the CSL phase. For more details, this topic is well-documented in the literature on topological superconductors; see, for example, Ref.~\cite{Read2000}. 
It is thus natural to deduce that, under a mean-field picture consisting of fluxes and matter fermions, this intuition can be extended to the regime with a moderate magnetic field, where the nature of the emergent quantum phase has remained elusive (Table~\ref{tab:tab1}). Since the magnetic field breaks time-reversal symmetry and does not commute with the compass interactions, fluxes can proliferate in the low-energy sector when the field becomes sufficiently strong, thereby trapping numerous Majorana fermions. This mechanism resembles the superposition among gauge superselection sectors induced by “gauge frustration” \cite{Eschmann19}, but in our case the flux fluctuations arise from the non-commuting magnetic‐field term.
As the flux density increases, such that the average separation $1/\sqrt{\text{flux density}}$ becomes comparable to the localization length of the trapped Majorana modes, these fermions begin to weakly couple, resulting in a gapless density of states. In the following section, we will discuss in detail the case of a moderate magnetic field and provide numerical validation in the thermodynamic limit.

In the aforementioned intuitive picture, when a magnetic field is applied such that the ground state features a proliferation of $\pi$-fluxes with an average separation small enough for the localized Majorana modes on the fluxes to overlap, a percolation-like process occurs. This overlap allows the otherwise localized Majorana modes to undergo a nucleation transition, forming a gapless spectrum. Consequently, this leads to the closure of the fermion gap, giving a gapless Majorana metal due to the presence of high-density $\pi$ fluxes, well-known in the context of topological superconductor of class D \cite{Senthil2000,Huse2012}.

\begin{figure}[t]
    \centering
    \includegraphics[width=\linewidth]{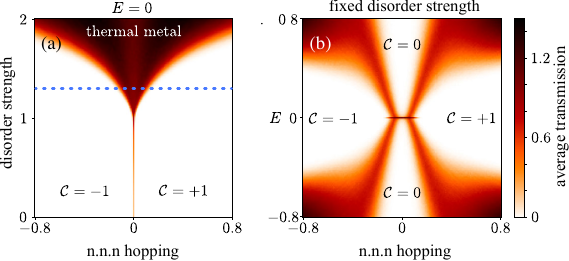}
    \caption{The emergence of quenched Majorana metal visualized in transmission. (a) Disorder-averaged transmission as a function of disorder strength and n.n.n. hopping $\lambda$, evaluated at zero energy. (b) Disorder-averaged transmission as a function of n.n.n. hopping amplitude $\lambda$ and energy $E$, evaluated at a fixed disorder strength cut shown as the horizontal dashed line in panel (a). Figure adapted from Ref.~\cite{Fulga20}. }
    \label{fig:thermalmetalquench}
\end{figure}

However, a crucial piece remains to be clarified in the theoretical framework of this claim: unlike the conventional thermal or quench-disordered Majorana metal, the IGP arises in the absence of explicit quenched disorder in the Hamiltonian. Instead, the disorder-like effect stems from strong quantum fluctuations in the gauge fields induced by the magnetic field. This is similar to the setup in disorder-free localization where the disorder is generated by dynamical fluctuation \cite{Knolle17}. Their difference and similarity are illustrated in Fig.~\ref{fig:quvscl}. Therefore, before delving into the theory of the quantum Majorana metal in the specific microscopic model, we first carefully distinguish between two relevant scenarios: 
\begin{enumerate}
    \item `Classical' Majorana metal (Class~D topological superconductor with quenched or thermally agitated flux disorder) sketched in Fig.~\ref{fig:quvscl}(a,b): A gapless Majorana metal formed due to the thermally agitated flux disorder or quenched disorder in bonds or sites \cite{do2017majorana,Knolle19,Knolle21,Knollethermal,KAO2021168506}; 

    \item Quantum Majorana metal sketched in Fig.~\ref{fig:quvscl}(c): A gapless Majorana metal that emerges from the coherent superposition of disordered flux configurations at zero temperature, i.e. due to the strong quantum fluctuations in the gauge field, without any explicit disorder in the Hamiltonian \cite{wang2024,penghao24,Feng2025}.
\end{enumerate}
To illustrate the Majorana metal in scenario (1), consider the following tight-binding model of Majorana fermions:
\begin{equation}
\label{eq:Majoranahopping}
H = \sum_{\langle jk\rangle} i t_{jk}c_{j}c_{k} + \sum_{\langle \langle jk\rangle \rangle} i t'_{jk}c_{j}c_{k} + {\rm H.c.}, 
\end{equation}
where $t_{jk} \propto u_{jk}$ and $t'_{jk} \propto \lambda\, u_{jl} u_{lk}$ represent random nearest-neighbor ($\langle jk\rangle$) and next-nearest-neighbor ($\langle \langle jk \rangle \rangle$) hopping amplitudes from a perturbative weak magnetic field (See also Sec.~\ref{sec:review}). In the case of uniform $t_{jk}=1$, this Hamiltonian corresponds to the free fermion sector of the Kitaev honeycomb model in its ground state. However, when $t_{jk}$ becomes random, the density of states (DOS) near zero energy develops a weak anti-localization divergence, a hallmark of the thermal Majorana metal phase. Specifically, near the particle-hole symmetry line at $E = 0$, the DOS was derived using a non-linear sigma model to exhibit the following logarithmic divergence \cite{Senthil2000}:
\begin{equation} \label{eq:rho}
    \rho(E) \sim -\ln E.
\end{equation}
This behavior has been extensively studied in the context of the Kitaev model at finite temperatures, where thermally agitated fluxes effectively act as a source of disorder in the hopping amplitudes $t_{jk}$ \cite{Knolle19,Knolle21,Knollethermal}.
On the other hand, the thermally excited flux disorder is also reminiscent of the Anderson localization \cite{Mirlin08}; or, in absence of any quench disorder, a dynamical Anderson localization induced by fermions' interaction with massive (slowly fluctuating) boson modes \cite{Fratini11,Fratini16,Fratini21,Hadi24}. Indeed, in the integrable limit where the fractionalization into Majorana fermions and static flux is exact, the fermion localization length takes the form
\begin{equation}
    \xi(E) \sim \frac{1}{\sqrt{E}} \exp(\frac{1}{4} \ln^2 \frac{1}{E}).
\end{equation}
Under strong quench disorder, this indicates localization physics at all non-zero energy scales, except at the particle-hole symmetry line $E = 0$ where a multi-fractal extended state is present due to weak anti-localization \cite{Fulga20}. This is illustrated in Fig.~\ref{fig:thermalmetalquench}.

Given the well-established case of thermal or quenched Majorana metals, a critical question remains to be addressed: Can similar Majorana metal physics emerge at zero temperature in a pristine system without quenched disorder, and what are its implications for spectral functions relevant to scattering experiments? This scenario corresponds to the aforementioned case (2) and Fig.~\ref{fig:quvscl}(c), which we refer to as the quantum Majorana metal.
The notion of emergent disorder in a translationally invariant system has been previously explored in the context of disorder-free localization \cite{Knolle17, Knolle17b, Cirac05}, where an extensive set of (approximately) conserved charges, e.g. static or slow vison fields, effectively serves as a source of disorder for the quantum sector comprising itinerant particles.

We emphasize that the emergent quantum Majorana metal cannot be adequately described by microscopic mean-field theories, which rely on self-consistent averaging. Notably, previous biased mean-field treatments of this problem \cite{Jiang2020, ZhangNatComm2022} predicted a finite energy gap, in contrast to the gapless spectrum observed in two-dimensional tensor network simulations \cite{wang2024, penghao24}. This limitation is well recognized in the study of high-temperature Hubbard models and dynamical localization in translationally invariant systems. For instance, even the most state-of-the-art dynamical mean-field theory fails to capture the correct low-energy features due to the presence of dynamical disorder \cite{Fratini21}. In comparison to unbiased numerical methods such as (i)PEPS, (i)DMRG or ED, this shortcoming arises from the inherent neglect of non-local interference effects and current vertex corrections in mean-field approximations \cite{Mravlje19}.  Therefore, to achieve a faithful description, it is essential to construct a mean-field ansatz that explicitly incorporates the effects of coherent disorder, which must thereafter be validated against unbiased numerical approaches such as iPEPS. We will discuss these results in detail in the following sections.

\subsection{Interplay between Majorana fermions and fluxes under moderate magnetic field} \label{sec:interplay}
The emergence of the IGP under a moderate magnetic field is primarily driven by the interplay between flux fluctuations and the Majorana Chern band. The transition at  $h_{c1}$  from the chiral spin liquid to the IGP can thus be understood as a nucleation transition of Majorana fermions in the presence of field-induced fluxes in the ground state. Previous studies have shown that the gap for $Z_2$ flux excitations decreases as the magnetic field increases \cite{Zhu_PRB_2018,hickey2019emergence}. Consequently, once the magnetic field induces a finite density of fluxes in the ground state, it facilitates the nucleation of Majorana fermions and enables their tunneling between neighboring sites, ultimately leading to the formation of a gapless phase. 

To confirm that the emergence of the IGP is driven by flux fluctuations, we can consider its contrapositive: if flux fluctuations are energetically suppressed, the IGP should be eliminated. A natural way to test this is by modifying the original Hamiltonian to include an additional energy penalty for flux excitations:
\begin{equation}
H_{\rm biased} = \sum_{\langle ij\rangle, \alpha} K_\alpha \sigma^\alpha_i \sigma^\alpha_j - \mu \sum_p W_p - h\sum_{i,\alpha} \sigma_i^\alpha,
\end{equation}
where the term  $- \mu \sum_p W_p$  with  $\mu > 0$  makes flux excitations energetically costly, thereby suppressing their fluctuations.
This provides a controlled way to probe the role of flux dynamics in stabilizing the IGP. By tuning $\mu$, we can interpolate between the Majorana metal in IGP, where fluxes strongly fluctuate under a moderate magnetic field, and a regime where fluxes are pinned, preventing the formation of a gapless Majorana metal.
If the IGP disappears as $\mu$  increases, this directly supports the argument that flux fluctuations are essential for its existence. It also circumvents the need for explicitly tracking individual flux which are not exactly conserved, instead leveraging an effective energy scale to probe their influence on the phase structure. 

\begin{figure}[t]
    \centering
    \includegraphics[width=\linewidth]{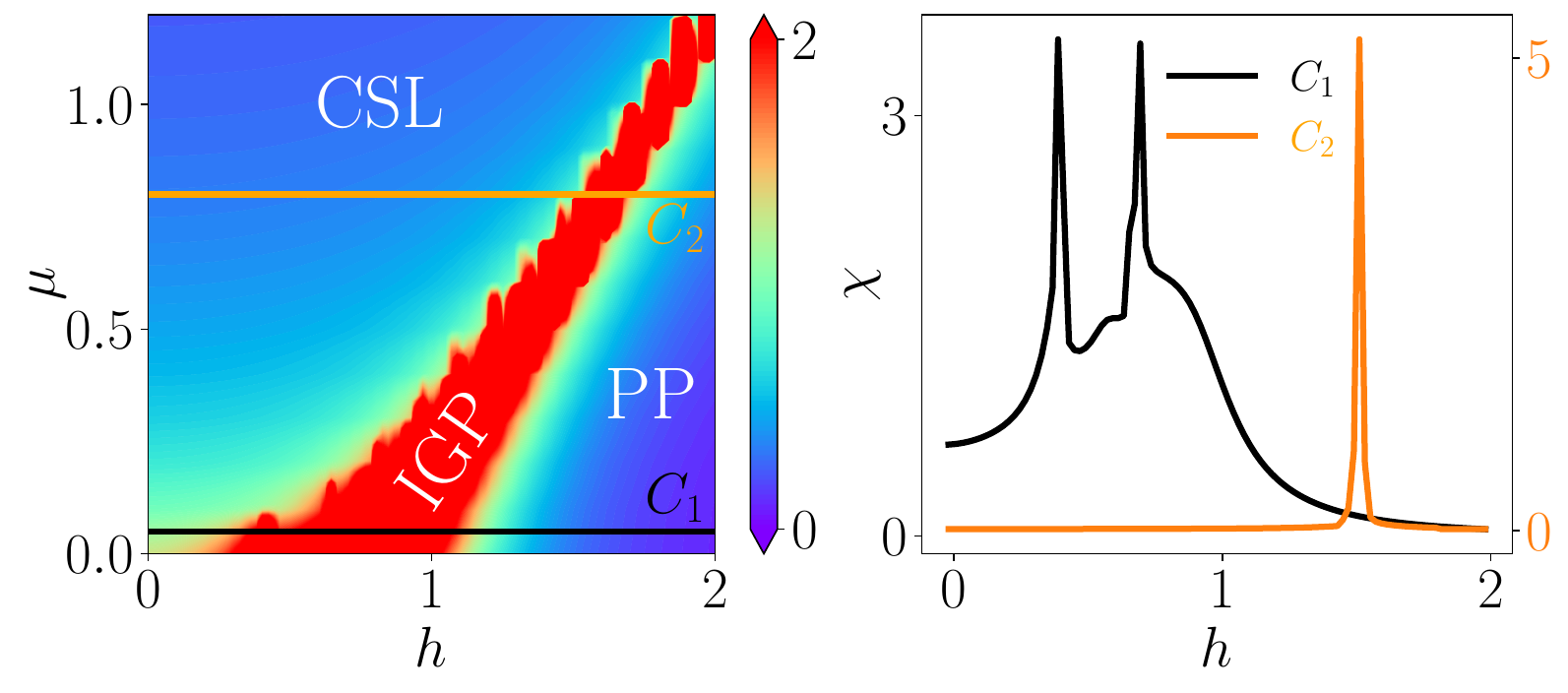}
    \caption{The concomitant presence of flux fluctuations and the IGP as a quantum Majorana metal. (Left) Phase diagram as a function of flux bias $\mu$ and magnetic field $h \parallel [111]$ measured by susceptibility $\chi(\mu,h) = N^{-1} \partial^2 E_{0} /\partial h^2$. The IGP ceases to exist when fluxes become energetically unfavorable. Data were obtained by exactly diagonalizing $H_{\rm biased}$ in a 24-site cluster ($3\times 4$ unit cells) with torus geometry.  (Right) $C_1$ and $C_2$ represent two constant-$\mu$ cuts in (Left). Figures are adapted from Ref.~\cite{wang2024}. }
    \label{fig:WpBias}
\end{figure}

As expected, the susceptibility results presented in Fig.~\ref{fig:WpBias} validate the aforementioned argument.
For  $\mu \simeq 0$, the ED results indicate that the intermediate phase, corresponding to the IGP, emerges under a finite magnetic field and persists over a finite range of $h$ before undergoing a confinement transition into the partially polarized (PP) phase. However, as shown in Fig.~\ref{fig:WpBias}, for larger values of $\mu$, where flux fluctuations are increasingly suppressed, the IGP spans a progressively narrower range of $h$ and eventually disappears at $\mu \gtrsim 0.6$. This highlights the crucial role of finite flux density in the formation of the IGP and suggests that any faithful variational ansatz must incorporate the two fractionalized degrees of freedom: itinerant Majorana fermions and fluctuating local $Z_2$ fluxes, even under a moderate magnetic field beyond perturbation theory.
Therefore, to understand the details of this mechanism and its observable consequences, it is necessary to formulate a suitable variational ansatz within the aforementioned quasi-particle framework that captures the essential interaction between Majorana fermions and fluxes.

Since Majorana fermions and $Z_2$ fluxes together form a complete basis for the Hilbert space, and the magnetic field induces coupling between the gauge and fermion sectors, it is suitable to consider an ansatz for the IGP in which these two sectors are inherently entangled. Thus, the ground state of the IGP can be expressed within this basis:
\begin{equation} \label{eq:mft}
    \ket{\Psi_{\text{IGP}
    }} = \sum_{\mathcal{F}} a_\mathcal{F} \ket{\mathcal{F}} \otimes \ket{\mathcal{M}_\mathcal{F}},
\end{equation}
where $\ket{\mathcal{F}}$ denotes a disordered $Z_2$ flux configuration; $\ket{\mathcal{M}_\mathcal{F}}$ denotes the Majorana fermion conditioned on $\ket{\mathcal{F}}$; and $a_\mathcal{F}$ is a complex scalar conditioned on $\mathcal{F}$, attributed to the interaction-induced hybridization between the two sectors. 
$a_{\mathcal{F}}$ is typically nonzero for any arbitrary flux configurations due to the field-induced mixing between the two sectors.
The concrete form of $a_\mathcal{F}$ is not essential as long as the two sectors remain highly entangled \cite{Knolle17,penghao24}. 
We emphasize that the ansatz in Eq.~\eqref{eq:mft} differs fundamentally from existing microscopic mean-field theories that attempt to explain the origin of the IGP in moderate fields by solving self-consistent equations for quadratic partons \cite{Jiang2020, ZhangNatComm2022}. These approaches inherently fail to capture the entanglement between the two fractionalized quantum sectors and do not treat fluxes as physical degrees of freedom with many-body entanglement among the six $\hat{u}_{ij}$ operators within a hexagon, and hence cannot be used for testing the Majorana metal theory.

%%%%%%%%%%%%%%%%%%%
%%%%%%%%%%%%%%%%%%%

For a given state in each summand of Eq.~\eqref{eq:mft}, translational symmetry is explicitly broken due to the disorder present in a specific instance of $\ket{\mathcal{F}}$. However, since the full Hamiltonian remains translationally invariant, we expect the true ground state to retain this symmetry, potentially with an enlarged unit cell. This expectation enables numerical validation using infinite-size tensor network methods and its comparison with reconstructed Majorana Fermi surface at zero energy, which will be discussed in later sections.
In the mean-field picture described by Eq.~\eqref{eq:mft}, this symmetry is naturally recovered by averaging over all flux configurations $\ket{\mathcal{F}}$. Through this averaging process, we recover a translationally invariant ground state $\ket{\Psi_{\rm IGP}}$ and average quasi-momenta, as all flux patterns related by translations appear in the linear combination with identical coefficients. $\ket{\mathcal{M}_{\mathcal{F}}}$ is approximately the ground state of $H_{\mathcal{M}}= \bra{\mathcal{F}}H\ket{\mathcal{F}}$, and $H_{\mathcal{M}}$ is given by Eq.~\eqref{eq:Majoranahopping} with sign-disorder in $t_{ij}$. Disorder in $t_{ij}$ gives the effective Majorana states as an ensemble average over disorder. Tracing out the flux sector gives the density matrix of Majorana:
\begin{equation} \label{eq:rhom}
    \rho_\mathcal{M} = {\rm Tr}_{\mathcal{F}} \ket{\Psi_{\text{IGP}}}\bra{\Psi_{\text{IGP}}} = \sum_{\{\mathcal{F}\}}\abs{a_\mathcal{F}}^2 \ket{\mathcal{M}_\mathcal{F}}\bra{\mathcal{M}_\mathcal{F}}.
\end{equation} 
This gives the same picture of a thermal Majorana metal in the matter fermion sector, whereby the Majorana ground state is extended with multi-fractality and exhibits a weak anti-localization divergence in its localization length \cite{Huse2012}.  
The extended nature of Eq.~\eqref{eq:rhom}, however, is not directly observable in magnetic systems prior to parton decomposition. This is because it cannot be probed through spin transport due to the absence of global spin conservation, and the extended lowest-energy states carry very little energy, making them inaccessible via energy conductivity measurements.

More relevant for measurements are the finite-energy states, as they directly contribute to dynamical response functions probed in scattering experiments. This motivates the calculation of the Majorana density of states (DOS) under flux disorder, which is directly linked to the spectral function. The precise relationship between the spin spectral function and the Majorana DOS will be discussed in the next section. Before that, it is instructive to first examine how the application of a magnetic field induces a gapless DOS and how the DOS is distributed across energy levels in the presence of disorder. 

\begin{figure}
    \centering
    \includegraphics[width=\linewidth]{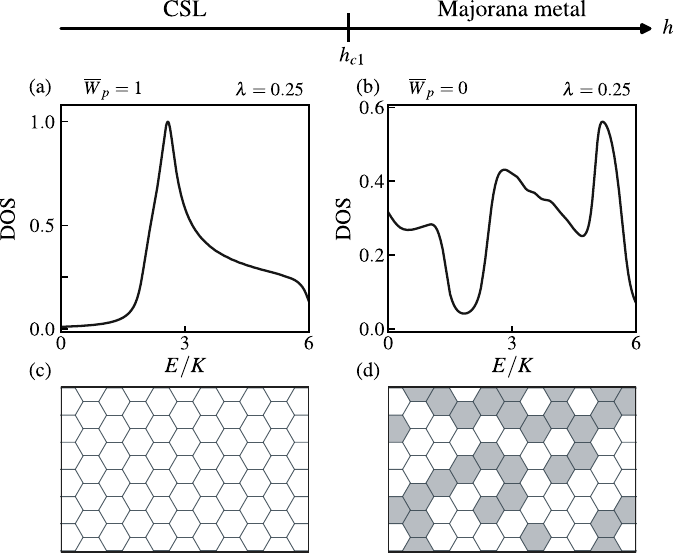}
    \caption{
    Emergence of a Majorana metal in the matter-fermion sector from the chiral spin liquid (CSL) under increasing magnetic field, beyond a critical field $h_{c1}/K \approx 0.45$.
    The top schematic indicates the CSL and Majorana-metal regimes as a function of field $h$.
    Panels (a,b) show the corresponding bulk Majorana density of states (DOS) under periodic boundary conditions, computed from the Majorana hopping model with fixed next-nearest-neighbor hopping $\lambda=0.25$.
    Panel (a) is the clean CSL sector with $\overline W_p=1$, while panel (b) includes flux-sign disorder with $\overline W_p=0$, producing finite low-energy DOS characteristic of the Majorana metal.
    Panels (c,d) show representative plaquette-flux configurations used for the two DOS calculations: white hexagons denote unflipped plaquettes and gray hexagons denote flipped plaquettes.
    }
    \label{fig:tbDOS}
\end{figure}

The external magnetic field does not directly enter the flux-Majorana representation; instead, its primary effect is captured indirectly through a time-reversal-symmetry-breaking perturbation, notably by introducing next-nearest-neighbor (n.n.n.) Majorana hoppings into $H_{\mathcal{M}}$. To quantify the effect of the magnetic field on the gauge sector, we define the ensemble-averaged flux expectation as
\begin{equation}
    \overline{W}_{p} = \frac{1}{N_{\varhexagon}}\sum_{\varhexagon}\sum_{\mathcal{F}} |a_{\mathcal{F}}|^2 \bra{\mathcal{F}}\hat{W}_{p}\ket{\mathcal{F}},
\end{equation}
where $N_{\varhexagon}$ is the total number of hexagonal plaquettes. Because the wavefunction ansatz $\ket{\Psi_{\rm IGP}}$ preserves translation and three-fold rotational symmetries, the expectation values $\langle u_{ij}\rangle$ and $\overline{W}_{p}$ are identical on each bond and plaquette, respectively.
Within the IGP, as the magnetic field increases, the density of flux excitations grows, reducing the flux expectation $\overline{W}_{p}$ from unity (flux-free state) toward zero (half-filled flux state). Numerical simulation of this emergent flux disorder can be done by introducing random sign-flips on link variables $u_{ij}$ with a predefined probability, mimicking the proliferation of flux excitations under a moderate magnetic field \cite{penghao24}. A similar numerical approach has previously been successfully employed to study the thermal Majorana metal phase of Kitaev honeycomb models at elevated temperatures, yielding excellent agreement with quantum Monte Carlo results~\cite{yoshitake2017}.
In Fig.~\ref{fig:tbDOS}, it is demonstrated explicitly how ensemble averaging over random flux configurations leads to a transition from a gapped CSL to a gapless Majorana metal phase with diverging fermion DOS near zero energy.

%%%%%%%%%%%%%%%%%%%
%%%%%%%%%%%%%%%%%%%

\subsection{Majorana spectral function and spin response} \label{sec:D}

The quantum Majorana-metal ansatz in Eq.~\eqref{eq:mft} provides a useful way to connect the iPEPS results with an effective analytical description. In the integrable Kitaev limit, a local spin operator necessarily changes both the flux and Majorana sectors, so the spin dynamics is a composite flux-plus-Majorana response, as reviewed in Sec.~\ref{sec:review}. In the intermediate-field quantum Majorana metal, however, the situation is different. The wave function already contains a dense superposition of flux sectors, and the fluxes are slow compared with the itinerant Majorana fermions. Perturbative estimates and numerical simulations indicate that the field-induced vison bandwidth is much smaller than the Majorana bandwidth, or equivalently that the vison effective mass is much larger than the Majorana effective mass, $m_v\gg m_f$~\cite{Aprem21,Baskaran2023,Feng2025}. Thus, over low but finite energy scales and intermediate time scales, the flux background may be treated as quasi-static, while the spin response can retain direct information about the Majorana spectrum. The ultimate asymptotic low-frequency regime may still depend on the full vison dynamics, but the quasi-static-flux approximation captures the finite-energy response relevant to current tensor-network calculations and experiments of finite energy scales.

We consider the spin structure factor of the intermediate gapless phase,
\begin{equation} \label{eq:s1w}
\begin{aligned}
S_1(\mathbf{k},\omega)
&=
\frac{1}{N}\sum_{ij,\alpha}
e^{-i\mathbf{k}\cdot(\mathbf{r}_i-\mathbf{r}_j)} \\
&\quad \times
 \int dt\, e^{-i\omega t}  \expval{{\rm IGP}|
\sigma_i^\alpha(t)\sigma_j^\alpha(0)
|{\rm IGP}} .
\end{aligned}
\end{equation}
Using the ansatz of Eq.~\eqref{eq:mft}, the correlator can be related to Majorana propagation in an ensemble of flux backgrounds. For notational simplicity, we suppress the spin component index in the following; the detailed derivation can be found in Refs.~\cite{penghao24,Feng2025}. Inserting Eq.~\eqref{eq:mft} gives
\begin{equation} \label{eq:s1t}
    \begin{split}
        \expval{\sigma_i(t)\sigma_j(0)}
        =
        \sum_{\mathcal{F},\mathcal{F}'}
        a_{\mathcal{F}'}^* a_{\mathcal{F}}
        \langle
        \mathcal{F}',\mathcal{M}_{\mathcal{F}'}
        |
        \sigma_i(t)\sigma_j(0)
        |
        \mathcal{F},\mathcal{M}_{\mathcal{F}}
        \rangle .
    \end{split}
\end{equation}
Acting with a spin operator locally flips the corresponding $Z_2$ link variable and therefore changes the two adjacent plaquette fluxes, while also inserting a Majorana operator. Under the separation of slow flux and fast Majorana dynamics, the relevant matrix element can be approximated as
\begin{equation} \label{eq:apprx1}
\begin{aligned}
&\langle
\mathcal{F}',\mathcal{M}_{\mathcal{F}'}
|
\sigma_i e^{-iHt}\sigma_j
|
\mathcal{F},\mathcal{M}_{\mathcal{F}}
\rangle
\\
&\quad \approx
\delta_{\mathcal{F}'_i,\mathcal{F}_j}
\langle
\mathcal{M}_{\mathcal{F}'}
|
c_i e^{-i\tilde{H}_{\mathcal{F}_j}t}c_j
|
\mathcal{M}_{\mathcal{F}}
\rangle .
\end{aligned}
\end{equation}
Here $\mathcal{F}_j$ denotes the flux configuration obtained from $\mathcal{F}$ after the spin operator at site $j$ flips the adjacent link, and $\mathcal{F}'_i$ is defined analogously for the bra state. The Kronecker symbol enforces the matching of the flux sectors after the two spin insertions, in keeping with the flux orthogonality described in Sec.~\ref{sec:bondf}. The Hamiltonian $\tilde{H}_{\mathcal{F}_j}$ is the Majorana Hamiltonian conditioned on the corresponding quasi-static flux background.
\begin{figure*}[t]
    \centering
\includegraphics[width=0.98\textwidth]{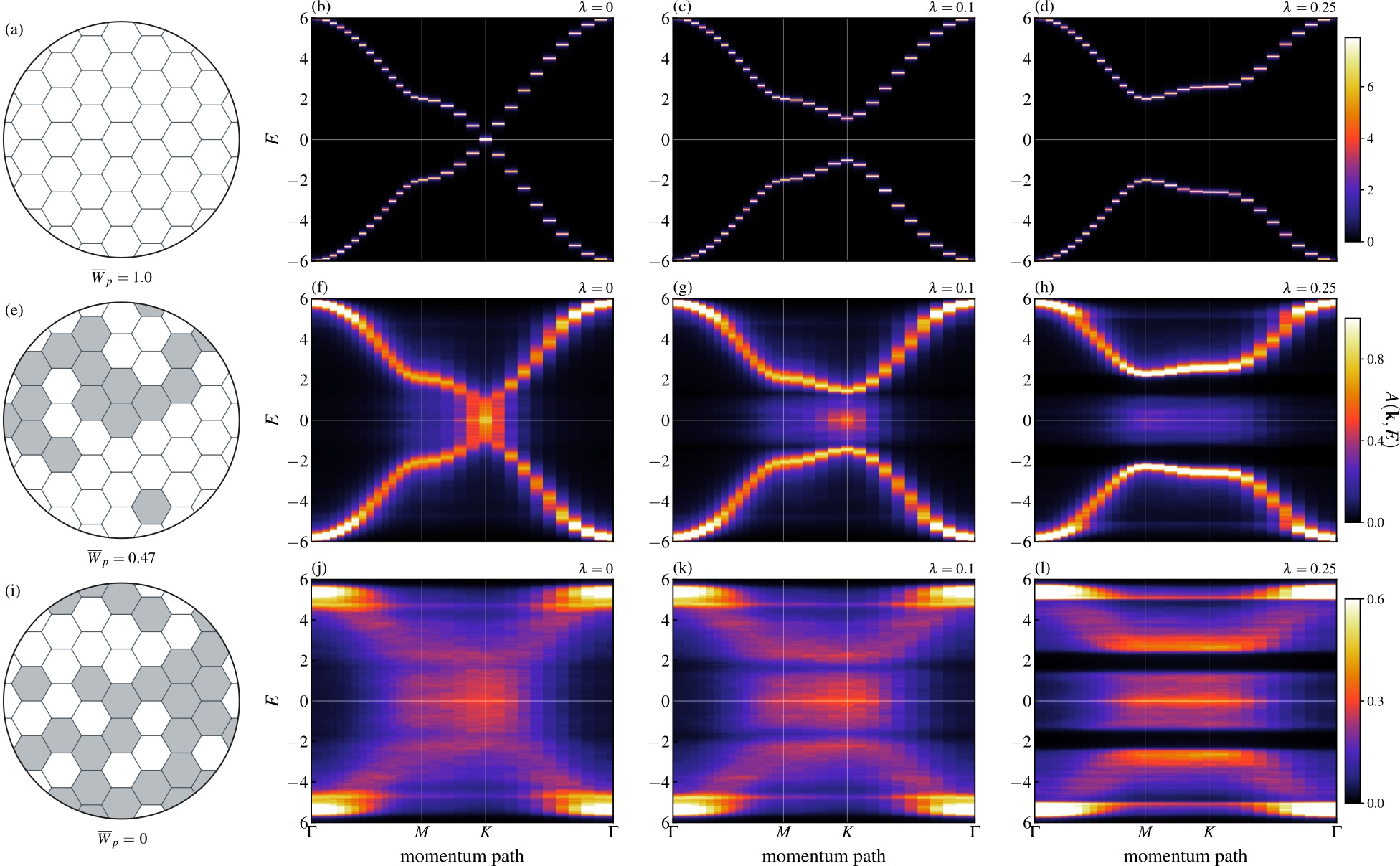}
\caption{
Center-of-mass-averaged momentum-resolved Majorana spectral function of the honeycomb Kitaev model with flux disorder and next-nearest-neighbor hopping $\lambda$.
Panels (a,e,i) show representative flux configurations used in the ensemble average, with filled hexagons denoting flipped plaquette fluxes and white hexagons denoting unflipped plaquettes. The three rows correspond to increasing $\pi$-flux density, in keeping with the corresponding disorder-averaged flux $\overline W_p$.  Panels (b--d), (f--h), and (j--l) show the corresponding disorder-averaged spectral function $A(\mathbf{k},E)$ along the $\Gamma$--$\rm M$--$\rm K$--$\Gamma$ path for $\overline W_p = 1.0$, $0.47$, $0$; and $\lambda=0$, $0.1$, $0.25$, respectively.  In the clean limit (b--d), the low-energy modes are centered near the $\rm K$ point for weak next-nearest-neighbor hopping, while for larger $\lambda$ the lowest-energy spectral weight shifts toward $\rm M$.  With weak flux disorder (f--h), gapless spectral weight first develops near $\rm K$ for small $\lambda$ and near $\rm M$ for larger $\lambda$; at stronger flux disorder (j--l), low-energy weight appears near both $\rm M$ and $\rm K$, with the weight at $\rm K$ slightly stronger. These demonstrate, at an effective tight-binding level, the emergence of Majorana-metallic states from gapped chiral states due to strong fluctuations in $Z_2$ fluxes.
}
    \label{fig:specfunc}
\end{figure*}
For a dense and broadly distributed flux ensemble, the local difference between
$\mathcal{F}$ and $\mathcal{F}_j$ is not expected to affect the long-wavelength,
disorder-averaged spectral response. If the weights $|a_{\mathcal{F}}|^2$ are
taken to be nearly uniform within a fixed flux-density sector, Eq.~\eqref{eq:s1t}
reduces to
\begin{equation}
\begin{split}  \label{eq:correlator1}
     \expval{\sigma_i(t)\sigma_j(0)}
     \approx
     \sum_{\mathcal{F}}
     |a_{\mathcal{F}}|^2
     \mel{\mathcal{M}_{\mathcal{F}}}
     {c_i(t)c_j(0)}
     {\mathcal{M}_{\mathcal{F}}} .
\end{split}
\end{equation}
This is the adiabatic approximation familiar from the spin response in the
integrable limit~\cite{Knolle2014,Knolle15,Feng2025}, now applied to the
dense-flux intermediate phase.

Equation~\eqref{eq:correlator1} is the main motivation for the following. 
The left-hand side is precisely the real-space, time-dependent spin correlator whose Fourier transform gives the dynamical spin structure factor $S_1(\mathbf{k},\omega)$, while the right-hand side is an ensemble average of Majorana propagators evaluated in quasi-static flux backgrounds. 
Thus, within the adiabatic-flux approximation, the low-energy spin response can be obtained by Fourier transforming a flux-averaged Majorana propagator. 
The only subtlety is that a single flux configuration $\mathcal F$ generally breaks translation symmetry, so crystal momentum is not a good quantum number sector by sector. 
However, the full flux ensemble restores translation symmetry on average, which motivates defining a center-of-mass-averaged, momentum-resolved Majorana spectral function. 
This is the quantity that can be directly compared with the momentum-resolved spin response $S_1(\mathbf{k},\omega)$ obtained in the spin basis.
For compactness we introduce the flux-ensemble average
\begin{equation}
\begin{split}
\left\langle O_{\mathcal F}\right\rangle_{\mathcal F}
=
\sum_{\mathcal F} p_{\mathcal F} O_{\mathcal F},  \;\;
p_{\mathcal F}
= |a_{\mathcal F}|^2 ,
\end{split}
\label{eq:fluxaverage}
\end{equation}
where, in the high-density flux regime, $p_{\mathcal F}$ may be approximated
by an equal-weight ensemble within a fixed flux-density sector. We also define
the fixed-flux Majorana Green function $G_{\mathcal F}(i,j;t)
\equiv
\mel{\mathcal{M}_{\mathcal F}}
{c_i(t)c_j(0)}
{\mathcal{M}_{\mathcal F}}$. 
Then Eq.~\eqref{eq:correlator1} can be written schematically as
\begin{equation}
\label{eq:spincorr}
\expval*{
\boldsymbol{\sigma}_i(t)\!\cdot\!
\boldsymbol{\sigma}_j(0)}
\sim
\left\langle
G_{\mathcal F}(i,j;t)
\right\rangle_{\mathcal F}.
\end{equation}
Here the site index $i$ includes both a Bravais-cell coordinate and an internal sublattice index. Thus, in frequency space, the spin response is reduced to a flux-ensemble average of fixed-flux Majorana spectral functions. We define the Majorana spectral function in a fixed flux background $\mathcal F$. Writing a site as $(\mathbf r,\alpha)$, where $\mathbf r$ labels
a Bravais unit cell and $\alpha$ labels the internal sublattice degree of freedom, the traced fixed-flux spectral function is
\begin{equation}
A_{\mathcal F}
(\mathbf r_1,\mathbf r_2;\omega)
=
\sum_{n,\alpha}
\phi_{n,\mathcal F,\alpha}(\mathbf r_1)
\phi^{*}_{n,\mathcal F,\alpha}(\mathbf r_2)
\delta(\omega-E_{n,\mathcal F})
\label{eq:specfuncF}
\end{equation}
Here $\phi_{n,\mathcal F,\alpha}(\mathbf r)$ is the $\alpha$-sublattice
component of the $n$th eigenstate of the Majorana Hamiltonian
$\tilde H_{\mathcal F}$, and $E_{n,\mathcal F}$ is its eigenenergy. The
flux-averaged real-space spectral function is then
\begin{equation}
\begin{split}
A(\mathbf r_1,\mathbf r_2;\omega)
=
\left\langle
A_{\mathcal F}
(\mathbf r_1,\mathbf r_2;\omega)
\right\rangle_{\mathcal F}.
\end{split}
\label{eq:specfunc}
\end{equation}
Although each fixed flux configuration $\mathcal F$ breaks translation
symmetry, the full flux ensemble is translation invariant on average.
Momentum is therefore not a good quantum number of an individual
$\tilde H_{\mathcal F}$, but it becomes meaningful after center-of-mass
averaging. For a fixed $\mathcal F$, define
\begin{equation}
\begin{split}
\overline{A}_{\mathcal F}
(\boldsymbol{\rho};\omega)
&\equiv
\frac{1}{N}
\sum_{\mathbf R}
A_{\mathcal F}
(\mathbf R+\boldsymbol{\rho},\mathbf R;\omega),
\end{split}
\label{eq:cmaverage}
\end{equation}
where $\boldsymbol{\rho}=\mathbf r_1-\mathbf r_2$ is the relative coordinate
and $\mathbf R$ is the center-of-mass coordinate. To obtain its momentum
representation, expand the fixed-flux eigenstate in the plane-wave basis of the
underlying translation-invariant lattice: $\phi_{n,\mathcal F,\alpha}(\mathbf r)
=
\frac{1}{\sqrt{N}}
\sum_{\mathbf k}
c^{\,n,\mathcal F}_{\alpha\mathbf k}
e^{i\mathbf k\cdot\mathbf r}$. 
Substituting this decomposition into Eq.~\eqref{eq:cmaverage} gives
\begin{equation}
\overline{A}_{\mathcal F}
(\boldsymbol{\rho};\omega)
=
\frac{1}{N}
\sum_{n,\alpha,\mathbf k}
e^{i\mathbf k\cdot\boldsymbol{\rho}}
\left|
c^{\,n,\mathcal F}_{\alpha\mathbf k}
\right|^2
\delta(\omega-E_{n,\mathcal F}). 
\label{eq:specfuncderive2}
\end{equation}
Fourier transforming with respect to the relative coordinate then yields the
average-momentum spectral function for this fixed flux pattern,
\begin{equation}
\begin{split}
A_{\mathcal F}(\mathbf k,\omega)
&=
\sum_{\boldsymbol{\rho}}
e^{-i\mathbf k\cdot\boldsymbol{\rho}}
\overline{A}_{\mathcal F}
(\boldsymbol{\rho};\omega)
\\
&=
\sum_{n,\alpha}
\left|
c^{\,n,\mathcal F}_{\alpha\mathbf k}
\right|^2
\delta(\omega-E_{n,\mathcal F}) .
\end{split}
\label{eq:specfunckF}
\end{equation}
Finally, the average-momentum-resolved Majorana spectral function
entering the approximation to $S_1(\mathbf k,\omega)$ is
\begin{equation}
\begin{split}
A(\mathbf k,\omega)
=
\left\langle
A_{\mathcal F}(\mathbf k,\omega)
\right\rangle_{\mathcal F}.
\end{split}
\label{eq:specfunck}
\end{equation}
In practice, this final average is taken over flux configurations that are
inequivalent under lattice translations; translations of a given configuration
are already accounted for by the center-of-mass average in
Eq.~\eqref{eq:cmaverage}.

The importance of Eqs.~\eqref{eq:specfunckF}--\eqref{eq:specfunck} is that they
provide a direct bridge between a gauge-invariant spin response and the
Majorana spectral function in the quantum Majorana-metal regime. This is
qualitatively different from the integrable Kitaev limit, where a local spin
flip necessarily creates a flux pair together with a Majorana excitation, so
the spin structure factor is a composite flux-plus-fermion response with no
simple one-to-one correspondence to either the flux spectrum or the Majorana
band structure alone. In the intermediate-field Majorana metal, the fluxes are
already proliferated and form a dense, slowly fluctuating background. After
averaging over this quasi-static flux ensemble, the spin correlator effectively
tracks the propagation of Majorana fermions in disordered flux sectors. Thus
the low-energy continuum in $S_1(\mathbf{k},\omega)$ is not merely a broad
incoherent feature; within this picture it is a finite-energy fingerprint of
the disorder-averaged Majorana density of states and momentum-resolved
Majorana spectral function. This relation makes the spin response a much more
informative probe of the Majorana metal than in the flux-free integrable
regime, while still respecting the fact that only gauge-invariant combinations
are experimentally observable.

Figure~\ref{fig:specfunc} presents the averaged Majorana spectral function  $A(\mathbf{k},\omega)$ over the fermion sectors conditioned on different flux fillings $\overline{W}_p$, characterized by the expectation value of the flux operator $\overline{W}_p$, and the TR-breaking next-nearest-neighbor (n.n.n.) Majorana hopping  $\lambda$.
The first row [panels (a--d)] of Fig.~\ref{fig:specfunc} shows the spectral function of the non-interacting Majorana Hamiltonian in a flux-free uniform gauge field. At $\lambda = 0$, the well-known Dirac nodes appear at the  $\rm K$  points, which become gapped upon introducing the TR-breaking perturbation. Notably, the n.n.n. perturbation evaluates to zero at the  $\rm M$  points, leading to a shift in the lowest-energy modes toward the  $\rm M$  points as $\lambda$ increases, as observed in Fig.~\ref{fig:specfunc}(c,d).
The second row [panels (e--h)] and the third row [panels (i--l)] of Fig.~\ref{fig:specfunc} display the averaged spectral function of Majorana fermions over random flux configurations at sparse ($\overline{W}_p = 0.47$) and dense ($\overline{W}_p = 0.0$, half-filling) flux fillings, respectively. These disordered flux configurations give rise to gapless modes near both the  $\rm M$  and  $\rm K$  points, which is responsible for logarithmic scaling of the low-energy density of states of a Majorana metal [Fig.~\ref{fig:tbDOS}(b)]. For the sparse flux filling shown in the second row of Fig.~\ref{fig:specfunc}, a gapless continuum emerges immediately above zero energy [Fig.~\ref{fig:specfunc}(g,h)], while the higher-energy branches remain sharp. However, at approximately half-filling in the random flux sector (third row), the Majorana bands become significantly broadened, leading to a nearly featureless spectral function across all energies. 
Notably, the presence of next-nearest-neighbor (n.n.n.) hopping significantly alters the spectral weight distribution at finite energy, as shown in Fig.~\ref{fig:specfunc}(j–l). In the TR-symmetric case ($\lambda = 0$), no finite-energy gap opens, and the $k$-integrated spectral function closely resembles that of a thermal Majorana metal \cite{Knollethermal}. However, for nonzero $\lambda$, as shown in Fig.~\ref{fig:specfunc}(k,l), a mini-gap emerges around  $E/K \sim 2$. This feature could serve as a useful diagnostic for identifying the quantum Majorana metal numerically, a point we will elaborate on in the following discussion.
We point out that the precise mechanism underlying the emergence of this mini-gap remains an open question for future research. However, qualitatively, it may be understood as a consequence of the TR-breaking term $\lambda$, which in a $p+ip$ superconductor generally suppresses low-energy states. This depletion of states at low energies causes the flux-induced Majorana resonance to shift, leading to a more concentrated spectral weight at lower energies and the appearance of a mini-gap at finite energy.

It is important to emphasize that the random-flux-induced gapless continuum shown in Fig.~\ref{fig:specfunc} is not a result of spin fractionalization in the microscopic spin model but rather a continuum formed by the already fractionalized Majorana fermions that constitute the gapless quantum Majorana metal. In the following, we will discuss how this continuum can be verified through unbiased iPEPS and single-mode approximations within tensor network approaches.

The approximation in Eq.~\eqref{eq:correlator1} and its resolution in average momentum in Eq.~\eqref{eq:specfunck} are central to identifying the IGP as a quantum Majorana metal, because it relates the gauge-invariant spin response to the disorder-averaged Majorana correlator. In contrast to the integrable Kitaev limit, where a local spin flip necessarily creates both fluxes and Majorana excitations and therefore gives a composite response, the dense and slow flux background in the IGP allows the spin dynamics to track the Majorana spectral function more directly.

\begin{figure}
    \centering
    \includegraphics[width=\linewidth]{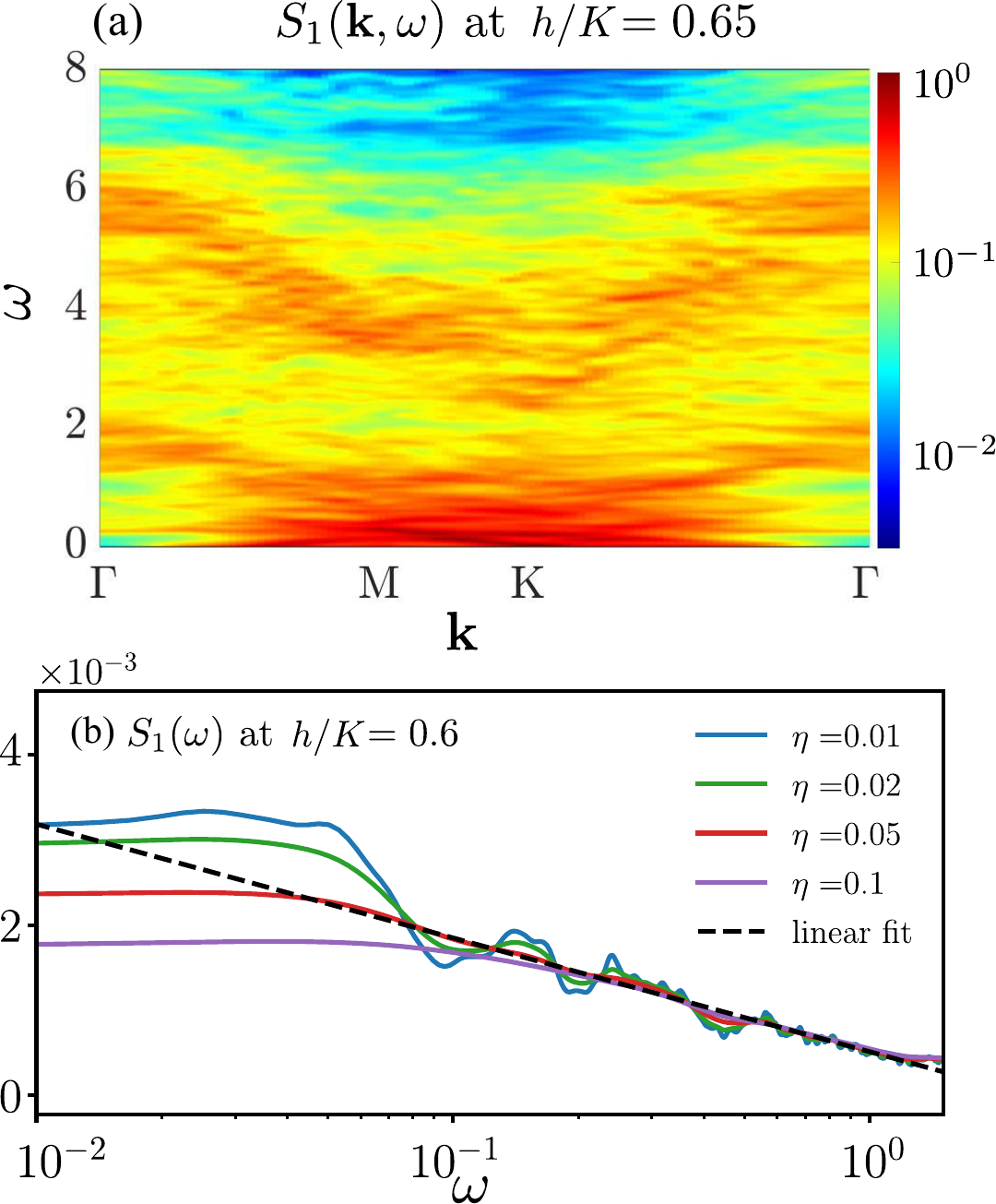}
    \caption{(a) Total single-spin spectrum $S_1(\mathbf{k},\omega)$ at $h/K=0.65$ inside the intermediate gapless phase, i.e. the quantum Majorana-metal regime, along the momentum path $\Gamma$--$M$--$K$--$\Gamma$. The spectrum shows a gapless continuum immediately above zero energy, separated from the upper branch by a region of strongly suppressed spectral weight around $\omega/K\sim 2$. (c) Normalized momentum-integrated spectra $S_1(\omega)$ at $h/K=0.6$ for different broadening factors $\eta$. The logarithmic scaling is consistent with a Majorana-metal response. All data are obtained by iPEPS with bond dimension $D=5$. Figures adapted from Ref.~\cite{wang2024}.}
    \label{fig:igp_ipeps}
\end{figure}

Having examined the Majorana sector on the right-hand side of Eq.~\eqref{eq:correlator1}, a direct test of the ansatz,
$\ket{\Psi_{\text{IGP}}}=\sum_{\mathcal{F}}a_{\mathcal{F}}\ket{\mathcal{F}}\otimes\ket{\mathcal{M}_{\mathcal{F}}}$,
is to compute the corresponding spectrum directly in the spin basis, i.e., the left-hand side of Eq.~\eqref{eq:correlator1}. Figure~\ref{fig:igp_ipeps}(a,b) shows iPEPS results for $S_1(\mathbf{k},\omega)$ at $h/K=0.60$ and $h/K=0.65$ inside the putative quantum Majorana-metal phase, or IGP. The spectra exhibit a broad, smeared low-energy continuum, consistent with the tight-binding Majorana calculation in a finite-density flux background shown in Fig.~\ref{fig:specfunc}(k,l). No observable gap appears in $S_1(\mathbf{k},\omega)$ in Figs.~\ref{fig:igp_ipeps}(a,b). In particular, the strong spectral weight immediately above zero energy near the $\rm M$ and $\rm K$ points reflects the gapless fermionic sector expected for a Majorana metal shown in Fig.~\ref{fig:specfunc}(k,l).

Beyond this low-energy consistency, the iPEPS spectra also reveal a finite-energy separation between two spectral branches centered around $\omega/K \sim 2$, most clearly visible in Fig.~\ref{fig:igp_ipeps}(a). This separation has a simple interpretation from the random-flux Majorana picture. In the clean flux-free chiral spin liquid with a nonzero TR-breaking term, the bulk Majorana bands are gapped, and the lower edge of the bulk band lies at an energy of order $\omega/K\sim 2$, as illustrated in Fig.~\ref{fig:specfunc}(d). A $Z_2$ flux binds a Majorana zero mode inside this clean bulk gap. When the flux density becomes large, these flux-bound Majorana zero modes hybridize with one another and broaden into a smeared in-gap band centered near zero energy, producing the gapless continuum shown in Fig.~\ref{fig:specfunc}(h,l). However, this hybridized zero-mode band still has a finite bandwidth: it fills the low-energy part of the clean bulk gap, but does not immediately merge with the remnant high-energy bulk Majorana bands. As a result, an energy window with strongly depleted spectral weight remains around the original clean-limit band edge, near $\omega/K\sim 2$, as is also visible in the DOS shown in Fig.~\ref{fig:tbDOS}(b). The same structure appears in the spin-basis iPEPS spectrum in Fig.~\ref{fig:igp_ipeps}(a): the lower branch corresponds to the hybridized flux-bound Majorana zero-mode continuum, while the upper branch is the remnant of the bulk Majorana bands. This agreement between the spin-basis iPEPS spectra, Fig.~\ref{fig:igp_ipeps}(a), and the Majorana-basis random-flux calculation, Fig.~\ref{fig:specfunc}(k,l), provides further evidence that the IGP is well described as a quantum Majorana metal.

Further evidence comes from the momentum-integrated spectrum $S_1(\omega)$. If the low-energy spectral weight is approximately uniform in momentum, $S_1(\omega)$ should follow the low-energy density of states of the Majorana metal. The random-flux Majorana description predicts a logarithmically enhanced low-energy density of states, ${\rm DOS}(E)\propto \ln(1/E)$ as $E\rightarrow 0$, and this behavior is observed in the iPEPS results for several broadening factors $\eta$, as shown in Fig.~\ref{fig:igp_ipeps}(b). Even for the smallest $\eta$, no observable gap or trend toward gap opening is found down to the energy scale $\omega/K\sim 0.01$. This scale is comparable to, or smaller than, previous estimates of putative gaps obtained from parton mean-field theories, which do not capture the strong interference effects associated with emergent flux disorder.

We stress, however, that this comparison should be viewed as strong evidence for the finitely-low-energy and intermediate-time phenomenology of a quantum Majorana metal, rather than as a complete resolution of the ultimate universal ground-state theory. As discussed in Sec.~\ref{sec:reliability} and summarized in Table~\ref{tab:igpcriteria}, distinguishing the asymptotic low-frequency behavior of a highly entangled two-dimensional many-body state remains challenging for current tensor-network methods. In particular, iPEPS calculations are performed at finite bond dimension and finite real-time evolution, so they are most reliable at low but nonzero energy scales where the spectra can be directly compared with effective Majorana descriptions. The agreement between the spin response from iPEPS and the random-flux Majorana spectral function therefore supports the Majorana-metal interpretation of the IGP, but it should not be read as excluding all alternative scenarios at the deepest infrared scale. The same caveat holds for the forthcoming discussion of the spin-dimer response.  Establishing the ultimate universal limit will require further cross-checks among larger-bond-dimension real-time tensor networks, finite-temperature purification with dynamical fluxes, transport calculations, and material-specific extended Hamiltonians.

\begin{figure}
    \centering
    \includegraphics[width=\linewidth]{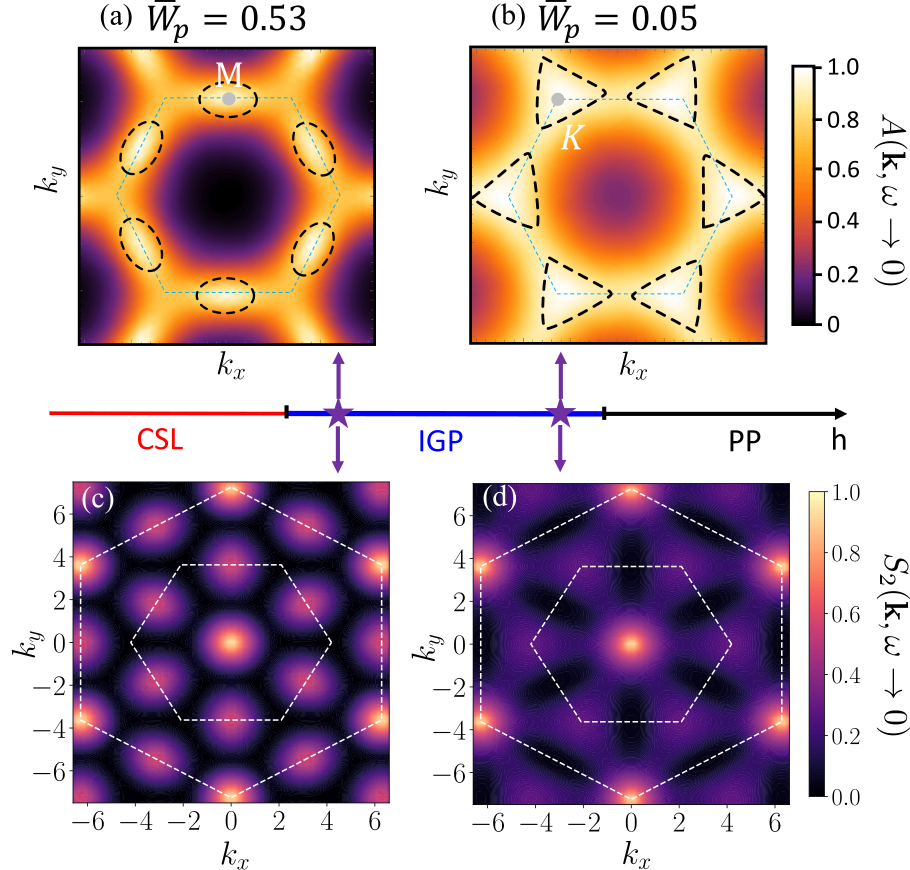}
    \caption{Putative Majorana FS and its response. (a,b) Majorana spectral function at zero energy on the FS, $A(\mathbf{k},\omega=0)$, in the intermediate gapless phase (IGP) sandwiched between the chiral spin liquid (CSL) phase and the polarized phase (PP), for (a) $\overline{W}_{p}=0.53$ and (b) $\overline{W}_{p}=0.05$, where $\overline{W}_{p}$ is the ensemble average of the $\pi$-flux density. $h$ represents the strength of the Zeeman field [c.f. Eq.~\eqref{eq_ham}]. $\lambda=0.25$ is used for the calculations. Black dashed lines mark out the putative Majorana FS within which the zero-energy spectral function has the largest intensity, and grey dashed lines represent the first Brillouin zone.
    (c,d) The zero-energy contour of the dimer spectrum $S_2(\mathbf{k}, \omega)$ at different flux fillings corresponding to different magnetic fields within IGP. Data obtained using four-Majorana correlation (r.h.s. of Eq.~\eqref{eq:dimer}) and the Majorana FS marked out by the black dashed lines in (a,b). The white dashed lines represent the first and the second Brillouin zones. Panels (a) and (b) are adapted from Ref.~\cite{penghao24}.}
    \label{fig:mfs}
\end{figure}

\subsection{Majorana Fermi surface and dimer response}
Since we have already established the gapless nature of the IGP in the previous section and demonstrated that it can be described as a Majorana metal in the Majorana fermion sector, where flux dynamics is relatively slow and can thus be neglected, it is natural to ask whether an emergent Fermi surface (FS) structure appears at zero energy in the Majorana fermion spectrum and its observable consequences in spin response functions.
Indeed, it has been proposed that the same phase could be described in terms of a Fermi surface of complex fermions, where the effective gauge theory is $U(1)$ instead of the $Z_2$ gauge field associated with real Majorana fermions \cite{Patel12199,hickey2019emergence}. However, while these pioneering works successfully captured the transition using unbiased numerical techniques such as ED and DMRG, the proposed $U(1)$ gauge structure remains a conjecture. The underlying assumption is that the pairing of canonical fermions in the CSL phase at low fields is suppressed under a moderate magnetic field, leading to a transition from a topological  $p$-wave spinon superconductor to a $U(1)$ metallic phase with a neutral spinon FS. However, the dynamical response in these studies was not tested with sufficiently high energy-momentum resolution, and the calculations were limited to finite-size ED clusters or DMRG cylinders, which may not fully capture the properties of an extended system.
In this section, we review the recent proposal on a putative Majorana-Fermi surface with a $Z_2$ gauge structure, producing results that are quantitatively in keeping with the unbiased numerical simulations. The first strength of this framework is its parsimony: the gauge structure of the IGP is naturally inherited from the CSL and does not require an untestable mechanism for transitioning to a $U(1)$ gauge field at higher fields instead of undergoing a confinement transition. More importantly, the Majorana metal yields a strikingly consistent dynamical response across both low and finite energies, as already seen in the spectral function analysis from the previous section, and a Majorana FS follows naturally.
We now turn to a discussion of the zero-energy Majorana Fermi surface in the IGP, viewed as a quantum Majorana metal phase. To verify this, we compare next-order correlations between Majorana fermions with unbiased iPEPS results for next-order (dimer) correlations in the spin model.

An estimation of the Majorana FS can be obtained via the averaged $A(\mathbf{k} \in {\rm BZ},\omega=0)$ over tight-binding Majorana fermions conditioned on different flux fillings set by $\overline{W}_{p}$. Here, the magnetic field enters the picture indirectly by affecting the average flux $\overline{W}_{p}$. 
As shown in Fig.~\ref{fig:mfs}(a,b), the presence of gapless states around $\rm M$ and $\rm K$ in the Majorana sector suggests a set of zero-energy modes corresponding to a set of definite momenta along the boundary of the first BZ, forming an effective Majorana FS. 
For sparse flux density under weak magnetic fields, zero-energy states are mainly found around the $\rm M$ point, as reflected by the strong signal near $A(\mathbf{k} \sim \rm M,\omega=0)$ shown in Fig.~\ref{fig:mfs}(a), see also Fig.~\ref{fig:specfunc}(h). However, for larger flux filling under stronger magnetic fields, it is $A(\mathbf{k} \sim \rm K,\omega=0)$ that has the largest weight, as shown in Fig.~\ref{fig:mfs}(b), see also Fig.~\ref{fig:specfunc}(l), and thus zero-energy states concentrate around $\rm K$ points for magnetic fields that correspond to dense flux fillings.  
This is also consistent with the low-energy spectral function reported in Ref.~\cite{penghao24}, where the strongest zero-energy signal occurs at different momentum for different magnetic fields. 
This shift in the zero-energy momenta can be qualitatively understood in the clean limit without flux disorder. In the flux-free uniform gauge, the band edge in the Majorana sector is at the $\rm M$ point for moderate n.n.n. hopping $\lambda$, which is distinct from weak n.n.n. hopping whereby the band edge remains at $\rm K$ point, as depicted in Fig.~\ref{fig:specfunc}(b,c). Hence for large enough n.n.n. hopping, intuitively, it is the $\rm M$ point that is more susceptible to be rendered gapless by random fluxes.

The spectral function at the Fermi energy provides information of the topology of the Majorana FS. Based on this, it is possible to verify the validity of the FS by checking the next-order dynamical correlations (four-Majorana correlation) given the Majorana FS topology, which can be compared to the next order dynamical spin correlation, i.e. the dimer-dimer correlation relevant for Raman and RIXS experiments. 
The connection between the spin dimer correlation and the four-Majorana correlation can be approximated by the same token of Eq.~\eqref{eq:spincorr} in Sec.~\ref{sec:D}, that is
\begin{equation}
    \expval*{\boldsymbol{\mathcal{D}}_i(t) \cdot \boldsymbol{\mathcal{D}}_j} \sim \langle\bra{\mathcal{M}_\mathcal{F}} c_{i}(t) c_{i+z}(t) c_j c_{j+z}\ket{\mathcal{M}_\mathcal{F}}\rangle_{\mathcal{F}} \label{eq:dimer}
\end{equation}
whose Fourier transformation is the dimer spectral function $S_2(\mathbf{k}, \omega) = {\rm F.T.}\{\expval*{\boldsymbol{\mathcal{D}}_i(t) \cdot \boldsymbol{\mathcal{D}}_j}\}$. Here $\mathcal{D}_j^\alpha = \sigma_{j}^\alpha \sigma_{j+z}^\alpha$ denotes the spin dimer operator, and $\expval{}_\mathcal{F}$ the ensemble average over flux disorder conditioned on a certain filling factor. The left- and right-hand sides of Eq.~\eqref{eq:dimer} can be computed by iPEPS and disorder-averaged tight-binding Majorana models respectively, which can thereafter be compared to each other for the validation of the Majorana FS.

%%%%%%%%%%%%%%%
\begin{figure}
    \centering
    \includegraphics[width=\linewidth]{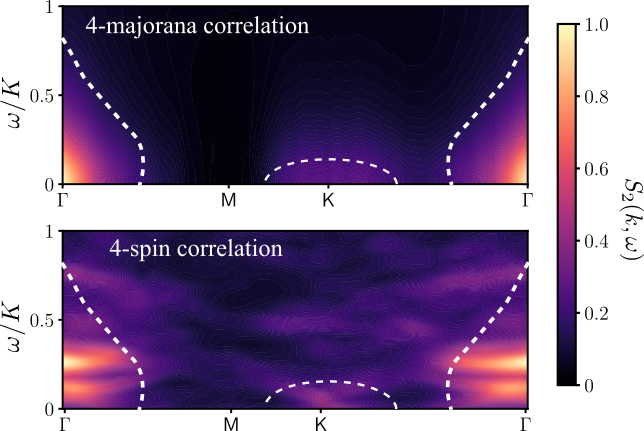}
    \caption{Comparison between the four-Majorana correlation (upper panel) and the dimer-dimer correlation in the intermediate gapless phase obtained by iPEPS (lower panel) along the momentum cut through high-symmetry points $\Gamma\rm M K \Gamma$. The white dashed lines are eye-guiding lines that enclose the energy-momentum region having the strongest intensities, i.e. those near $\rm K(K')$ and $\Gamma$ at the lowest energy; while signals around $\rm M$ point are negligible. Figure adapted from Ref.~\cite{penghao24}.}
    \label{fig:ipepsvsmft}
\end{figure}
%%%%%%%%%%%%%%%%%%%%%%%%%

As an example, consider the Majorana FS, which forms rounded triangular regions around the K points, as illustrated by the thick closed dashed lines in Fig.~\ref{fig:mfs}(b). This corresponds to a high density of random fluxes near half-filling.
For the right-hand side of Eq.~\eqref{eq:dimer}, the low-energy behavior can be estimated using a single-mode approximation around the soft modes near the $\rm K$  and $\rm K^{\prime}$  points. This allows for a straightforward calculation of the four-Majorana correlation function, which, in the Lehmann spectral representation, takes a similar form of Eq.~\eqref{eq:s2z}.  
The zero-energy cut for the four-Majorana correlation, $S_2(\mathbf{k}, \omega \rightarrow 0)$, which arises from the Majorana FS delineated by the black dashed lines in Fig.~\ref{fig:mfs}(b), is presented as a contour plot in Fig.~\ref{fig:mfs}(d). A highly discernible rotationally symmetric pattern emerges, where the Majorana FS covering the $\rm K$ and $\rm K'$ points produces a strong signal at the $\Gamma$ point, along with weaker intensity at $\rm K$ and $\rm K'$, while no signal appears at the $\rm M$ points.  
The absence of intensity forms a distinct paddle-like shape along the momentum path $\Gamma-\rm M-\Gamma$ in Fig.~\ref{fig:mfs}(d), reflecting the characteristic spectral weight distribution of the Majorana FS, which arises due to the presence of the finite density of random fluxes. 
Notably, the low-energy scattering signatures evolve with applied magnetic field, which helps explain why different studies predict distinct shapes and locations for the spinon Fermi surface \cite{Patel12199,Yuanming2018}. This variation arises because changing the flux density qualitatively alters the topology and position of the Majorana Fermi surface. Since the flux density itself depends on the magnetic field, both the Majorana Fermi surface and the associated low-energy scattering intensities become field-dependent -- analogous to field-driven Lifshitz transitions that interpolate between different gapless spinon bands' topology in Mott insulators \cite{Feng_gapless,Feng2020}. As shown in Fig.~\ref{fig:mfs}(a,c), at a lower field within the intermediate gapless phase, the Majorana Fermi surface appears around the M points rather than the K points, and the corresponding $S_2$ scattering intensity is concentrated near the $\rm M$ and $\Gamma$ points.

\begin{table*}[t]
\begingroup
\footnotesize
\setlength{\tabcolsep}{3.5pt}
\renewcommand{\arraystretch}{1.15}
\begin{tabular}{p{0.10\textwidth}p{0.14\textwidth}p{0.43\textwidth}p{0.25\textwidth}}
\hline
Lattice & Material & Experimental gapless signature & Thermal transport \\ 
\hline\hline

Honeycomb 
& $\alpha$-RuCl$_3$ 
& Field-induced gapless signatures reported by thermodynamics/NMR~\cite{Zheng17}; defect-induced Friedel-like oscillations in monolayers interpreted as neutral Fermi-surface scattering~\cite{Kohsaka24}$^\dagger$
& No sizable residual $\kappa_0/T$ reported in ultralow-$T$ thermal conductivity~\cite{nokappaKitaev18}
\\ \hline

Honeycomb 
& Na$_2$Co$_2$TeO$_6$ 
& Field-induced disordered regime with spinon-continuum-like spectral weight reported in INS~\cite{Gaoting24}$^\ddagger$
& No evidence for mobile magnetic quasiparticles beyond phonons in longitudinal thermal conductivity~\cite{Hong2024}
\\ \hline

Honeycomb 
& H$_3$LiIr$_2$O$_6$ 
& Enhanced low-energy thermodynamics, with $C/T$ reported to increase strongly at low $T$, together with NMR signatures of abundant low-energy states~\cite{Kitagawa2018}
& No established residual-$\kappa_0/T$ comparison
\\ \hline

Triangular 
& dmit-131 
& Sizeable residual linear term, $\gamma\sim 20~{\rm mJ\,K^{-2}\,mol^{-1}}$~\cite{Taillefer19}
& Vanishing or negligible residual magnetic $\kappa_0/T$~\cite{Taillefer19,Ni19}
\\ \hline

Triangular 
& NaYbS$_2$ 
& Small residual linear term, $\gamma\sim 1.1~{\rm mJ\,K^{-2}\,mol^{-1}}$~\cite{Zhang24}
& Vanishing or negligible residual $\kappa_0/T$~\cite{Zhang24}
\\ \hline

Triangular 
& NaYbSe$_2$ 
& Small residual linear term, $\gamma\sim 1.1~{\rm mJ\,K^{-2}\,mol^{-1}}$~\cite{Zhang24}
& Vanishing or negligible residual $\kappa_0/T$~\cite{Zhang24}
\\ \hline

Triangular 
& YbMgGaO$_4$ 
& Power-law magnetic specific heat, $C_m\sim T^{0.74}$~\cite{Xu16}
& No significant magnetic thermal conductivity~\cite{Xu16}
\\ \hline

\end{tabular}
\caption{
Comparison between experimental low-energy signatures and thermal transport in selected QSL candidate materials. 
The table highlights a recurring phenomenology: thermodynamic, spectroscopic, or real-space probes may indicate abundant low-energy excitations, while thermal transport finds little or no mobile magnetic heat conduction. 
This tension is sharpest in materials with a residual $\gamma$ term, strong power-law magnetic specific heat, or Fermi-surface-like experimental signatures. 
In $\alpha$-RuCl$_3$, however, the nature of the field-induced low-energy spectrum remains under active investigation: Fermi-surface-like interpretations should be contrasted with recent specific-heat results in ultraclean crystals that support Dirac-like Majorana dispersions~\cite{imamura2025}. 
A clean two-dimensional Dirac-like spectrum does not by itself require a finite $\kappa_0/T$. 
$^\dagger$For $\alpha$-RuCl$_3$, the Fermi-surface-like signatures and the residual-transport measurements are not from the same experimental setting. 
$^\ddagger$For Na$_2$Co$_2$TeO$_6$, field-angle-resolved specific heat has also been interpreted as suggesting nodeless low-energy excitations in the high-field phase~\cite{Fang25}, which should be viewed as a caveat to the spinon-continuum interpretation. 
Thermal-transport data and their interpretation remain controversial in several QSL candidates; see Ref.~\cite{matsuda2025}.
}
\label{tab:tab2}
\endgroup
\end{table*}

The energy-momentum-resolved spectrum is also computed for comparison with numerical results to verify the validity of the theory. The results are shown in Fig.~\ref{fig:ipepsvsmft} for both the four-Majorana correlation, obtained using the tight-binding ansatz, and the four-spin correlation, computed via iPEPS simulations under a moderate magnetic field. Two key features emerge from the analytical calculation shown in Fig.~\ref{fig:ipepsvsmft}(a): at the corresponding magnetic field, consistent with Fig.~\ref{fig:mfs}(b,d), the majority of the four-Majorana spectral weight is concentrated near the $\Gamma$ point, with a slightly weaker signal appearing near the $\rm K$ points; the spectral weight near the $\Gamma$ point extends to a higher energy compared to that near the $\rm K$ point, as indicated by the dashed lines in Fig.~\ref{fig:ipepsvsmft}(a).  
This can be directly compared to the numerical dimer-dimer correlation function obtained via iPEPS, which corresponds to a direct evaluation of the left-hand side of Eq.~\eqref{eq:dimer} in the spin basis, as presented in Fig.~\ref{fig:ipepsvsmft}(b).
The iPEPS results exhibit a strikingly similar spectral structure at low energies compared to the analytically derived four-Majorana spectrum. The strongest intensities in the dimer-dimer spectral function are observed near the $\Gamma$ point, followed by slightly weaker signals at $\rm K$, with negligible intensity near $\rm M$; and a similar spectral distribution for $\omega > 0$ matches that from the tight-binding theory in Fig.~\ref{fig:ipepsvsmft}(a). Minor discrepancies between the Majorana-based analysis and the iPEPS results can likely be attributed to truncation errors in the iPEPS calculations and the omission of higher-energy states in the mean-field treatment. The consistency between these two independent approaches provides strong evidence for the presence of a Majorana Fermi surface through higher-order scattering processes. This finding not only reinforces the quantum Majorana metal picture of the IGP but also suggests that such emergent gapless excitations should be experimentally observable. We hope that future Raman and RIXS experiments on field-stabilized Kitaev QSL candidates will be able to directly probe these signatures and further validate this theory.

\subsection{Emergent disorder and thermal conductivity}
The gaplessness of the aforementioned field-induced phase naturally evokes the familiar expectation that low-energy fermionic excitations should leave signatures in thermodynamics, spectroscopy, or real-space response, and, if sufficiently mobile, should also contribute to longitudinal heat transport. The most direct thermodynamic example is a finite low-energy density of states, which can produce a linear-in-temperature magnetic specific heat, $C_m\sim \gamma T$ with $\gamma>0$, or more generally an enhanced low-temperature $C_m/T$. Via the Einstein relation, one might then naively expect a finite residual thermal conductivity $\kappa_0/T>0$ as $T\to0$ if the same low-energy excitations are itinerant heat carriers. Indeed, Monte Carlo simulations of the gapless Kitaev spin liquid predict a finite residual thermal conductivity in the zero-temperature limit~\cite{Nasu17}, and several candidate systems with neutral-Fermi-surface-like phenomenology exhibit low-energy thermodynamic signatures suggestive of abundant fermionic excitations~\cite{Pengcheng21,Hong2024}.

\begin{figure*}[t]
    \centering
    \includegraphics[width=0.9\linewidth]{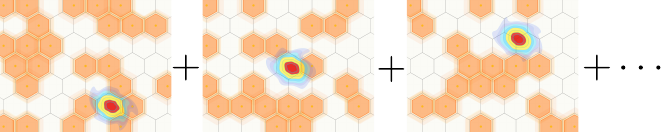}
\caption{
Schematic illustration of coherent flux disorder in the emergent gauge sector and transient Majorana localization. 
Each term in the superposition represents a distinct $Z_2$-flux/vison configuration $\ket{\mathcal F}$ in Eq.~\eqref{eq:mft}; within a given sector, the quasi-static visons act as an effectively quasi-static disorder background for the itinerant Majorana fermions. 
The orange shaded plaquettes denote visons, with diffuse halos indicating their slow quantum dynamics under finite magnetic fields, while the colored density profile represents a Majorana wave packet that is transiently localized in that particular flux background. 
Although each flux sector locally breaks translation symmetry, the full quantum superposition of such sectors does not correspond to quenched disorder in the microscopic Hamiltonian. 
This picture is intended for low but finite energies or intermediate time scales, where the vison mass is large compared with the Majorana mass. 
At the deepest infrared scale, visons can acquire a small mobility, so the ultimate universal transport regime may differ from this quasi-static coherent-disorder description.
}
\label{fig:coherent_disorder}
    \label{fig:coh_dis}
\end{figure*}

Experimentally, however, a broader and more puzzling pattern has emerged. In several honeycomb, triangular, and kagome QSL candidates, thermodynamic, spectroscopic, or real-space probes indicate abundant low-energy excitations, for example residual or enhanced $C_m/T$, power-law magnetic specific heat, spinon-continuum-like spectral weight, or Fermi-surface-like oscillatory signatures, while longitudinal thermal transport often finds little or no residual magnetic contribution, $\kappa_0/T\to0$. This apparent mismatch does not always have the same meaning in every material: for instance, a clean two-dimensional Dirac-like spectrum need not produce a finite $\kappa_0/T$, while a Fermi-surface-like or large-$\gamma$ signal would more directly suggest mobile low-energy quasiparticles. Nevertheless, the recurring coexistence of strong low-energy signatures with weak heat transport is one of the central phenomenological motivations for considering emergent disorder, flux scattering, localization, or other mechanisms that suppress quasiparticle mobility without eliminating low-energy spectral weight. Representative examples are summarized in Table~\ref{tab:tab2}.

The absence of mobile magnetic heat carriers suggests that the low-energy excitations inferred from thermodynamics or spectroscopy may be strongly scattered, localized, or phase separated by extrinsic defects or disorder~\cite{James25}. 
However, in a fractionalized spin liquid there is also an intrinsic route to anomalously slow transport that does not require quenched disorder in the microscopic Hamiltonian. The coherent-disorder wave function in Eq.~\eqref{eq:mft} is a quantum superposition of many $Z_2$-flux configurations. Each individual flux sector $\ket{\mathcal F}$ locally looks like a disordered gauge background for the itinerant Majorana fermions, while the full superposition need not break the translation symmetry of the Hamiltonian; this averaged translation symmetry has been exploited in computing the average-momentum-resolved spectral function previously in Sec.~\ref{sec:D}. This physical picture is illustrated schematically in Fig.~\ref{fig:coherent_disorder}: the orange plaquettes represent slowly fluctuating visons, and the colored profile shows a Majorana wave packet that can be transiently localized within a given flux sector.

The key ingredient here is mass-imbalanced fractionalization. If one fractionalized excitation is much heavier than the other, as expected for slow visons and light Majorana fermions, the heavy sector can act as an effectively quasi-static disorder background over experimentally or numerically accessible time scales. The resulting dynamics formally resembles disorder-free localization: the Hamiltonian remains translation invariant, but the light fractionalized particle propagates in a self-generated, slowly fluctuating flux landscape. This description should be understood as applying to low but finite energies, or equivalently to intermediate time scales. At the deepest infrared scale the visons can acquire a small mobility \cite{Aprem21}, so the ultimate universal transport regime may differ from the quasi-static coherent-disorder regime.

It has been shown that Majorana fermions in the presence of random flux disorder can be localized in different energy windows, and can become completely localized in thin ladder geometries at zero temperature~\cite{Brenig21}. 
These results provide a useful reference point for the field-induced Majorana metal, where the fluxes are not externally quenched but are generated by quantum fluctuations. Since the fluxes in field-induced phases appear to have very slow or glassy dynamics~\cite{Baskaran2023,Jin23}, the Majorana sector conditioned on a given flux background can be strongly localized at least over a long prethermal window. This can lead to anomalous energy transport even when the spectrum remains gapless~\cite{Lucas25}.
For example, recent studies of non-equilibrium quench dynamics in the integrable Kitaev QSL~\cite{Rademaker19,Heyl21} have shown very slow spreading of correlation functions from a trivial initial state.

While the situation is analytically understood in the integrable limit with quench- or thermally excited flux disorder~\cite{Zekun,Fulga20}, the extent to which field-induced flux fluctuations in the quantum Majorana metal localize low-energy fermions, without quenched or extrinsic disorder~\cite{Schiulaz2014,DeRoeck2014,Papic2015,Moore16,DarkwahOppong2020}, remains largely unexplored in QSL context. 
This calls for a detailed theoretical framework for localization, subdiffusion, or anomalously slow heat transport in the putative quantum Majorana metal, together with numerical calculations of thermal conductivity at very low temperatures and long times.
Progress in this direction has recently been made for the Kitaev model under a moderate magnetic field in a quasi-one-dimensional ladder geometry, which provides a controlled illustration of dynamical localization of Majorana fermions by quantum fluctuations of quasi-static visons~\cite{Feng2025}. 
The quasi-one-dimensional setting has a short localization length and is therefore amenable to a controlled demonstration of self-localization from mass-imbalanced fractionalization. Extending this mechanism to the two-dimensional thermodynamic limit remains an important open problem.

\begin{figure*}
    \centering
    \includegraphics[width=0.95\linewidth]{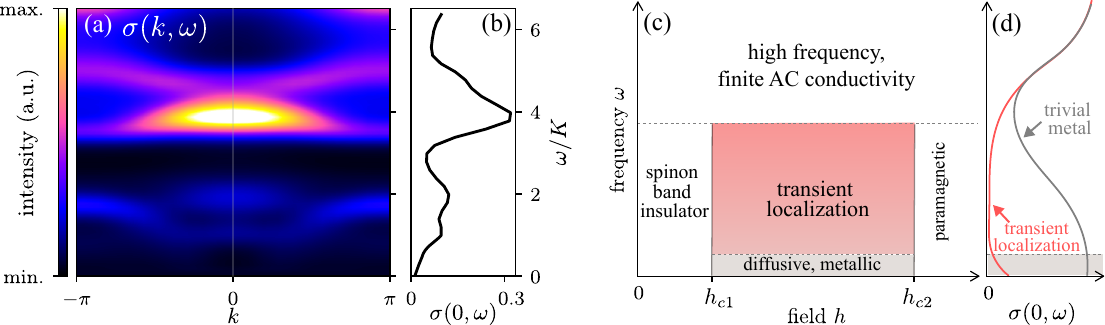}
    \caption{(a) Dynamical energy conductivity $\sigma(k, \omega)$ at intermediate magnetic field $h/K=0.8$ in a quasi-one-dimensional Kitaev spin ladder. (b) A cut of $\sigma(k, \omega)$ at $k=0$. In the low-energy, long-wavelength DC limit, the energy conductivity is strongly suppressed due to low-energy fluctuating visons. (c) Schematic dynamical energy conductivity of the AFM Kitaev model in a magnetic field, computed at zero temperature. At ultra-low energies the system is extended and diffusive due to non-integrability, whereas just above the universal regime mass-imbalanced fractionalization allows heavy visons to transiently localize light Majorana fermions, leading to a strong suppression of heat transport at low but finite energy scales. (d) Representative frequency cut of the dynamical energy conductivity $\sigma(\omega)$ at intermediate field, highlighting the strongly reduced response at finite frequencies above the universal diffusive regime (red solid line), in sharp contrast to a conventional diffusive metal exhibiting a pronounced Drude peak in the DC limit (gray solid line). Panels (c) and (d) are adapted from Ref.~\cite{Feng2025}.   }
    \label{fig:localization}
\end{figure*}

In connection with the strongly suppressed heat transport underlying the aforementioned thermal paradox, the dynamical energy conductivity can be defined through the energy-current linear response. Introducing a local energy-density operator $\mathcal{E}(r,t)$ and an energy-current operator $J(r,t)$ via the discrete continuity equation
\begin{equation}
\frac{d}{dt}\mathcal{E}(r,t)
+\left[J_{r,r+1}(t)-J_{r-1,r}(t)\right]=0,
\end{equation}
one obtains in momentum--frequency space
\begin{equation} \label{eq:jjnn}
C_J(k,\omega) = \frac{\omega^2 C_\mathcal{E}(k,\omega)}{2-2\cos k},
\;\;
\sigma(k,\omega) \sim \frac{C_J(k,\omega)}{\omega}
\end{equation}
where $C_\mathcal{E}(k,\omega)=\langle \mathcal{E}_k(\omega)\mathcal{E}_{-k}(\omega)\rangle$ and $C_J(k,\omega)=\langle J_k(\omega)J_{-k}(\omega)\rangle$ are the energy--energy and current--current spectral functions, respectively, and $\sigma(k,\omega)$ denotes the dynamical energy conductivity \cite{Luttinger64}. In the energy window accessible to dynamical MPS simulations, as shown in Fig.~\ref{fig:localization}(a,b), $\sigma(k,\omega)$ is strongly suppressed at all momenta for small $\omega$. Such suppression of the long-wavelength energy conductivity in a gapless system is a hallmark of transient localization by dynamical disorder in effectively mass-imbalanced systems \cite{Giorgio06,Heller24,Hadi24}. We emphasize, however, that this conclusion concerns the intermediate-time and finite-frequency regime directly accessed by the MPS calculation \cite{Feng2025}. The maximum reachable evolution time sets a finite low-frequency resolution, $\delta \omega \sim 1/t_{\max}$, and therefore the data cannot resolve arbitrarily small energy scales. This limitation is especially severe in gapless systems, where the entanglement entropy grows rapidly during real-time evolution, making the true asymptotic long-time limit prohibitively expensive to access. Physically, one such low-energy crossover scale is expected to be set by the extremely narrow bandwidth, or equivalently the very long relaxation time, of the heavy vison sector. Once this scale is resolved, the vison background is no longer strictly quasi-static, and a small but finite contribution to the dc energy conductivity may in principle be restored. This caveat is not specific to the ladder geometry; the same dynamical-vison timescale is also present in two dimensions. Thus the MPS result should be interpreted not as a proof of strictly vanishing conductivity at $\omega\to0$, but as evidence for a broad finite-energy regime in which Majorana energy transport is effectively localized by the slow vison background.

Beyond the quasi-one-dimensional Kitaev ladder, direct numerical simulations of two-dimensional systems are substantially more challenging, again due to the rapid growth of entanglement entropy in MPS-based time evolution. In the ladder geometry discussed above, transient localization and the associated suppression of energy transport provide a particularly sharp diagnostic: the short correlation/localization length allows for controlled real-time simulations and finite-size checks. In two dimensions, by contrast, localization is often marginal, and obtaining equally unambiguous evidence from the dynamical conductivity is much more difficult. Nevertheless, at finite energies one expects analogous transient-localization physics to persist. In the parton language, the Kitaev honeycomb model realizes an effective class-D topological superconductor. In addition to the dynamical-vison crossover discussed above, two dimensions contain a separate asymptotic low-energy crossover associated with class-D scaling itself. In the universal ultra-low-energy regime, this class exhibits weak anti-localization and multifractal extended states with a diverging localization length, which lies within the ultra-low-energy gray area in Fig.\ref{fig:localization}(c,d). However, at energies slightly above this universal regime, which are more relevant to experimentally and numerically accessible windows, the weak-anti-localization divergence is cut off, and the localization length rapidly drops to microscopic scales, as denoted by the pink region in Fig.\ref{fig:localization}(c) and the red solid line in Fig.~\ref{fig:localization}(d). Consequently, within the finite energy and temperature window accessible in experiments, a strong suppression of thermal conductivity due to transient localization is also anticipated in two dimensions. The possible recovery of conductivity in the strict ultra-low-energy limit may therefore be understood as arising from asymptotic crossovers beyond the finite-time numerical window, while leaving intact the broad finite-energy suppression relevant to the thermal-transport phenomenology. We also note that the dynamical conductivity diagnostics discussed here are obtained from zero-temperature real-time correlators. A direct finite-temperature calculation of the thermal conductivity, including the temperature dependence of the vison dynamics and the associated crossover scales, remains an important problem for future work.

\subsection{Discussion and outlook}
\subsubsection{Signatures of the gapless quantum spin liquids}
The quantum Majorana metal framework naturally accounts for a variety of experimental and numerical observations in candidate gapless $Z_2$ QSLs. Moderate magnetic fields drive a field-dependent flux filling that generates an emergent Majorana Fermi surface, whose characteristic scattering profile in the Majorana spectral function provides a concrete basis for interpreting dynamical probes such as inelastic neutron scattering or Raman spectroscopy (Fig.~\ref{fig:mfs}). Because the geometry of this Fermi surface is set by the underlying $\pi$-flux density tuned by the applied field, real-space Friedel-type oscillations of neutral fermions acquire a field-dependent wavelength. This mechanism thus provides partial support for the recent observation of Friedel-like oscillations in monolayer $\alpha$-RuCl$_3$, where the oscillation periodicity shifts with in-plane field strength; similarly, as discussed in Ref.~\cite{baskaran2015majorana}, the de Haas van Alphen quantum oscillations of neutral fermions can still be observed as long as Majorana fermions remain gapless, albeit a concrete microscopic picture for the correspondence between gapless Majorana fermions and the oscillation remains to be further explored.

On the thermodynamic front 
at low temperatures, the analytical result for the DOS and the specific heat of a 2D TR-broken Majorana metal scales as
\begin{equation} \label{eq:cv}
\rho (E) \sim \ln (1/T),\;\;C_V(T)\sim T \ln\bigl(1/T\bigr)
\end{equation}
as computed in Ref.~\cite{Senthil2000}. 
However, this behavior cannot be observed directly in the pure Kitaev honeycomb chiral spin liquid due to the presence of dynamical gauge field and the resulting finite flux gap $\Delta_f\approx0.26K$.  It is only when a thermal Majorana metal is induced at $k_B T>\Delta_f$ (at the order of $10$ Kelvin in solid-state candidates), that the low‐temperature specific heat becomes approximately linear in $T$ \cite{Nasu15}, which is indistinguishable from Eq.~\eqref{eq:cv} due to the slow logarithmic divergence above the universal regime. 
In the field‐induced gapless quantum Majorana metal, where the flux gap has already closed and $Z_2$ fluxes fluctuate even in the $T=0$ ground state, there is no intervening energy scale $\Delta_f$ to cross.  One should therefore observe the intrinsic scaling of Eq.~\eqref{eq:cv}, or at least an approximate linear‐in‐$T$ specific heat lower than the energy scale of $\Delta_f$.

Recent numerical simulations of extended Kitaev models under a moderate out-of-plane magnetic field have reported an approximate linear-in-$T$ specific heat, consistent with the linear scaling of a Majorana metal \cite{Han21,zhou2025,Wei2025}; while experiments on ultraclean $\alpha$-RuCl$_3$ under an \emph{in-plane} field find gapless behavior with $C_V\propto T^2$, suggestive of Dirac fermion excitations \cite{Kumpei24,imamura2025}. These results raise the important question: how the low-temperature specific heat evolves when an out-of-plane field component is applied. Measurements of specific heat under tilted field on $\alpha$-RuCl$_3$, would provide a crucial test for emergent quantum Majorana metal theory and translation invariant $Z_2$ gauge disorder in real materials.

\subsubsection{Universal or non-universal}
A large part of modern thinking about quantum phases—especially fractionalized ones, relies on universal low-energy field theories~\cite{wen_book}. In this paradigm, the asymptotic infrared structure, encoded in gauge theory descriptions and continuum quasiparticles, is expected to control thermodynamics and transport. This is the language in which Dirac spin liquids, U(1) spinon Fermi surfaces, and Majorana metals are usually discussed.

\begin{figure}
    \centering
    \includegraphics[width=\linewidth]{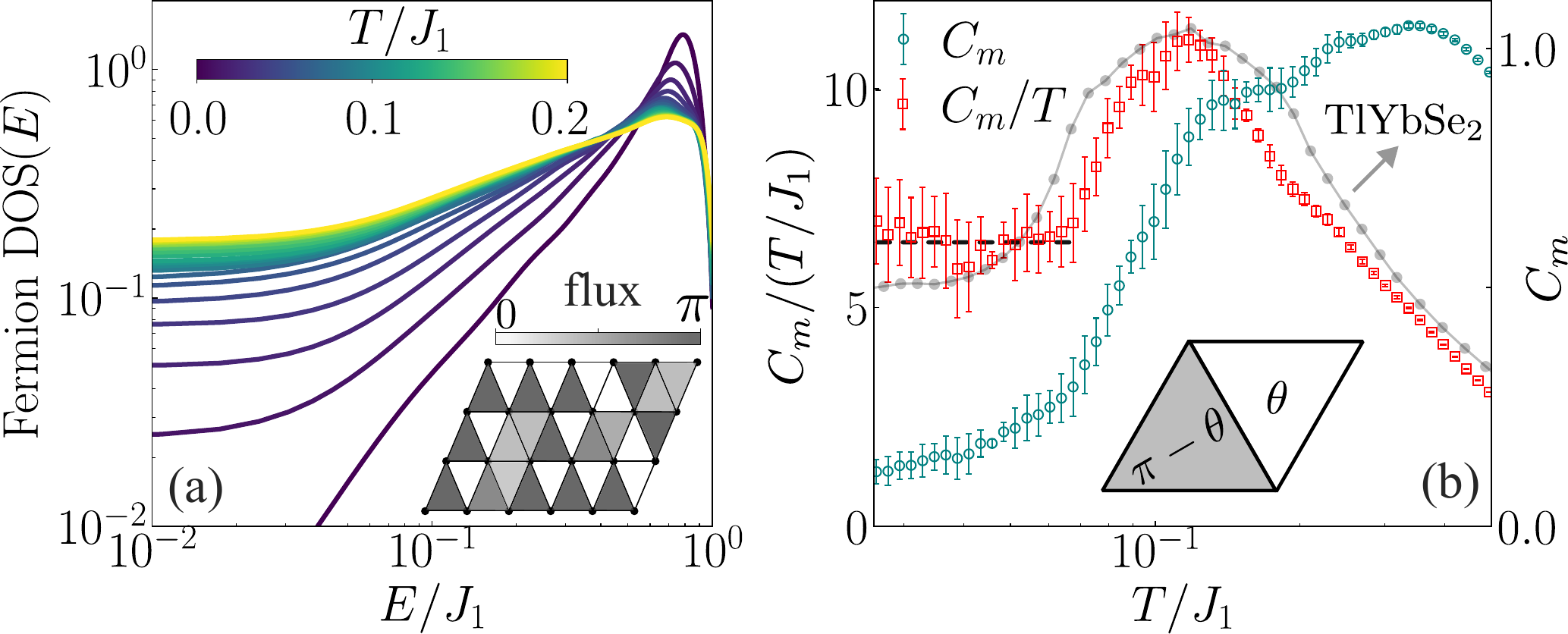}
    \caption{Specific heat and fermion DOS in the rare-earth delafossite compound TlYbSe$_2$, the candidate material for U(1) spin liquid on frustrated triangular lattice. (a) Fermionic spinon density of states (DOS) as a function of temperature. The temperature sets the density of excited gauge fluxes, see inset for an  illustration of a flux configuration. (b) Specific heat presented as $C_m/T$ and $C_m$, with the dashed horizontal line as a guide to the eye for the linear specific heat.  The experimental data is denoted in gray dots. The inset shows a local flux variation with phase angle $\theta$ on top of the staggered flux configurations of the ground state Dirac QSL. Figure adapted from Ref.~\cite{belbase2025}. }
    \label{fig:u1qsl}
\end{figure}

However, many experiments do not access the true asymptotic regime, instead, they probe a finite window of energies and temperatures, where non-universal microscopic details and emergent disorder can qualitatively reshape observable responses without contradicting the underlying universal field theory. Our analysis of transient localization in Kitaev systems provides a concrete illustration of this point.
In the presence of mass-imbalanced fractionalization, heavy visons act as a quasi-static random background for light Majoranas on experimentally relevant time scales. This self-generated “intrinsic disorder” can induce transient localization: the Majorana fermions appear localized or subdiffusive over long but finite times, leading to a strong suppression of dynamical thermal conductivity, even though the low-energy theory might nominally allow extended or weakly antilocalized states. In quasi-one-dimensional geometries (ladders), this effect is especially sharp due to the short localization length and is numerically accessible; in two dimensions it is more delicate but can still dramatically reduce heat transport within the finite frequency/temperature window probed in real materials.

Furthermore, the linear-in-$T$ specific heat is often taken as a hallmark of itinerant fermionic spinons. Yet even in the Kitaev honeycomb model, the situation is more subtle. At intermediate temperatures, thermally excited $Z_2$ fluxes strongly reconstruct the spinon spectrum and produce an apparent linear contribution over a broad window, despite the fact that the strict $T\to0$ limit is governed by a different universal structure. This intuition can be further extrapolated in particle-hole symmetric fermions without pairing, where the DOS was also found to be enhanced at low energy due to random disorder in real hopping amplitudes \cite{Olexei02,das2025}; and the U(1) Dirac QSLs. In the latter case, at finite temperatures the U(1) fluxes also fluctuate, leading to the enhancement of fermion density of states at finitely low energy scales. This possibility has recently been reported in Ref.~\cite{belbase2025}, where the rare-earth delafossite compound TlYbSe$_2$ has been found to harbor a robust linear temperature dependence of the heat capacity accompanied by the complete absence of long-range order at low fields at finite temperature scales such that strong thermal $U(1)$ flux fluctuation greatly enhances the low-energy fermion DOS. 
Nevertheless, it alone need not imply a stable spinon Fermi surface or a particular universal scaling of a spin liquid; it instead reflects a non-universal regime dominated by gauge fluctuations and proximate fractionalization physics of a quantum spin liquid.
Recent numerical results have also reported a finite monopole density induced by Zeeman magnetic field in U(1) QSL models, i.e. $J_1$-$J_2$ Heisenberg model on triangular lattice \cite{bader2025,budaraju2026}, where impact of flux fluctuation on the low-energy fermion DOS and localization physics remains to be explored in future studies.

The broader message is that many puzzling experimental observations, such as an apparent thermal paradox where spectroscopy indicates gapless excitations but $\kappa/T$ is strongly suppressed, may naturally reside in this non-universal, finitely low-energy regime. Rather than viewing such data as inconsistent with fractionalization, one should interpret them as windows into the interplay between emergent particles, slow gauge-sector dynamics, and self-induced disorder. Careful distinction between universal IR predictions and measurable finite-energy behavior is therefore essential when diagnosing fractionalized phases of matter.

\subsubsection{The limits of translation-invariant mean-field theories} \label{sec:mftlimits}
The standard theoretical description of fractionalized phases of matter is built on parton constructions and associated gauge theories formulated in the universal low-energy limit~\cite{wen_book}. In this framework, one typically assumes a translationally invariant emergent gauge background and characterizes the phase in terms of long-wavelength quasiparticles, such as canonical spinons, Majorana fermions, photons in emergent QED$_3$, which govern thermodynamic and transport properties. This approach has provided a powerful unifying language for a broad class of candidate quantum spin liquids and related systems.

The results discussed in this report indicate, however, that such translationally invariant, continuum descriptions can be incomplete due to properties that are sensitive to spatially random fluctuations in gauge fields \cite{Aavishkar23}. In particular, Kitaev-type materials in a magnetic field offer a concrete setting where the interplay between fractionalization and emergent disorder becomes essential: Under field-induced flux proliferation, the system is more appropriately viewed as an ensemble of gauge backgrounds, which recovers translation invariance by superposition, rather than a single uniform flux sector by self-consistent mean-field assumption. This generates an effective disordered class-D Majorana problem, leading to gapless spectra and large low-energy density of states in the universal regime.
This observation highlights an important limitation of self-consistent mean-field and other self-averaging schemes commonly employed in the study of fractionalized phases. By construction, such approaches replace the spatially fluctuating gauge configuration with an averaged background. This may explain why self-averaging mean-field treatments with a biased translation invariance in similar models fail to capture the emergent gapless phase \cite{ZhangNatComm2022}, whereas unbiased numerical methods that include field-induced gauge fluctuations do observe such gapless spectrum \cite{wang2024}.

Another consequence of translation-invariant self-consistent mean fields is that they suppress the nonlocal quantum interference effects responsible for Anderson-type localization and related anomalous transport phenomena at finite energy scales. 
Hence they cannot capture the coherent‐disorder–induced metallic spectrum nor the resulting localization or anomalous transport \cite{Pastor2003,Hadi24}.
This point can be made explicitly simple by considering the return probability $R(t)=\bigl|\langle j|e^{-iHt}|j\rangle\bigr|^2$ as an order parameter for localization. 
Here $\ket{j}$ denotes a localized fermion at site $j$, and the return probability $R(t)$ measures the system's memory of the state $\ket{j}$ under time evolution: $R(t)$ remains finite even at late times in a fully localized phase, whereas it decays to zero in an extended phase \cite{Anderson58}.
This can be made explicit by examining a single disorder realization.  For a localized state $\ket{j}$, the return amplitude $A(t)$ is defined as
\begin{equation}
A(t)=\langle j|e^{-iHt}|j\rangle=\sum_{k} |\langle k|j\rangle|^2\,e^{-iE_k t},    
\end{equation}
where $\{|k\rangle\}$ are the eigenstates of $H$ with eigenvalues $E_k$. Because the state is localized, only a few eigenstates have significant weight. This makes dephasing highly unlikely on the right-hand side. For a localized system, $A(t)$ and $R(t) = |A(t)|^2$ do not decay to zero. 
However, upon averaging over flux configurations labeled by $n$ with weights $p_n$, the disorder-averaged return amplitude
\begin{equation}
   \overline{A(t)}=\sum_n p_n \int dE\,\rho_n(E)\,e^{-iEt}
\end{equation}
where we used $A_n(t) = \int dE\,\rho_n(E)\,e^{-iEt}$, with $\rho_n(E)=\sum_k|\langle k|j_n\rangle|^2\delta(E-E_k)$ the local density of states.  The ensemble-averaged local density of states $\rho_n(E)=\sum_n p_n\rho_n(E)$ becomes continuous over the quasiparticle bandwidth $D$.  Assuming a uniform DOS distribution and particle-hole symmetry, the return probability becomes
\begin{equation}
R(t)
=\Bigl[2\sin(\frac{Dt}{2})/(Dt)\Bigr]^2    
\end{equation}
which rapidly decays to zero within $t\sim2\pi/D$.  Thus, the averaged signal loses memory on $1/{\rm bandwidth}$ timescales, and the self-consistent mean-field theories assuming translation-invariance of gauge fields are incapable of capturing localization due to intrinsic self-generated disorder.

Therefore, capturing both the long‐lived local return amplitude in each sector and the slow decay of the average $R(t)$ requires methods that preserve long-range interference beyond mean‐field self‐averaging, such as those of Ref.~\cite{Pastor2003}; or by unbiased numerical methods such as Lanczos algorithm or MPS dynamics.
This also explains why self-averaging treatments in similar models \cite{ZhangNatComm2022} fail to capture the emergent gapless phase due to flux fluctuations, whereas unbiased numerical methods that include field-induced gauge fluctuations do observe such gapless spectra \cite{wang2024,penghao24}.

%%%%%%%%%%%%%%%%%%%%
\begin{figure*}[t]
    \centering
    \includegraphics[width=0.98\linewidth]{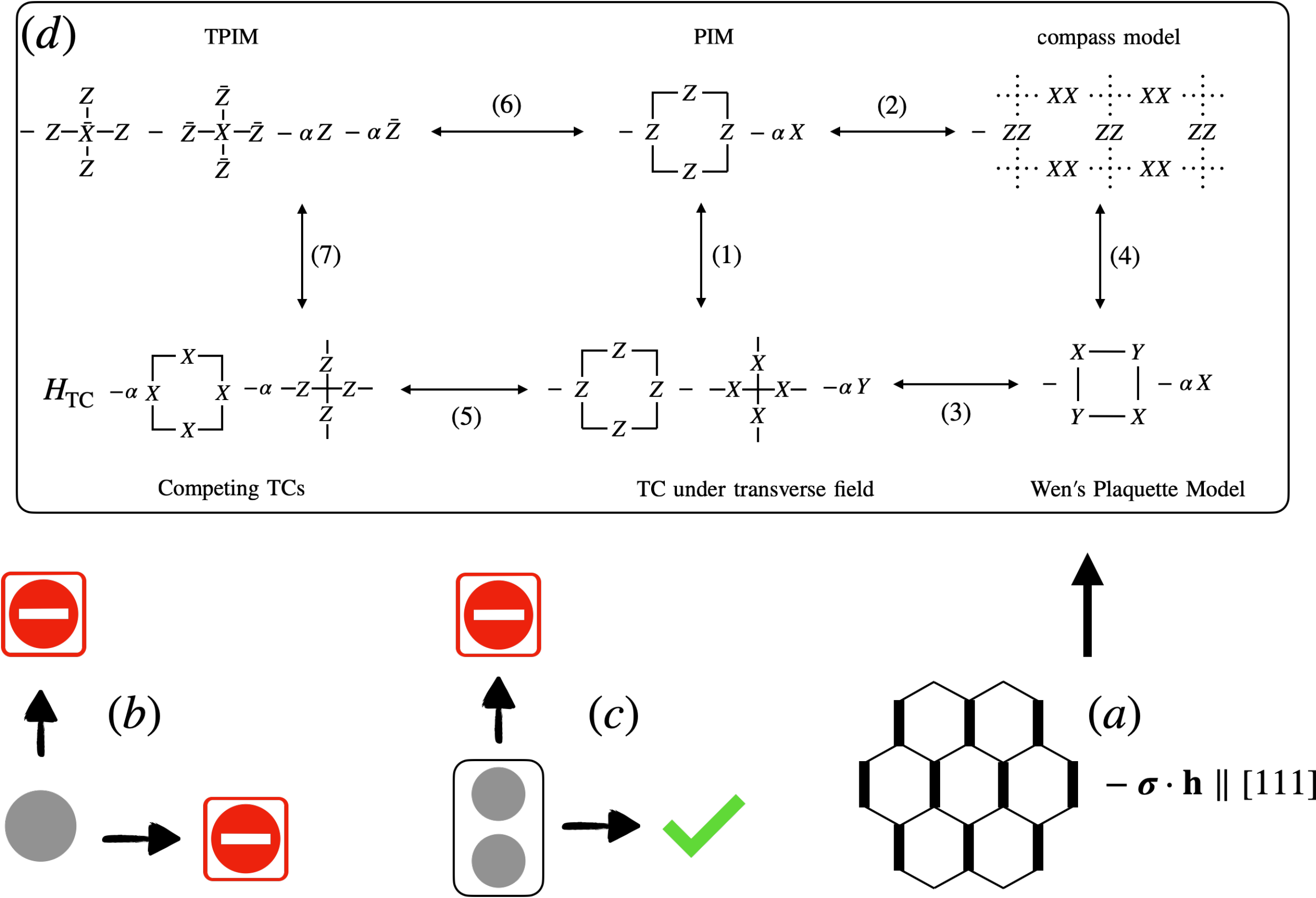}
    \caption{An overview of sub-dimensional dynamics in Kitaev’s Abelian QSL or toric code (TC) under a weak magnetic field. In the bottom row: (a) an illustration of the model setup: the anisotropic Kitaev honeycomb model under a weak out-of-plane magnetic field. The anisotropy in the compass interaction is represented by thick vertical bonds, leading to the  A  phase in Kitaev’s phase diagram. (b–c) Illustration of constrained dynamics: a single particle remains immobile, whereas a two-particle composite gains partial mobility, allowing it to move along a fixed one-dimensional direction. In the top row: (d) a family of models exhibiting subsystem symmetries and constrained quasiparticle dynamics, which are equivalent to the anisotropic Kitaev honeycomb model under a weak out-of-plane magnetic field and can be mapped onto each other. The transformations relating these models include (1) duality defined in Ref.~\cite{Vidal09}; (2) duality defined in Ref.~\cite{Zohar05}; (3) a unitary rotation \cite{kitaev2006anyons}; (4) composition of (2,1,3); (5) duality defined in Ref.~\cite{Feng24}; (6) transformation given in Ref.~\cite{You2018}; and (7) in  Ref.~\cite{Feng24}. }
    \label{fig:sets}
\end{figure*}
%%%%%%%%%%%%%%%%%%%%

\section{Sub-dimensional fractonic dynamics by weak magnetic field} \label{sec:dim}
From this section we report on the recent progress on another class of field-induced physics in Kitaev QSLs: the emergence of dimensional reduction or sub-dimensional physics. 
Emergent sub-dimensional dynamics and dynamical constraints in quantum many-body systems have been the subject of intense study in recent years, particularly under the framework of fracton or fractonic physics \cite{Sagar2016,Nandkishore_ARCMP_2019}. These systems exhibit restricted mobility of excitations due to conservation laws or gauge constraints, leading to unconventional dynamical behavior.
Interestingly, phenomenologically similar models, where dynamics are also constrained, have been explored in various other contexts, including topological quantum glass \cite{Claudio05, Claudio12}, dimensional reduction at quantum criticality, and systems with subsystem symmetries \cite{Nussinov2015,You2018}. A key feature of such constrained dynamics is that they give rise to exceptionally sharp dynamical signatures for quasiparticle excitations.
This raises the exciting possibility that emergent sub-dimensional or constrained dynamics, unique to the fractionalized excitations of QSL models, could serve as distinct dynamical fingerprints for identifying QSLs in experiments. By leveraging these unique constraints, one may be able to establish sharper experimental probes in the ongoing search for QSL phases.

\begin{table}[t] \label{tab:motion}
\centering
\caption{Schematic comparison of dynamical responses in the Abelian phase of the Kitaev honeycomb model and in a conventional Heisenberg magnet. In the Kitaev Abelian QSL, the gauge constraint renders the two-spin-flip excitation fractonic, leading to sharp, immobile Abelian-anyon modes, in contrast to the extended two-magnon continuum of the Heisenberg magnet.}
\label{tab:tab3}
\vspace{2mm}
{\renewcommand{\arraystretch}{1.35}
\setlength{\tabcolsep}{3pt}
\footnotesize
\begin{tabular}{|p{0.28\columnwidth}||p{0.29\columnwidth}|p{0.29\columnwidth}|}
\hline
&
\centering\textbf{One-spin-flip response}
&
\centering\textbf{Two-spin-flip response}
\tabularnewline
\hline\hline
Kitaev Abelian QSL under weak field
&
\centering Majorana--flux continuum
&
\centering Sharp fractonic modes of Abelian anyons
\tabularnewline
\hline
Heisenberg magnet under field
&
\centering Sharp magnon modes
&
\centering Two-magnon continuum
\tabularnewline
\hline
\end{tabular}
}
\end{table}

In this section, we review recent findings on emergent sub-dimensional dynamics in the Abelian phase of Kitaev spin liquids, where the Abelian anyons excited by multi-spin flips exhibit sharp dispersion under the perturbation of Zeeman fields, with the composite fermion preferentially moving coherently along one-dimensional directions. This behavior arises due to sub-extensive symmetries, leading to sharp signatures of anyon bound states in scattering experiments, in sharp contrast to those of trivial Heisenberg magnets where multi-spin-flip excitations generally give a featureless magnon continuum. The sharp comparison is sketched in Table~\ref{tab:tab3}, whose content will be elaborated in Sec.~\ref{sec:khmy} and Sec.~\ref{sec:dimerspec}.

The combination of this constrained motion with the external magnetic field’s ability to separate the dynamics of different fractionalized particles provides a promising pathway for experimentally detecting these anyons.
In the following texts, we begin by briefly reviewing the simplest system exhibiting subextensive symmetries: the Plaquette Ising model (PIM), which hosts sub-dimensional dynamics of excitations, and can be mapped to many other models that exhibit spatially constrained dynamics (See Fig.~\ref{fig:sets}(d) for a collection of these models and their relations). The Plaquette Ising model under a transverse field can be mapped onto the Toric Code model in a transverse field, where it is the composite anyons that exhibit one-dimensional dynamics. We then discuss how similar behavior emerges in the anisotropic limit of the Kitaev honeycomb model under a weak magnetic field, demonstrating how sub-dimensional dynamics naturally arise in these systems.

\subsection{Plaquette Ising model} \label{sec:pim}
\begin{figure*}[t]
    \centering
    \includegraphics[width=0.98\linewidth]{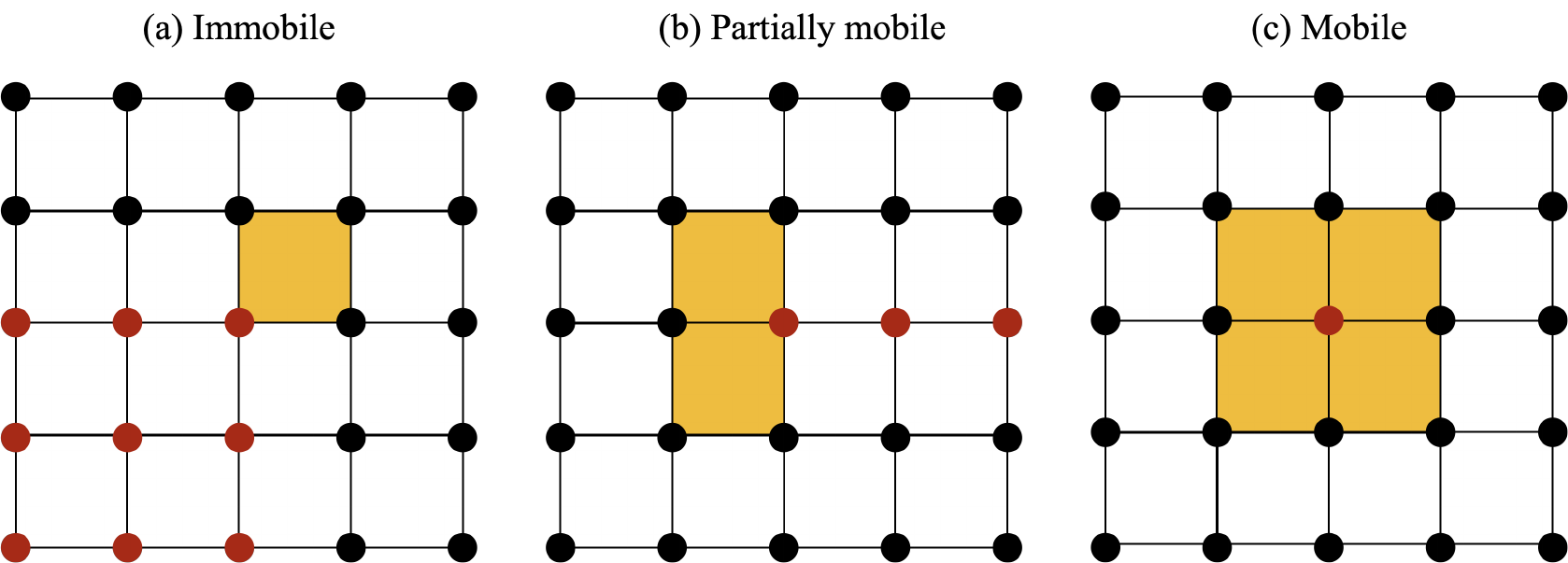}
    \caption{An illustration of the Plaquette Ising model and its low-energy excitations. The spin degrees of freedom are represented by black dots, while excitations correspond to plaquettes where the product of the four vertices equals $-1$, highlighted in orange. (a) A single-face flip represents the lowest-energy excitation, but it involves an infinite number of flipped spins (red dots) extending into the lower-left corner. This excitation remains immobile under a transverse field, as its movement requires a perturbation of infinite order. (b) A two-face flip, the next lowest-energy excitation, can be generated by flipping an entire line of spins. This excitation exhibits partial mobility, constrained to a fixed one-dimensional direction, and can move via a first-order process. (c) Flipping a single spin creates an excitation consisting of four flipped plaquettes. This excitation is fully mobile, as its propagation requires only a second-order process.}
    \label{fig:pim}
\end{figure*}

For clarity we start from the simplest spin toy model that exhibits sub-dimensional physics. 
The Plaquette Ising model that we are going to briefly review in this section is also known as the Xu-Moore model, originally proposed as a model describing $p+ip$ superconducting arrays \cite{Xu2004,Xu2005}, and it has become one of the standard illustrative pictures for fractonic dynamics \cite{niggemann2025b,niggemann2025}. The model is defined on a square lattice with qubits living on the vertices, as shown in Fig.~\ref{fig:pim}. The basic building block of the Hamiltonian is the four-point plaquette interaction on each face:
\begin{equation} \label{eq:pim}
    H_{\rm PIM} = -\sum_p F_p - h \sum_i \sigma_i^x, ~~F_p = \prod_p \sigma_{i\in p}^z
\end{equation}
For $h = 0$, the ground state of Eq.~\eqref{eq:pim} is given by a uniform configuration of plaquettes $\prod_p \sigma_{i\in p}^z = 1, ~\forall p$. For $h \neq 0$, 
the model is self-dual, that is, equivalent under $h \leftrightarrow h^{-1}$ with a first-order phase transition at $h =1$ \cite{Xu2004,Xu2005}. 
Under open boundary conditions, there are an exponential number of ways in arranging spin configurations such that the ground-state condition $\prod_p \sigma_{i\in p}^z = 1, ~\forall p$ can be fulfilled. These correspond to a sub-extensive number of $Z_2$ symmetries on $L_x$ rows and $L_y$ columns $U_m = \prod_{n \in\rm cols} \sigma_{m,n}^x,~ U_n = \prod_{m \in\rm rows} \sigma_{m,n}^x$, 
where $m,n$ are row and column indices bounded by $L_x, L_y$ respectively. Note that flipping all column is the same as flipping all rows, so there is one less $Z_2$ symmetry than the total number of lines in the lattice ${\rm GSD} = 2^{L_x + L_y - 1}$, 
and the total Shannon entropy is proportional to the size of the boundary.

%%%%%%%%%%%%%%%%%%%%
%%%%%%%%%%%%%%%%%%%%

The sub-extensive symmetries naturally divide the Hilbert space into a sub-extensive number of blocks, where sub-dimensional dynamics can be expected. 
To see the constrained mobility explicitly, let us consider one-~, two- and four-face-flip excitations in Eq.~\eqref{eq:pim} at perturbative transverse field $h$. Under open boundary conditions, the one-face excitation can be made only by flipping an extensive number of spins  defined on vertices. 
As shown in Fig.~\ref{fig:pim}(a), a single-face flip represents the lowest-energy excitation with an excitation gap of $\Delta_1 = 2$. However, it involves an infinite number of flipped spins, marked by red dots in Fig.\ref{fig:pim}(a), extending across all spins to the lower-left of the face-flip excitation. This one-face flip excitation is analogous to a kink excitation, which remains immobile under a weak transverse field, as its movement requires a perturbation of order  $O(N)$, making it sub-extensively suppressed.
A two-face flip, shown in Fig.~\ref{fig:pim}(b), is the next lowest-energy excitation with an energy gap of $\Delta_2 = 4$. This excitation can be generated by flipping an entire line of spins, granting it partial mobility constrained to a fixed one-dimensional direction determined by the flipped red spins in Fig.~\ref{fig:pim}(b). It can propagate via a first-order process but cannot change direction without leaving behind a single-face flip excitation. 
Finally, Fig.~\ref{fig:pim}(c) illustrates an excitation with energy $\Delta_4 = 8$, where flipping a single spin creates an excitation consisting of four flipped plaquettes. Unlike the previous cases, this excitation is fully mobile, as its propagation requires only a second-order process. 
The coexistence of immobile, partially mobile, and fully mobile quasiparticles is reminiscent of the topological quantum glass \cite{Claudio05, Claudio12} and fractonic physics such as that in the X-cube model. However, in this case, the constrained dynamics arise from subsystem symmetries rather than the more robust fracton topological order. Nevertheless, the emergent sub-dimensional dynamics serve as a distinctive feature of quasiparticles governed by subsystem symmetries.
In the following sections, we explore the connection between the toric code model under a transverse field and its exact mapping to the PIM in Sec.~\ref{sec:tcy}, whereby the constraint dynamics is reflected in the partial mobility of composite Abelian anyons. Furthermore, given that the anisotropic limit of the Kitaev honeycomb model maps to the toric code model at fourth-order perturbation theory, it is expected that the Abelian QSL phase will exhibit discernible dynamical features associated with sub-dimensional dynamics, which we discuss extensively in Sec.~\ref{sec:khmy}.

\begin{figure*}[t]
\centering
\centering
\includegraphics[width=0.9\textwidth]{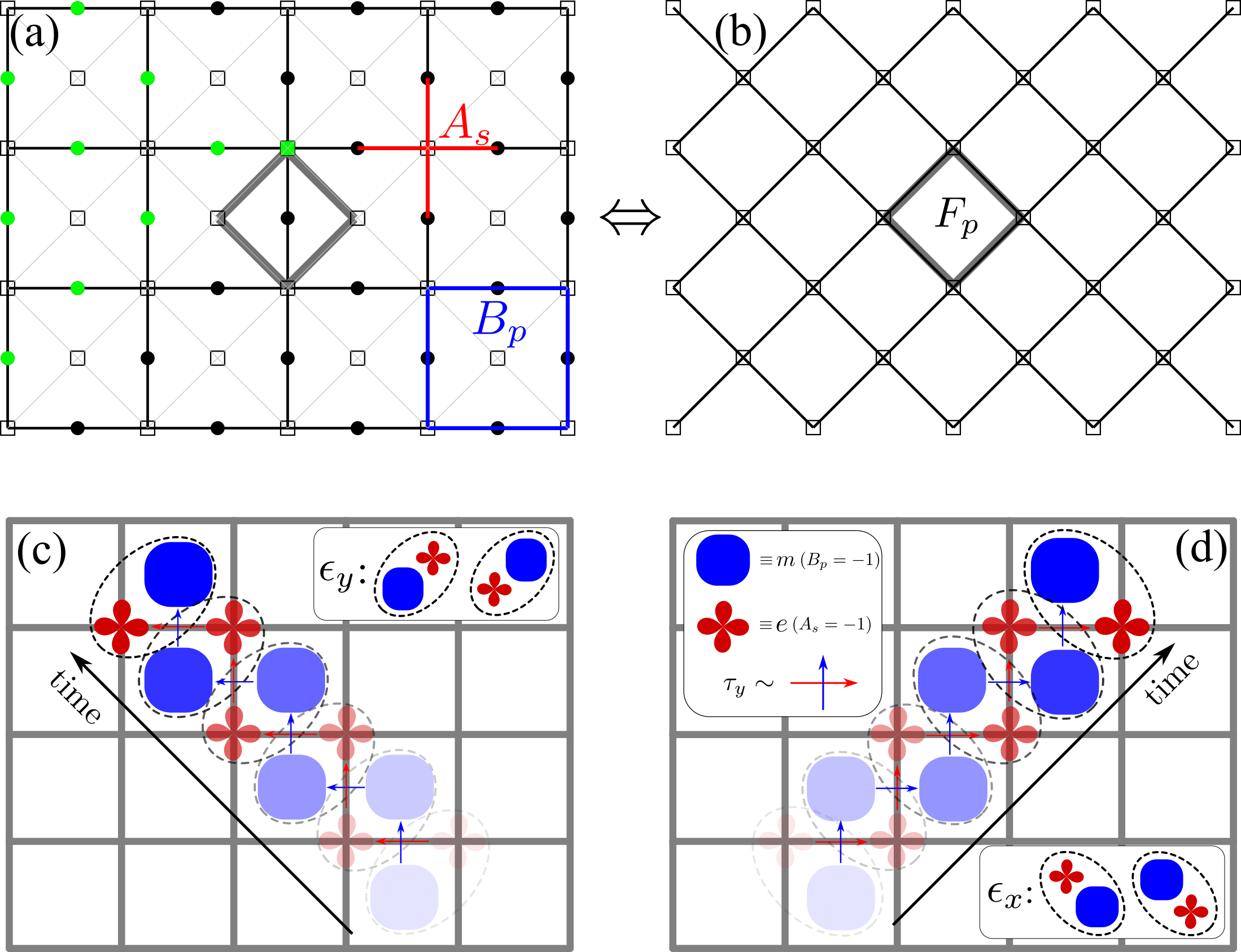}
\caption{Original square lattice $\Lambda$, marked by thick lines and filled circles, of TC on which plaquettes $p$ and stars $s$ are defined. Dual lattice $\tilde{\Lambda}$ is denoted by thin lines and empty squares living on the centers of faces and vertices. The green squares and circles illustrate the duality transformation: $\tau_j^z = \prod_{j>i} \sigma_i^y$, where the notation $j>i$ defines the set of all sites  $i\in \Lambda$ whose both coordinates are smaller than those of  $j\in \tilde{\Lambda}$. Panels (c) and (d) are adapted from Ref.~\cite{Feng2023}.  }
\label{fig:lat1}
\end{figure*}

\subsection{Toric code under transverse field} \label{sec:tcy}
Now let us take a step beyond the trivial example from the previous section toward Kitaev's $Z_2$ quantum spin liquids. We begin by relating the toric code (TC) model under a transverse $\sigma^y$ field to the previously discussed Plaquette Ising Model. This mapping was first introduced in the context of TC in \cite{Vidal09} and later found useful for connecting more intricate topological models protected by subsystem symmetries to competing toric codes \cite{Fenghidden23}.  
Consider the toric code model subjected to a magnetic field \cite{kitaev2003fault}:
\begin{equation}
H = -\sum_{s} A_s - \sum_{p} B_p - h\sum_i \sigma^y_i.
\label{eq:tcy}
\end{equation}
with $A_s = \prod_{j\in s} \sigma_j^x$ and $B_p = \prod_{j\in p} \sigma_j^y$ defined on vertices and faces, respectively. 
The external magnetic field along the $y$-direction has a qualitatively different impact on the phase diagram and dynamics compared to a field along $x$ or $z$. The latter cases are well known to induce confinement and condensation of anyons, leading to phase transitions characterized by the breakdown of topological order. Specifically, a field in the $x$- or $z$-direction directly couples to the anyons and modifies their statistical properties, eventually driving the system into a trivial paramagnetic phase. In contrast, a transverse $\sigma^y$ field preserves subsystem symmetries, allowing for emergent sub-dimensional dynamics.  
The sub-dimensional dynamics of Eq.~\eqref{eq:tcy} can be readily understood through the following mapping:
\begin{equation}
A_s = \sigma_j^x, \quad B_p = \sigma_j^x, \quad \prod_{i>j} \sigma_i^y  = \sigma_j^z.
\end{equation}
Here, the operators $A_s$, $B_p$, and $\prod_{j>i} \sigma_i^y$ on the left-hand side are defined on the original toric code lattice, represented by black dots in Fig.~\ref{fig:lat1}(a), where degrees of freedom reside on links. Meanwhile, the Pauli operators on the right-hand side are supported on the dual lattice, marked by hollow squares in Fig.~\ref{fig:lat1}(a).  
The last mapping, $ \prod_{j>i} \sigma_i^y $, is illustrated in green in Fig.~\ref{fig:lat1}(a). This indicates that a $\sigma_j^z$ operator, defined on the dual lattice (the green square), is equivalent to the product of all spins within the triangular region extending immediately to its left (the green dots) on the original TC lattice. Under this mapping, it is straightforward to see that a single-site Zeeman field $\sigma_i^y$ in the TC model—such as the spin encircled in the thick gray square in Fig.~\ref{fig:lat1}(a)—is mapped to the product of four spins $F_p = \prod_{j\in p} \sigma_j^z$
on the dual lattice, corresponding to the thick gray square in Fig.~\ref{fig:lat1}(a). Consequently, the TC Hamiltonian in Eq.~\eqref{eq:tcy} transforms into $\tilde{H} = -h \sum_p F_p - \sum_j \sigma_j^x$
which is self-dual \cite{Xu2004}. Applying another self-duality transformation, we see that the TC model under a transverse field, as defined in Eq.~\eqref{eq:tcy}, maps exactly to the PIM in Eq.~\eqref{eq:pim}, discussed in the previous section. Since the PIM is known to exhibit emergent sub-dimensional dynamics, this mapping directly implies that the toric code under a transverse $\sigma^y$ field harbors similar constrained dynamics.  

Besides connecting Eq.~\eqref{eq:tcy} to the simplest PIM model with sub-dimensional dynamics, in fact, this behavior can also be understood directly from the TC model. The local application of $\sigma_j^y$ excites four fluxes, i.e., two $\epsilon = e \times m$ fermions. As a result, the lowest-order process induced by $\sigma_j^y$ can, at best, allow a single $\epsilon = e \times m$ excitation to move along a fixed one-dimensional direction dictated by the initial state, or chosen by slightly breaking the lattice symmetry locally, as illustrated in Fig.~\ref{fig:lat1}(c,d). This mechanism is closely related to the construction of fractonic particles, where elementary excitations with finite mobility appear only as composite objects. Furthermore, similar constrained dynamics can emerge under perturbations other than the transverse Zeeman field, as long as the underlying structure of the effective theory remains valid \cite{Fenghidden23}. A summary of these models with emergent sub-dimensional physics relevant for TC is summarized in Fig.~\ref{fig:sets}(d). 

At the level of stabilizer code models, the emergent dynamical constraint can be easily understood by mapping to the simplest PIM, as illustrated in Fig.~\ref{fig:sets}(d). A natural next step is to investigate whether similar dynamical constraints manifest in QSL models. Notably, the anisotropic Kitaev honeycomb model is directly related to the TC when one of the three bond interactions is significantly stronger than the other two.
In particular, in the Abelian phase of the Kitaev QSL, the matter Majorana fermions are fully gapped, opening the possibility of directly detecting anyons by probing their constrained dynamics without hybridization from free Majorana excitations. This provides a sharp and unambiguous signature of a single type of fractionalized particle, offering a cleaner experimental fingerprint for identifying QSLs.
In the next section, we explore this connection in detail and present unbiased numerical results demonstrating emergent dynamical constraints, which naturally arise under the application of an out-of-plane magnetic field in the honeycomb QSL model.

\subsection{Constraint anyon mobility in the anisotropic Kitaev honeycomb model} \label{sec:khmy}
Consider the Kitaev honeycomb model with $K_z/K > 2$, where $K = K_x = K_y$ and subjected to [111] magnetic field:
\begin{align}
    H = \sum_{\langle ij\rangle_{\alpha \in \{x,y\}}} K {\sigma^\alpha_i \sigma^\alpha_j}  + K_z\sum_{\langle ij\rangle_z} \sigma^z_i \sigma^z_j - \mathbf{h}\cdot\sum_{i}\boldsymbol{\sigma}_i 
    \label{eq:ham_abelian}
\end{align}
At $ h = 0 $, it is known that Eq.~\eqref{eq:ham_abelian} reduces to the toric code model at fourth-order perturbation theory, where the anti-aligned spins on a dimer become the new effective degrees of freedom \cite{kitaev2006anyons}. Now, we consider the case where $ h \neq 0 $ and derive the effective Hamiltonian in the strong $ K_z/K $ limit, incorporating both the Kitaev interaction and the Zeeman field. 
In this limit, the low-energy degrees of freedom are obtained by considering only the interactions along the $ z $-bonds, given by $ K_z \sigma_A^z \sigma_B^z $, which results in a ground-state doublet $ \{|\uparrow\downarrow\rangle,~|\downarrow\uparrow\rangle\} $ and excited states $ \{|\uparrow\uparrow\rangle,~|\downarrow\downarrow\rangle\}$. The ground-state manifold, before including the perturbation $h$, is spanned by eigenstates of the effective dimer pseudospin operator, defined as $ \tau^z = (\sigma_A^z - \sigma_B^z)/2 $.  
Since a single-spin flip induced by the Zeeman field changes the ground-state manifold to the high-energy excited sector (e.g., flipping $ |\uparrow\downarrow\rangle $ to $ |\uparrow\uparrow\rangle $), the leading perturbation of the Zeeman field in the dimer basis appears at second order in perturbation theory, yielding a prefactor proportional to $h^2/K_z$.  
Applying degenerate perturbation theory, the effective Hamiltonian for the dimer degrees of freedom $ \tau^\alpha $ is obtained to leading order as  
\begin{figure}
    \centering
    \includegraphics[width=0.8\linewidth]{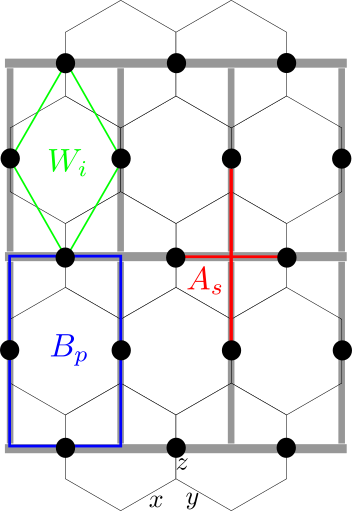}
    \caption{Mapping of the anisotropic Kitaev model of $\sigma$ qubits on the honeycomb lattice to an effective square lattice denoted by thick gray lines, of spin dimers denoted by black bullets. The two-level dimer degrees of freedom $\tau$ defined on links are given by the two $\sigma^\alpha$ spins on the $z$ bonds. The effective stabilizers under perturbations are marked in different colors. }
    \label{fig:tclattice}
\end{figure}
\begin{align}
    H_{\rm eff} = -J_{\rm TC}\sum_{i} W_i - \frac{2h^2}{K_z} \sum_i \tau^x_i,
    \label{eq_tcham}
\end{align}
where $W_i$ and $ J_{\rm TC} $ are given by  
\begin{align}
    W_i \equiv \tau^z_{i+n_1} \tau^z_{i-n_2} \tau^y_i \tau^y_{i+n_1-n_2}, \quad J_{\rm TC} \equiv \frac{K^4}{16 |K_z|^3},
\end{align}  
and $n_1$ and $n_2$ are the unit vectors of the triangular dimer Bravais lattice, see Fig.~\ref{fig:tclattice}. 
The first term in Eq.~\eqref{eq_tcham} corresponds to Wen’s plaquette interaction, which, after a unitary rotation, transforms into the toric code model \cite{PhysRevLett.90.016803, kitaev2006anyons}. The second term arises due to the application of a $[111]$ magnetic field.
The standard TC stabilizers can then be recovered by performing a unitary transformation on the horizontal and vertical bonds as follows:  
\begin{itemize}
    \item For horizontal bonds: $\{\tau^x, \tau^y, \tau^z\} \rightarrow \{\tau^y, \tau^x, -\tau^z\}$,  
    \item For vertical bonds: $\{\tau^x, \tau^y, \tau^z\} \rightarrow \{\tau^y, \tau^z, \tau^x\}$.
\end{itemize} 
The definitions of horizontal and vertical bonds are as illustrated in Fig.~\ref{fig:tclattice}. 
Applying this transformation to Eq.~\eqref{eq_tcham}, the effective Hamiltonian takes the form:  
\begin{equation}
    H_{\rm eff} = -J_{\rm TC} \left[\sum_{s} A_s + \sum_{p} B_p \right] - \sum_i \frac{2h^2}{K_z} \tau^y_i,
    \label{TChpert}
\end{equation}  
where $A_s$ and $B_p$ are the star and plaquette operators of the standard TC model, now perturbed by a Zeeman field in the $y$ direction. 
Notably, while the $W_i$ terms transform into $A_s$ and $B_p$ in an alternating pattern under the unitary transformation, the $\tau_i^x$ operator is identically rotated to $\tau_i^y$.  
Equation~\eqref{TChpert} is equivalent to Eq.~\eqref{eq:tcy} from the previous section and, in turn, to the transverse-field plaquette Ising model defined in Eq.~\eqref{eq:pim}, which exhibits sub-dimensional dynamics featuring immobile, partially mobile, and fully mobile particles under weak transverse field, as has been discussed in Sec.~\ref{sec:pim} and Sec.~\ref{sec:tcy}. The same sub-dimensional behavior is also present in $H_{\rm eff}$ and the microscopic honeycomb model defined in Eq.~\eqref{eq:ham_abelian}. When subjected under a larger Zeeman field, due to the Xu-Moore duality in the system \cite{Vidal09, Dusuel_PRL_2011}, a perturbation in the $y$ direction in the TC leads to a first-order transition at $\frac{K^4}{16 K^3_z} = \frac{2 h^2}{K_z}$,
which corresponds to a critical field strength of $h_c = \frac{K^2}{\sqrt{32} K_z}$ in the microscopic honeycomb model. 
At this transition, the system undergoes a phase change from a topologically ordered state to a $y$-polarized state of Eq.~\eqref{TChpert}, which corresponds to a VBS state on the honeycomb lattice in Eq.~\eqref{eq:ham_abelian} at large field strengths \cite{Feng2023}, as shown in Fig.~\ref{fig:phasesall}(b).
Note that although the full microscopic Hamiltonian defined in Eq.~\eqref{eq:ham_abelian} does not explicitly preserve exact Xu–Moore duality, such duality emerges effectively in the low-energy sector describing the TC \cite{Nussinov10}. Predictions based on this duality remain valid provided the anisotropy $K_z/K$ is sufficiently large and the magnetic field $h/K$ is weak compared to the field required for fully polarizing the honeycomb model. Beyond this regime, the emergent duality breaks down, and the VBS state (blue region in Fig.~\ref{fig:phasesall}(b)) is replaced by a trivial polarized state (gray region in Fig.~\ref{fig:phasesall}(b)).

In the regime where $h^2/K_z \ll 1$, the effective Hamiltonian $H_{\rm eff}$ remains gapped and is thus stable against small magnetic fields. While the excitations of this gapped Abelian QSL are typically described in terms of the bosonic Ising electric charge $e$, which resides on vertices where $A_s = -1$, and the magnetic charge $\rm M$, associated with plaquettes where $B_p = -1$, it is often useful to express the theory in terms of an equivalent set of anyonic excitations.  
Specifically, an alternative description of this Ising gauge theory can be formulated using the composite boson $2\epsilon = 2e \times 2m$, the composite fermion $\epsilon = e \times m$, and either one of the bosonic electric ($e$) or magnetic ($\rm M$) charges. This formulation provides a more direct way to summarize the dynamical effects of the magnetic field on the anyon excitations.  
In connection to Sec.~\ref{sec:pim}, these three types of excitations correspond, respectively, to the fully mobile four-face flip, the partially mobile two-face flip, and the immobile one-face flip excitations in the PIM.
Since the number of Abelian anyons is conserved modulo $2$, and a single immobile anyon cannot be created by local perturbation, we focus on the manifestation of the partially mobile $\epsilon$ and the fully mobile $2\epsilon$ particles.  

\begin{figure*}[t]
\centering
\centering
\includegraphics[width=0.95\textwidth]{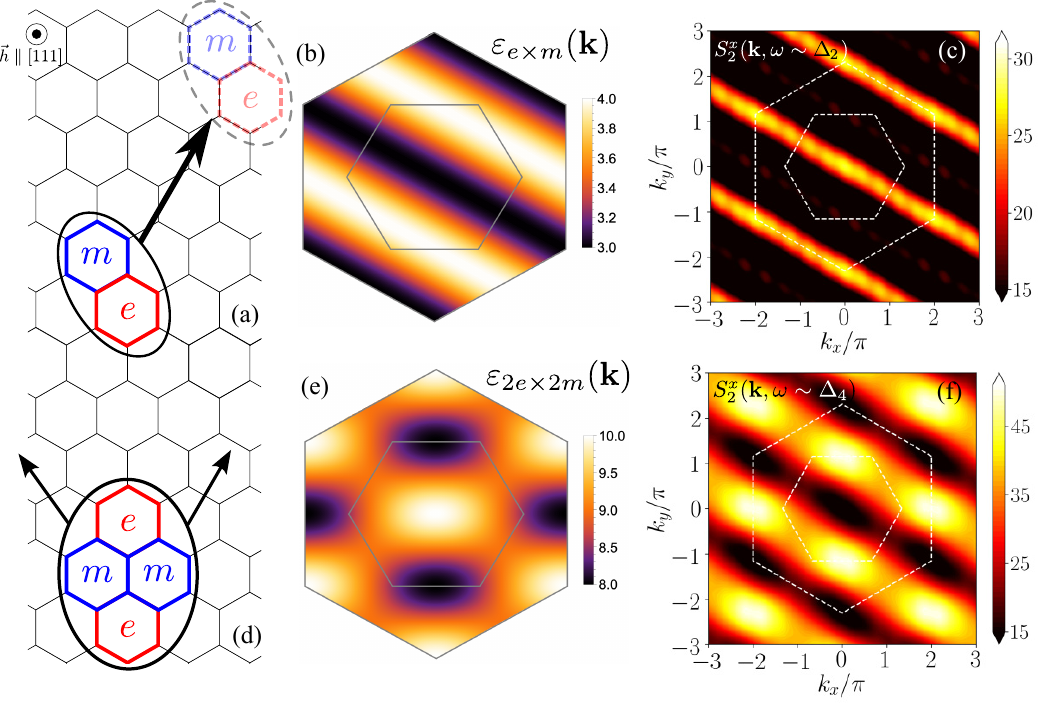}
\caption{
Magnetic field-induced dispersion of composite anyons in the Abelian QSL phase.  
(a) Illustration of the one-dimensional mode of the composite fermionic anyon $ \epsilon = e \times m $ induced by a weak out-of-plane magnetic field on the honeycomb lattice.  
(b) The dispersion $ \varepsilon(\mathbf{k}) $ of the two-anyon composite $ \epsilon $ within the second Brillouin zone (BZ). The gray solid lines denote the first and second BZ, and the color bar is in arbitrary units, emphasizing the one-dimensional structure.  
(c) Equal-energy slice of the two-spin-flip dynamical structure factor $ S_2^x(\mathbf{k}, \omega) $ at an energy near the gap of the $\epsilon $ particle, $ \omega = \Delta_2 $ The one-dimensional dynamical response reflects the low-energy soft modes of $ \epsilon $ shown in (b). Data obtained by DMRG with $96$-site ($12\times 4$ unit cells) cylinder geometry.   
(d) Illustration of the two-dimensional mode of the composite anyon $ 2\epsilon = 2e \times 2m $ induced by a weak out-of-plane magnetic field on the honeycomb lattice.  
(e) The dispersion $ \varepsilon(\mathbf{k}) $ of the four-anyon composite $ 2\epsilon = 2e \times 2m $ within the second BZ. The color bar is in arbitrary units, emphasizing the two-dimensional structure in contrast to (b).  
(f) Equal-energy slice of the two-spin-flip dynamical structure factor $ S_2^x(\mathbf{k}, \omega)$ at an energy near the gap of the $ 2\epsilon $ particle, $\omega = \Delta_4 \approx 2\Delta_2$. Data obtained by DMRG with $96$-site ($12\times 4$ unit cells) cylinder geometry. The two-dimensional dynamical response and the large weight near $\rm M$ points reflect the low-energy soft modes of $2\epsilon$ band near the $\rm M$ points, as seen in (e). Figures adapted from Ref.~\cite{Feng2023}.
}
\label{fig:tc24}
\end{figure*}

While all $2\epsilon$ and $\epsilon$ have gapped flat bands for $h=0$, the second term in Eq.~\eqref{TChpert}, provides one-dimensional dispersions in $\epsilon$, as is shown in Fig.~\ref{fig:lat1}(c,d), along the diagonal or off-diagonal directions $n_1$ or $n_2$ set by the initial excitation. To the leading order, the one-dimensional dispersions of $\epsilon$ fermion in the two-flux sector are
\begin{align}
    \varepsilon_\pm(\mathbf{k}) \sim 4J_{\rm TC} - \frac{4h^2}{K_z} \cos(\frac{1}{2} k_x \pm \frac{\sqrt{3}}{2} k_y) \label{eq:dispersion}
\end{align}
where $4J_{\rm TC}$ is the gap to an $\epsilon = e\times m$ excitation in absence of the magnetic field, and the subscript $\pm$ labels two different initial arrangements of $\epsilon$ shown in Fig.~\ref{fig:lat1}(c) and Fig.~\ref{fig:lat1}(d). Depending on whether the initial local excitation is $\epsilon_x$ or $\epsilon_y$, the transport directions are rotated by $90$ degrees on the square lattice compared with each other. The one-dimensional mobility of $\epsilon$ and its low-energy soft modes at $ \sqrt{3} k_x \pm 3 k_y = 4 n \pi $ are illustrated in Fig.~\ref{fig:tc24}(a,b).

In contrast to the constrained partial mobility of the $\epsilon$ excitation, the four-anyon composite $2\epsilon = 2e \times 2m$ is fully mobile without constraints, analogous to the unconstrained motion of the four-face flip excitations in the PIM shown in Fig.~\ref{fig:pim}(c).  
Notably, the $2\epsilon$ excitation acquires its two-dimensional mobility through a second-order process in the effective Hamiltonian Eq.~\eqref{TChpert}. Specifically, the motion of $2\epsilon$ occurs via the following sequence: first, the four-anyon composite is annihilated by the application of $\tau^y$ at the center of the excitation, forming an intermediate vacuum state devoid of any excitations. This is followed by the creation of another four-anyon excitation through the application of $\tau^y$ at a neighboring site. In the microscopic honeycomb model described by Eq.~\eqref{eq:ham_abelian}, this corresponds to a fourth-order process in $O(h^4/K_z^2)$.  
The resultant dispersion of the $2\epsilon$ anyons (four-face flip) is given by  
\begin{equation} \label{eq:dispersion2}
    \varepsilon(\mathbf{k}) \sim 8J_{\rm TC} + \frac{8h^4}{K_z^2} \sum_{\pm} \cos\left(\frac{1}{2} k_x \pm \frac{\sqrt{3}}{2} k_y\right),
\end{equation}
which allows the particle to move coherently along both the $n_1$ and $n_2$ directions, and the lowest-energy soft mode appears at the $\rm M$ point, as illustrated in Fig.~\ref{fig:tc24}(d,e) under arbitrary units.  

From Eq.~\eqref{eq:dispersion} and Eq.~\eqref{eq:dispersion2}, we see that the application of an out-of-plane magnetic field to the anisotropic Kitaev honeycomb model induces qualitatively different leading-order dynamics in the Abelian anyons and their composites. Given that the itinerant matter Majorana fermions are gapped at an energy scale of $O(K_z/K)$, which is significantly larger than the anyon energies of $O(K^4/K_z^3)$ in the anisotropic coupling regime of Eq.~\eqref{eq:ham_abelian}, it is expected that the distinct dynamical features of Abelian anyons and their various composites can be observed directly, without hybridization with Majorana excitations that are virtually spectators at very high energy.  
Furthermore, since the system hosts both fracton-like one-dimensional dispersion and fully mobile excitations at different energy slices, scattering experiments should be able to resolve these features separately. This would provide definitive evidence of fractionalization into Abelian anyons and the underlying $Z_2$ lattice gauge theory.  
In the next section, we explore how these signatures explicitly manifest in the dynamical structure factor of microscopic spin dimers (see also Table.~\ref{tab:tab3}), which is directly relevant for Raman and RIXS experiments.

\subsection{Anyon dispersion in dimer spectral function} \label{sec:dimerspec}
To verify the proposed picture of constrained and unconstrained anyon composites using unbiased numerical calculations,  
ED and DMRG has been applied to compute the dynamical structure factor of spin dimers \cite{Feng2023}:  
\begin{equation} \label{eq:dimerstructure}
    S_2^{\alpha}(\mathbf{k}, \omega) = - \frac{1}{\pi} \mathfrak{Im} \left[ \mel{\Psi} {{\mathcal{D}}_{\mathbf{k}}^\alpha \frac{1}{\omega - \mathcal{H} + i \eta} {\mathcal{D}}_{-\mathbf{k}}^\alpha} {\Psi} \right],
\end{equation}  
where we defined ${\mathcal{D}}^{\alpha}_i = \sigma_i^\alpha \sigma_{i+z}^\alpha$ is the two-spin operator defined on a $z$ bond, $H$ is the total microscopic Hamiltonian defined in Eq.~\eqref{eq:ham_abelian} including the $K_z$ anisotropy and the Zeeman field, $\eta$ is a small positive spectral broadening factor, and $\mathfrak{Im}$ takes the imaginary part. The subscript in $S_2^\alpha$ indicates a two-spin flip excitation, distinguishing it from $S_1^\alpha$, which corresponds to single-spin-flip excitations used in previous sections.  
We focus on the $\alpha = x$ channel, with the dimer operator defined on the $z$ bonds, ensuring that in the low-field limit, the application of $\mathcal{D}^x$ creates four fluxes above the ground state, sufficient for probing the dynamics of both $\epsilon$ and $2\epsilon$ particles,  while acting as a fermion number operator $1-2n_i = c_i c_{i+z}$ in the high-energy Majorana sector.

\begin{figure}[b]
    \centering
    \includegraphics[width=\linewidth]{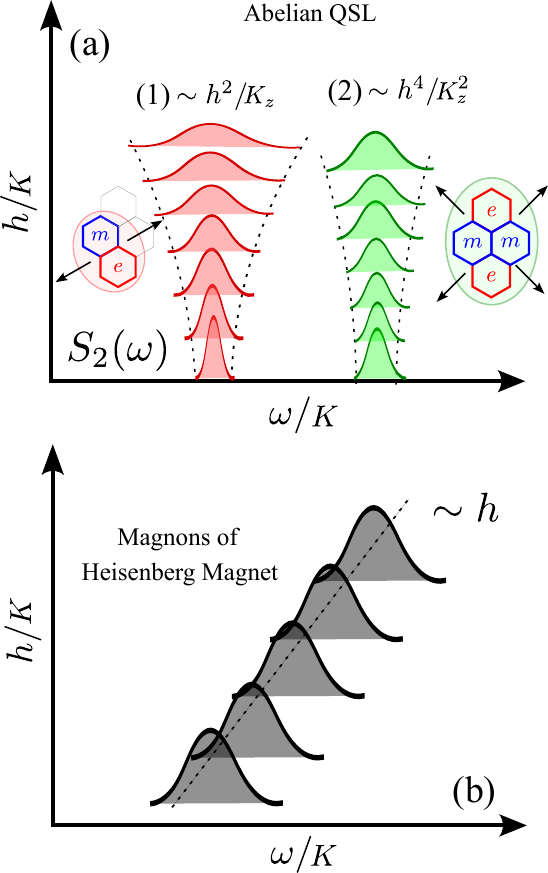}
    \caption{(a) An illustrative sketch of the rescaled spectrum $S^{x}_2(\omega)$ for the operator $\mathcal{D}^x = \sigma_j^x \sigma_{j+z}^x$ at $K_z/K=2.5$ for different values of $h$. The dashed black lines are eye-guiders that mark the evolution of anyon bandwidths as a function of $h$ in the perturbative limit. Spectral peaks along lines (1) and (2) correspond to the dispersion of $\epsilon$ and $2\epsilon$ shown on the right panel. (b) The same for Heisenberg magnet, where the dynamical spectral function scales linearly with a Zeeman field and the bandwidth of magnons remains unchanged.}
    \label{fig:ed}
\end{figure}

Figure~\ref{fig:ed} presents the ED results for the momentum-integrated spectral function, defined as $S_2^x(\omega) = \int_{\rm BZ} S_2^x(\mathbf{k}, \omega) d \mathbf{k}$. 
Since the operator ${\mathcal{D}}^{x}_i$ excites exactly four Abelian anyons (or equivalently, two $\epsilon$ particles) at zero magnetic field, a prominent spectral peak appears at $\omega/K \sim 4J_{\rm TC}$. At zero field, the two-anyon composite $\epsilon$ remains invisible due to its vanishing overlap with $2\epsilon$.  
As the magnetic field is introduced, the anyons begin to acquire mobility. The two constituent $\epsilon$ particles within the four-anyon excitation become partially mobile, and the overlap between the $2\epsilon$ composite boson and the partially mobile $\epsilon$ deviates from zero. This explains why the lowest-energy peak is initially weak at small field but becomes more pronounced as the field increases.  
The spreading of the spectral peak from zero field can be attributed to the increasing bandwidth of the dispersing $2\epsilon$ and $4\epsilon$ composite anyons.
The cartoon profile of the black dashed lines in Fig.~\ref{fig:ed}(a) elucidates the nature of the perturbative processes of the bandwidth spreading responsible for anyon mobility. The $\epsilon$ band width follows an $O(h^2/K_z)$ scaling, as predicted by Eq.~\eqref{eq:dispersion}, and is marked by the set of black dashed lines (1) in Fig.~\ref{fig:ed}(a). Meanwhile, the $2\epsilon$ composite scales as $O(h^4/K_z^2)$ according to Eq.~\eqref{eq:dispersion2}, exhibiting a much smaller broadening in its bandwidth at small field, as indicated by the black dashed lines (2) in Fig.~\ref{fig:ed}(a).   
The presence of these two distinct spectral peaks and their characteristic scaling with $h$ provides rudimentary numerical evidence that the two-spin-flip process in the presence of an external magnetic field is capable of resolving different anyon composites with potentially distinct mobilities.  This is in sharp contrast to the spectral function of a trivial Heisenberg magnet, see Fig.~\ref{fig:ed}(b), where the bandwidth of magnons are not affected by Zeeman fields.

To explicitly observe the sub-dimensional dynamics of $\epsilon$ and, in contrast, the full mobility of $2\epsilon$, requires a much finer momentum resolution than what is accessible in small ED clusters. This was achieved in \cite{Feng2023} using a Krylov-space DMRG algorithm for finite-energy correlation vectors in Eq.~\eqref{eq:dimerstructure}. The results are presented in Fig.~\ref{fig:tc24}(c,f) for $\epsilon$ and $2\epsilon$, respectively.  
Figure~\ref{fig:tc24}(c) shows the equal-energy slice of the two-spin-flip dynamical structure factor $S_2^x(\mathbf{k}, \omega)$ at an energy near the gap of the $\epsilon$ particle, $\omega = \Delta_2 \sim 4 J_{\rm TC}$. The prominent one-dimensional dynamical response directly reflects the low-energy soft modes of $\epsilon$ shown in Fig.~\ref{fig:tc24}(b).  
A similar analysis for $\omega = \Delta_4 \sim 8 J_{\rm TC}$ reveals that the $2\epsilon$ excitation enjoys full mobility along all directions. This is evident in the two-dimensional modulation of $S_2^x(\mathbf{k}, \omega)$ in Fig.~\ref{fig:tc24}(f), in sharp contrast to the constrained motion seen in Fig.~\ref{fig:tc24}(c). Furthermore, by comparing the perturbative dispersion of $2\epsilon$ in Fig.~\ref{fig:tc24}(e), one finds a striking consistency with $S_2^x(\mathbf{k}, \omega)$, which reflects its lowest-energy modes at the $\rm M$ point.  The numerical evidence along with analytical insights provide a clear separation of these mobility characteristics and suggests that scattering experiments such as Raman and RIXS could provide direct evidence in sub-dimensional dynamics of Abelian anyons and the underlying $Z_2$ lattice gauge theory in potential anisotropic Kitaev materials.

\subsection{Outlook: dynamical signatures of composite anyons}
Besides the anisotropic spin structure factor due to the emergent sub-dimensional dynamics of composite anyons as discussed in the previous section, 
based on the constraint dynamics of composite anyons, another promising direction is exploiting non-equilibrium dynamics to identify anyons and other fractionalized excitations in quantum spin liquids, with particular emphasis on regimes where one-dimensional constraints give rise to emergent 1D anyon dispersion and effective fermionic braiding in pump–probe experiments—mechanisms that are fundamentally distinct from the usual two-dimensional braiding of self-bosons or mutual semions. As illustrated by our dimensional-reduction theory, such constraints produce characteristic, tunable anyon-winding patterns and thus predictable non-equilibrium and nonlinear dynamical signatures.
One concrete protocol is pump–probe spectroscopy \cite{Parameswaran23,mcginley2022}: by driving a topologically ordered system with two pulses separated by a variable delay, anyons created by the first pulse braid in space–time with those generated by the second, imprinting a statistical phase onto the nonlinear response. Under nontrivial mutual statistics, the high-order response functions exhibit phase-interference patterns that grow in time rather than decaying exponentially as in purely bosonic media. 
In particular, recent simulations \cite{yang2025} have validated the nonlinear pump–probe technique by capturing mutual‐semion braiding statistics in the toric code, thus establishing a powerful tool for probing anyonic phases. However, that work did not address the one‐dimensional mutual‐fermion braiding statistics that emerge when a weak magnetic field is applied along specific directions — an effect of dimensional reduction unique to the spin-orbital frustrated quantum spin liquids. Because these 1D fermionic anyons obey fundamentally different interference along space-time trajectories, the early-time scaling of dynamical susceptibility would produce qualitatively distinct nonlinear susceptibilities than those with 2D dynamics. Varying the magnetic‐field angle in pump–probe experiments therefore offers a promising path to tune the effective dimension of anyon dispersion, directly observe 1D mutual‐fermion braiding, and differentiate it from the 2D semion case.

\section{Sub-dimensional dynamics by \\ strong magnetic field} \label{sec:dim2}
In the previous sections, we reported the recent progress on the emergence of sub-dimensional physics under a perturbative magnetic field, which induces inherent partial mobility in anyon pairs. This behavior is dictated by the underlying gauge theory, where local or gauge-like symmetries are broken down to subsystem symmetries. However, sub-dimensional physics can also arise through alternative mechanisms, where modes along certain directions are selectively condensed or depleted, effectively suppressing inter-line or inter-layer couplings. A paradigmatic example is the well-known phenomenon of dimensional reduction near a quantum critical point \cite{Fisher2006, Fisher2007, Batista2008}, as observed in the three-dimensional frustrated insulator BaCuSi$_2$O$_6$. In this system, a crossover to two-dimensional behavior occurs near a critical magnetic field, providing an experimentally verified instance of emergent sub-dimensional dynamics beyond gauge-theoretic origins.
In an analogous manner, a similar dimensional reduction can occur in the two-dimensional Kitaev QSL under a critical magnetic field. Specifically, near a critical field, the system can undergo a crossover from a fully two-dimensional entangled state to effectively decoupled one-dimensional spin chains \cite{Feng24}. This emergent one-dimensionality is a direct consequence of the unique spin-orbital coupling inherent in the Kitaev model, where spatial directions and exchange interactions are intertwined. Intuitively, this suggests that a tuned external magnetic field can selectively suppress inter-chain spin-orbital exchanges: for example, explicitly breaking the rotational symmetry by a weak directional magnetic field can introduce a strong anisotropic response at intermediate time scales \cite{Dai25}. 
Remarkably, a much more drastic crossover into completely 1D effective theory can be achieved under moderate field, 
effectively isolating the 2D lattice into a set of quasi-1D subsystems. Unlike the gauge-theoretic mechanisms discussed earlier, this dimensional decoupling arises from the competition between magnetic field and the intrinsic anisotropic spin liquid correlations. 
In the following sections, we present both analytical arguments and unbiased numerical results to establish this crossover.

\begin{figure*}[t]
    \centering
    \includegraphics[width=\linewidth]{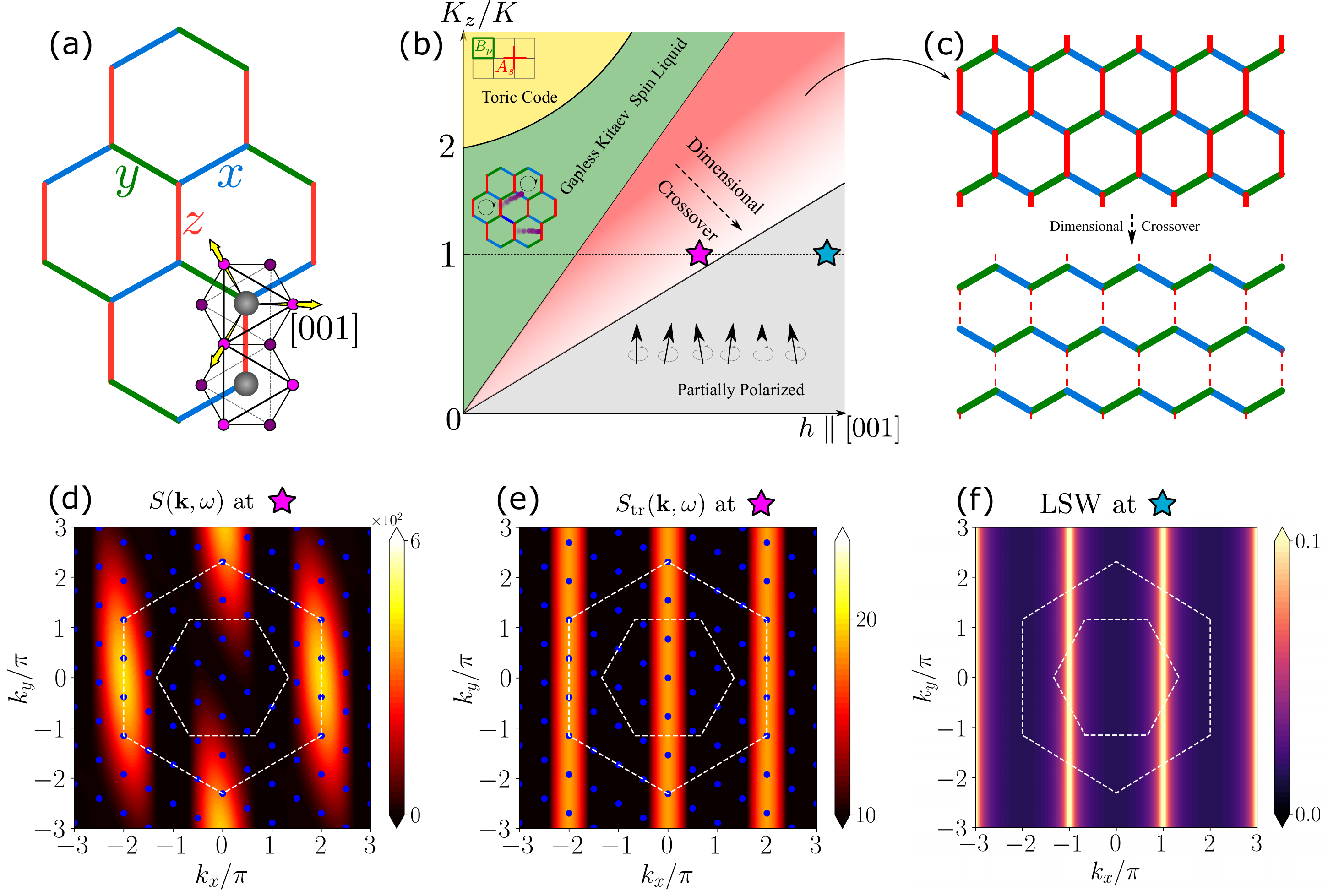}
    \caption{Magnetic-field-induced dimensional crossover in the Kitaev honeycomb model at a quantum critical point. (a) Bond-dependent spin exchange in the Kitaev material with an edge-sharing octahedra structure. Ligands are represented by small purple circles, while the pseudo-spins relevant for Kitaev interactions are defined on the large gray circle. The yellow arrows defines the local axis of spins. The [001] magnetic field points from a hexagonal vertex to one of its ligands. (b) The schematic phase diagram as a function of the anisotropy in the compass exchange and the [001] magnetic field. Four phases are identified with distinct colors: toric code in yellow; the gapless nodal Kitaev spin liquid in green; an emergent gapless phase under moderate [001] field in the red-to-white color gradient, depicting a dimensional crossover as a function of magnetic field; and the partially polarized phase in gray. (c) A cartoon illustration for the emergent one-dimensional physics under [001] field, i.e. the red-to-white region in (b). The inter-spin-chain coupling asymptotically vanishes as the field reaches the quantum critical point, e.g. as marked by the purple star in (b), whereby the quasi-particle mobility across $z$ bonds becomes negligible. 
    (d) The low-energy cut of the total dynamical structure factor $S(\mathbf{k}, \omega)$ near the critical point marked by the purple star in (b), and (e) of the reduced dynamical structure factor $S_{\rm tr} (\mathbf{k}, \omega)$ at the same point by tracing out sublattice degrees of freedom. The first and second Brillouin zones are marked by white dashed lines. The absence of modulation in $k_x$ showcases the reduced dimensionality in (c) under moderate magnetic field. Data obtained by $24$-site ED under PBC. The blue dots on top of the interpolated structure factors mark the available momenta in the finite-size cluster. (f) The lower magnon band of the partially polarized phase obtained by linear spin wave (LSW) theory, also featuring a one-dimensional dispersion. Figures adapted from Ref.~\cite{Feng24}. 
     }
    \label{fig:dimredhz}
\end{figure*}

\subsection{One-dimensional response near the critical point}
We now discuss the phenomenon of dimensional reduction near quantum criticality at high magnetic fields. Consider the Kitaev model under a [001] Zeeman field:  
\begin{align}
    H = \sum_{\langle ij\rangle_{\alpha \in \{x,y\}}} K \sigma^\alpha_i \sigma^\alpha_j  + K_z\sum_{\langle ij\rangle_z} \sigma^z_i \sigma^z_j  - h\sum_{i} \sigma^z_i,
    \label{eq_spinHam}
\end{align}
where the exchange interactions are antiferromagnetic ($K, K_z > 0$), and $\alpha \in \{x,y,z\}$.  
Notably, in contrast to a [111] field, which induces a gapped chiral spin liquid phase, a [001] field does not open a gap in the Majorana sector. At small [001] field strengths, the spin liquid phase—denoted by the green region in Fig.~\ref{fig:dimredhz}(b)—remains gapless, with spin-spin correlations exhibiting power-law decay \cite{Kitaev2011}. However, at larger magnetic fields, beyond the reach of perturbation theory, the physics changes dramatically.  

Early studies of Majorana topological transitions driven by a moderate [001] magnetic field in the isotropic limit ($K_x = K_y = K_z$) using mean-field theories in both the Majorana and complex fermion representations have been reported in \cite{Nasu2018, Liang2018}. These studies, based on different mean-field ansätze, demonstrated the persistence of gaplessness in an intermediate phase. While this phase bears similarities to the intermediate gapless phase induced by a [111] magnetic field, as discussed in previous sections, its underlying physical description is fundamentally distinct. These works revealed the presence of one-dimensional gapless nodal lines at intermediate magnetic fields, albeit followed by a regime where two-dimensional modulations re-emerge in the mean-field bands before the system ultimately undergoes confinement into the partially polarized phase. This suggests that the response and dynamics under a moderate magnetic field can exhibit an effective dimensionality distinct from that of the underlying Hamiltonian.  
This behavior is somewhat expected due to the spin-orbital coupling inherent in the Kitaev model and the fact that the three-fold spatial rotational symmetry of the pure Kitaev honeycomb model is explicitly broken by the application of a [001] field. However, the extent to which the system effectively transitions into a completely one-dimensional regime, where it behaves as a collection of decoupled compass spin chains, is striking and presents a potential positive evidence of Kitaev spin liquids.  
More concrete numerical evidence supporting this dimensional reduction has been provided recently by exact ED and DMRG studies \cite{Feng24}. These results demonstrate a stronger dimensional reduction near the critical magnetic field than previously found in mean-field studies, and report a completely one-dimensional spectral function at the end of the intermediate gapless phase (near the transition to the partially polarized phase), as indicated by the purple star in Fig.~\ref{fig:dimredhz}(b). A brief summary of the model setup and key numerical results is shown in Fig.~\ref{fig:dimredhz}.

The most direct evidence of this dramatic dimensional crossover is found in the dynamical spin structure factor. To connect with the parton mean-field theory, which will be discussed later in Sec.~\ref{sec:pmft}, it is useful to compute both the total dynamical structure factor, $S(\mathbf{k}, \omega)$, and the reduced structure factor, $S_{\rm tr}(\mathbf{k}, \omega)$, obtained by tracing out the sublattice degrees of freedom. The symmetrized dynamical structure factors, computed via ED at a low-energy cutoff near zero energy, are presented in Fig.~\ref{fig:dimredhz}(d,e).  
A striking feature of these results is that the intensity peaks are aligned parallel to the $k_y$ direction, indicating that the spectral function is predominantly governed by the zigzag compass chains. Additionally, the concentration of spectral weight in $S(\mathbf{k}, \omega)$ around the second Brillouin zone boundary suggests that the flux degrees of freedom in the spin liquid phase remain robust under the applied magnetic field, leading to short-range spin correlations due to the approximate orthogonality of flux states. This phenomenon closely resembles the intermediate phase observed under a [111] field, where the $Z_2$ gauge structure persists at moderate magnetic fields.  
The reduced structure factor, $S_{\rm tr}(\mathbf{k},\omega)$, shown in Fig.~\ref{fig:dimredhz}(e), further highlights the emergent one-dimensional nature of the intermediate phase. The intensity modulation is almost entirely along the $k_x$ direction near zero energy, with only minor variations along $k_y$ at higher energies, reflecting weak interchain coupling between the zigzag chains. Notably, the strongest spectral intensity in $S_{\rm tr}(\mathbf{k},\omega)$ occurs near $k_x = 0 \pmod{2\pi}$ for all $k_y$, consistent with the fact that the zigzag compass chain can be mapped to a critical transverse field Ising model (TFIM) with fermion dispersion given by \cite{Brzezicki2007} $\varepsilon(k_x) \propto \sqrt{2 - 2\cos{k_x}}$. 
While DMRG is less effective in capturing dynamical properties due to the diverging correlation length near the critical point, static structure factor calculations on moderate-sized cylinders, as shown in Ref.~\cite{Feng24}, exhibit similar one-dimensional characteristics to ED, even at lower magnetic fields further from the transition.  
Interestingly, this one-dimensional character persists even after the confinement transition into the partially polarized phase. As shown in Fig.~\ref{fig:dimredhz}(f), and further confirmed numerically by DMRG in \cite{Feng24}, the linear spin wave theory in the partially polarized phase still exhibits a completely one-dimensional magnon band.  
These numerical results reveal a striking sub-dimensional response induced by a strong magnetic field. In contrast to the previous sections, where sub-dimensional behavior was tied to the underlying gauge structure, here it arises from the energetic suppression of inter-chain excitations through an explicitly applied external magnetic field, and a larger anisotropic energy scale $K_z/K$ would require a larger magnetic field to achieve the crossover, as illustrated in the expanding intermediate phase towards the upper right shown in Fig.~\ref{fig:dimredhz}(b). Moreover, this dimensional reduction is even more pronounced than what was predicted by previous mean-field theories, potentially caused by omitting the superconducting pairing between canonical fermions that is essential to the $Z_2$ structure \cite{Liang2018,Feng24}. In the following section, we review the $Z_2$ mean-field ansatz which required to fully capture this drastic dimensional crossover.

\subsection{Parton mean-field theory} \label{sec:pmft}
Now we briefly review the parton mean-field theory which accounts for the dimensional reduction in the numerical results. In the Majorana/fermionized view, the honeycomb model maps to free matter fermions in a static $Z_2$ gauge background—equivalently a spinless $p+ip$ superconductor after Jordan–Wigner/fermionization—so band geometry and Lifshitz-type crossovers are transparent. Concretely, the Jordan-Wigner mappings of Ref.~\cite{Xiang07} and the stabilizer formalism of  Ref.~\cite{Kells2009,Kells2009b} make the $p+ip$ structure explicit, while slave-fermion extensions and the augmented parton MFT incorporate Heisenberg and $\Gamma$ perturbations as well as magnetic fields, reproducing the exact Kitaev QSL at zero field yet remaining tractable beyond integrability \cite{Subhro12,Knolle2018b}. Parton mean field theory is thus ideal for the emergent dimensional crossover: the gradual depletion of bond fermions reshapes the itinerant band structure without invoking flux proliferation. This is somewhat akin to a process of parton Lifshitz transition without opening or closing a gap due to the gradual depletion of a particular flavor of fermions \cite{Feng_gapless,Feng2020}. Its limitation is that self-averaged, translation-invariant ansätze cannot capture coherent-disorder effects (vison-induced scattering, localization); those require explicit gauge-fluctuation treatments or unbiased numerics. In the regime focused here, i.e. without strong spatial flux disorder near the critical point, the uniform-gauge MFT provides a compact, accurate description of the crossover physics.

To analyze this more effectively, it is convenient to formalize the mean field theory in the complex fermion representation by transforming the Kitaev Hamiltonian from the four-Majorana formulation into a $p+ip$ superconductor. This is achieved via the transformation $c_{i,A +\hat{z}} = i (f_i-f^\dagger_i), \quad c_{i,A} =  f_i + f^\dagger_i$,
where the Majorana operators $b^\alpha_{i}$ and $c_i$ satisfy the algebra $c^2_i=1$ and $\{c_i, c_j\}=2\delta_{ij}$. For convenience, we label the sublattice indices as $A$ and $B \equiv A + \hat{\alpha}$, with $\hat{\alpha} \in \{x,y,z\}$ denoting the lattice unit vectors.  
Under this notation, the hopping of Majorana particles along the $z$-bonds is now written as $c_{i,A}c_{i,A+\hat{z}} = i(2n^f_i-1)$.
Similarly, the static Majoranas can be expressed as bond fermions $\eta$, as has been discussed previously in Sec.~\ref{sec:bondf},
such that the static $b$ Majorana fermions are related to bond fermion occupation in terms of $b^\alpha_{i,A}b^\alpha_{i,A + \hat{\alpha}}= i(2n^{\alpha}_i-1)$.
Under these transformations, the Kitaev exchange interactions on the $z$- and $x$-bonds take the following forms:  
\begin{align}
    \sigma_{i,A}^z \sigma_{i,A+\hat{z}}^z &=  (2n^f_i-1) (1-2n^{z}_i), \label{eq:niz}\\
    \sigma_{i,A}^x \sigma_{i,A+\hat{x}}^x &= (1-2n^{x}_i)(f_i f_{i-\mathbf{n}_1}-f_if^\dagger_{i-\mathbf{n}_1}+ {\rm H.c.}), \label{eq:nif}
\end{align}
where the $K_y$ exchange interaction follows by exchanging the indices $\hat{x} \leftrightarrow \hat{y}$ and $\mathbf{n}_1 \leftrightarrow \mathbf{n}_2$.  
The external field along the $\alpha$ direction, acting on the $A$ sublattice, takes the form  
\begin{equation}
    h^\alpha (i b^\alpha_{i,A} c_{i,A}) = i h^\alpha (\eta_{i\alpha}+ \eta^\dagger_{i\alpha})(f_i + f^\dagger_i),
\end{equation}
which introduces hybridization between the canonical $f$ fermions and the bond fermions.  
Since the Majorana sector of the Kitaev model exhibits $p$-wave pairing \cite{Read2000}, we incorporate into our mean-field ansatz the superconducting correlators, Hartree terms, and bond fermion occupations along different bonds: $\phi = \expval{f_i f_j}, \quad \xi = \expval*{f_i^\dagger f_j}, \quad n_0^\alpha = \expval{n^\alpha}$,
where $\alpha = x,y,z$. In the flux-free sector, bond fermions are uniformly occupied with $n^z_\alpha=1$.  
Under this simplified formulation of parton mean-field theory, it becomes readily apparent under what conditions the system decouples into effectively disconnected chains with vanishing $z$-bond interactions, as suggested by previous numerical results. This can be understood by examining Eq.~\eqref{eq:niz}, where, within a mean-field picture, the exchange interaction along $z$-bonds vanishes if $\expval{n_i^z} \rightarrow 0.5$.

\begin{figure}[t]
    \centering
    \includegraphics[width=\linewidth]{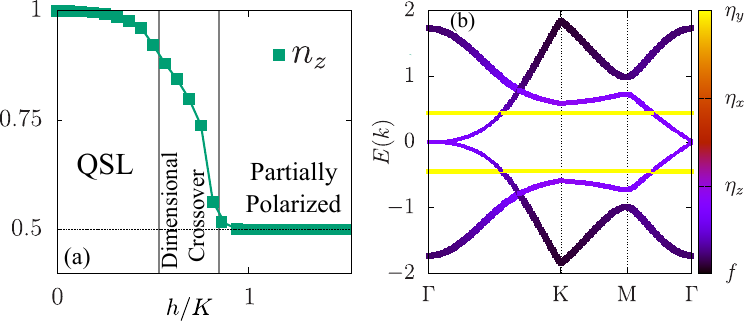}
    \caption{Results from parton mean-field theory. (a) The expectation of the occupation number of $z$ bond fermion at different magnetic fields and a fixed $K_z/K = 1$.  (b) Dispersion of different partons in the intermediate phase where the dimensional reduction occurs. The color bar denotes the orbital contribution of the states. Figures adapted from Ref.~\cite{Feng24}. }
    \label{fig:parton}
\end{figure}

Remarkably, the self-consistent solution of this mean-field theory produces precisely this half-filling condition for $n_i^z$ near the phase transition, as shown in Fig.~\ref{fig:parton}(a). The occupation number is initially $1$ in the nodal Kitaev spin liquid phase due to the flux-free condition but begins to rapidly deplete in the intermediate phase, where numerical results indicate a dimensional crossover, before eventually vanishing in the polarized phase beyond the confinement transition, where the parton decomposition is no longer valid. This behavior can be interpreted as a process of Lifshitz transition of partons \cite{Feng_gapless}, where the selective depletion of one species of parton effectively reduces inter-chain interactions. This is analogous to the well-known example in the emergent sub-dimensional behavior in frustrated insulator BaCuSi$_2$O$_6$, albeit in that case it is the selective condensation of bosonic modes that effectively eliminates inter-plane modes in the bulk material, leading to an effective reduction in dimensionality from three to two \cite{Batista2008}.  
Additionally, Fig.~\ref{fig:parton}(b) presents the dispersion of different partons in the intermediate phase, where the dimensional reduction occurs. The color bar indicates the orbital contribution of the states. A key takeaway from this result is that the zero-energy states appear at the $\Gamma$ point, which is consistent with the expectation that the gapless point of a one-dimensional $p$-wave superconductor, and its equivalent transverse field Ising model, occurs at $k_x = 0$, corresponding to the $\Gamma$ point.  
Further numerical validation may be obtained by fitting the central charge, although this remains a challenging task for two-dimensional models in DMRG. Initial attempts have been made in recent work \cite{Feng24}, where calculations on a moderate-sized cylinder suggest that the central charge is consistent with that of a free fermion system in one dimension.

\subsection{Outlook: Magnetic-field angle dependence }
Recent studies have proposed exploiting the magnetic‐field angle dependence of low‐energy responses to identify definitive signatures of Kitaev quantum spin liquids \cite{Kyusung22,imamura2025}. While prior work has elucidated how varying the field controls the Majorana gap and band topology, the pronounced dimensional reduction that emerges under moderate fields aligned along high‐symmetry axes remains underexplored. In three‐dimensional frustrated Mott insulators, similar field‐driven crossovers, e.g. from 3D to 2D and even to 1D, have served as clear markers of spin–orbital coupling \cite{Fisher2006,Fisher2007,Batista2008}. In the honeycomb Kitaev model, a field along the $[001]$ direction energetically suppresses dynamics along $z$ bonds and effectively decouples the system into anisotropic chains at large magnetic field, producing sub-dimensional dynamics and strongly anisotropic transport. 
The appearance of the sub-dimensional dynamics in spin-orbital liquid is not merely just an accidental quirk. 
In essence, it is because, in compass models spin and real space are locked in patterns due to spin-orbital coupling, and the field that couples to spin degrees of freedom would be reflected in the corresponding real-space dynamics. This allows in-situ tuning of dimensional reduction and anisotropic transport.
Angle‐resolved probes such as thermal conductivity or Raman spectroscopy should therefore exhibit dramatic anisotropies and abrupt dimensional crossovers as the field rotates through the $[001]$, $[010]$ or $[100]$ axis.

Programmable Rydberg‐atom arrays \cite{Semeghini2021} offer an especially clean platform to realize and probe this field‐induced dimensional reduction without the complications of non‐Kitaev exchange present in solid‐state materials \cite{Ruben22,Willian25}. By encoding spins as Rydberg excitations and engineering effective $x$, $y$, and $z$ spin-orbital couplings via laser detunings, one can apply a tunable longitudinal or transverse field that selectively weakens different bonds and drives the crossover to quasi‐1D behavior. Real‐time measurements in these arrays can also directly track the onset of sub-dimensional correlations and transport anisotropy, providing unambiguous evidence of field‐angle–tuned dimensional reduction in the Kitaev model.

\section{Concluding remarks} \label{sec:conclude}
Magnetic fields, besides playing a crucial role in stabilizing QSL phases in many candidate materials, serve as a versatile knob for probing fractionalized quantum matter. (1) From a fractionalization perspective, weak fields in the anisotropic Abelian limit produce a distinct, direction-dependent response of the $Z_2$ gauge fluxes, while moderate fields expose gapless Majorana modes in the dynamical spin structure factor without being obscured by the vison gap. These predictions can be directly tested in neutral-scattering experiments (e.g.\ inelastic neutron or Raman) and are well captured by a simple effective Majorana tight-binding model in the presence of a magnetic field. (2) In terms of quantum phases, a magnetic field can drive the system into a neutral quantum Majorana metal with finite zero-energy density of states akin to a filled-pocket Fermi surface. This scenario provides a concrete alternative to Dirac QSL or $U(1)$ spinon FS interpretations of recently observed gapless QSL candidates, and it predicts sub-diffusive behavior when translation-invariant flux disorder is induced by the field. (3) In view of dynamics, magnetic fields can act as part of the probe itself: when applied along special directions they induce emergent subdimensional dynamics by effectively decoupling Kitaev chains, and they generate field-tunable anyon motion whose nonlinear response provides a controlled benchmark for theory. This approach opens the door to observing novel fracton-like subdimensional physics of $Z_2$ flux excitations and to mapping out magnetic-field–dependent dimensional reduction in both time-domain and frequency-domain measurements.

We close this report by highlighting additional non-linear and non-equilibrium avenues for exploration beyond those discussed above.  Quantum spin liquids under magnetic fields offer a versatile playground and clean testbed for a wide range of experimental probes that can sharpen our understanding of fractionalization and validate candidate materials.  In particular, nonlinear spectroscopies, such as pump–probe and higher‐order response measurements, promise direct access to both the emergence of fractionalized Majorana and spinon excitations and their braiding statistics.  By tailoring the pump and probe field orientations, one can selectively excite noncommuting operators and distinguish trivial from nontrivial anyon exchanges in field‐induced phases.  When combined with tilted‐field calorimetry, noise spectroscopy, and angle‐resolved scattering, these nonlinear approaches will resolve the unique dynamical fingerprints of each magnetic‐field–driven spin liquid regime — whether it be the quantum Majorana metal, subdimensional decoupled chains, or fermion‐flux interplay — and thereby provide definitive tests of QSL behavior in real materials.

\begin{figure}[t]
    \centering
    \includegraphics[width=0.9\linewidth]{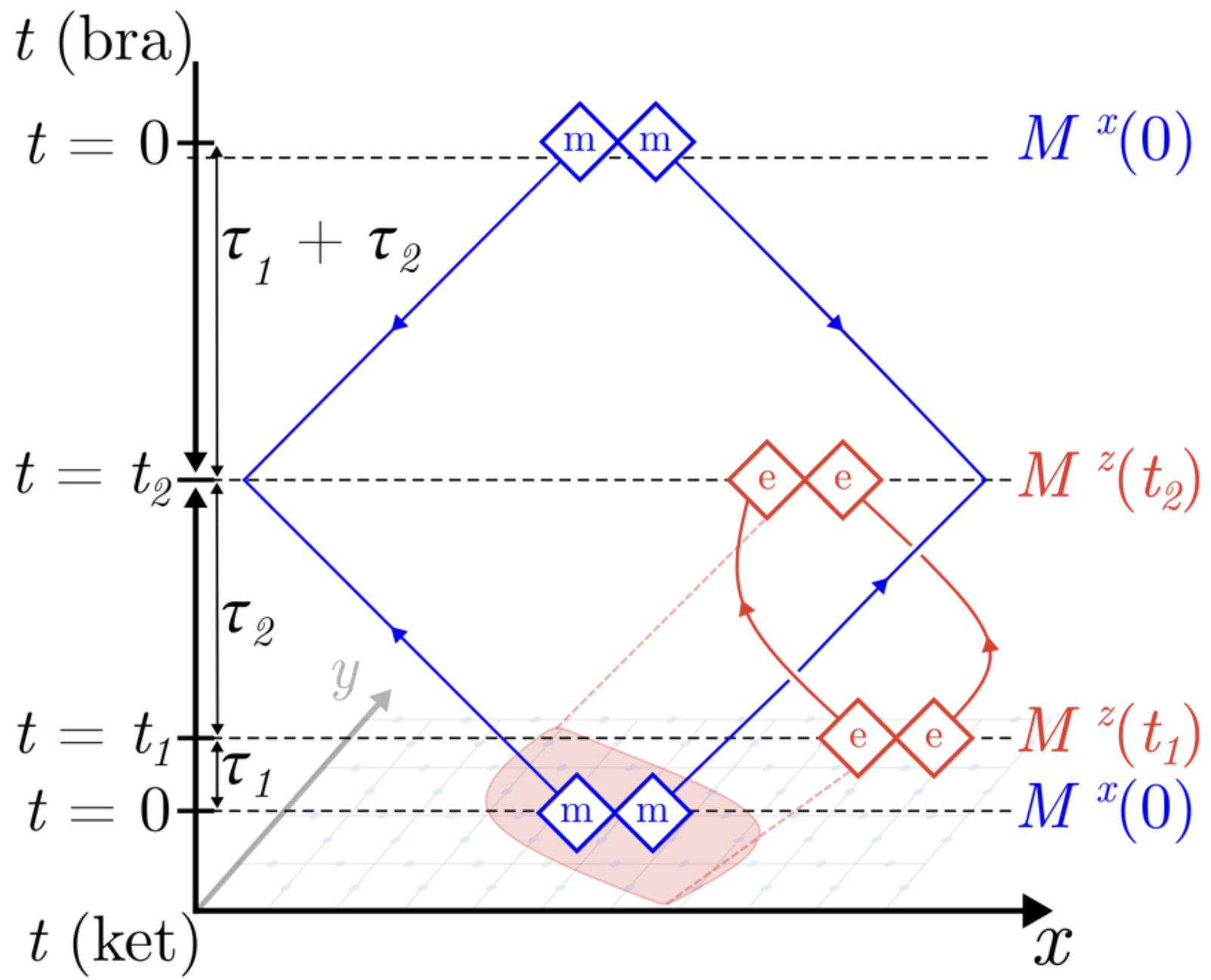}
    \caption{Schematic diagram of the anyon‐braiding processes corresponding to $\chi^{(3)}_{XZZ}$. At $t=0$, a pair of $\rm M$ anyons is created; at $t=t_1$, a pair of $e$ anyons is created and subsequently annihilated at $t=t_2$. The $e$ worldlines (red) form a closed loop that encloses the trajectories of the $\rm M$ (blue) anyons, and the probability of braiding, and hence the nonlinear response, scales with the area of the loop (shaded in red). Figure from Ref.~\cite{yang2025}.}
    \label{fig:pump}
\end{figure}

Recently, it has been proposed that nonlinear pump–probe spectroscopy can directly detect anyon braiding in lattice $Z_2$ gauge theory ~\cite{mcginley2022,Parameswaran23,yang2025}.  The key idea is to apply pump and probe pulses that couple via magnetic fields oriented along different axes, thereby interrogating non‐commuting spin operators.  In the toric‐code context, orthogonal fields excite mobile electric charge ($e$) and magnetic flux ($m$) that braid around each other with semion mutual statistics, whereas identical field orientations create only bosonic anyons ($e–e$, $m–m$).  As a result, the third‐order susceptibility $\chi^{(3)}_{XZZ}$, as illustrated in Fig.~\ref{fig:pump}, involves nontrivial $e–m$ braiding, hence exhibits a linear‐in‐time enhancement at early delays, in stark contrast to $\chi^{(3)}_{XXX}$ \cite{yang2025}.  Comparing these two signals therefore provides an unambiguous fingerprint of anyon statistics, pointing the way toward pump–probe experiments as a central tool for identifying fractionalization and braiding in QSLs, beyond the broad, often inconclusive features seen in single‐spin‐flip inelastic neutron scattering. These pump–probe braiding signatures have been verified in idealized models, such as the toric code or transverse‐field Ising chains, but their application to more realistic Kitaev and extended‐Kitaev models, as well as to actual material candidates in field‐induced QSL regimes, remains unexplored.
For example, the magnetic‐field‐induced sub-dimensional dynamics in Kitaev and related models can also be probed via pump–probe methods.  Braiding in 2+1D versus effective 1+1D regimes yields distinct temporal scalings of the nonlinear response, allowing one to map out field‐angle–tuned crossovers from 2D to quasi‐1D behavior. 
Moreover, closely related to time-resolved nonlinear pump–probe spectroscopy is frequency-resolved 2D coherent spectroscopy \cite{Cundiff13,Dorfman16}. Enabled by the ability to access multitime correlation functions sensitive to interactions between excitations, 2D coherent spectroscopy can resolve characteristic sharp signals of fractionalized modes that remain a fuzzy continuum in linear response theory \cite{GiBaik23,srivastava2025,Watanabe}. Unlike conventional linear spectroscopy, coherent spectroscopy reveals not only the linear response of spin flips but gives more direct access to the interplay of intrinsic excitations of magnets. It thus was theoretically proposed that such interplay can be used to identify the presence of fractionalized particles \cite{Armitage19,Yuan21} and quantum spin liquids \cite{Nandkishore21,Wonjune20,Langari23}. For a more comprehensive technical treatment, see Refs.~\cite{Dorfman16,Armitage19,Hamm_Zanni_2011} and references therein.

These magnetic‐field‐induced phase transitions and dynamics call for detailed exploration in other non-equilibrium platforms such as programmable Rydberg‐atom arrays. Recent works have shown how to engineer the Kitaev honeycomb and related compass models under both in‐plane and out‐of‐plane fields~\cite{Semeghini2021,Ebadi2021,Ebadi22,Ruben22}, entirely avoiding the large non‐Kitaev couplings of solid‐state platforms. Time‐resolved measurements in these arrays would thus offer clean tests of sub-dimensional decoupling, anyon braiding, and the putative quantum Majorana metal, and allow out‐of‐equilibrium preparation of $Z_2$ flux configurations to directly observe Majorana localization and quench dynamics. Together, these experimental and synthetic techniques will transform our ability to probe, verify, and ultimately harness fractionalized excitations in quantum spin liquids. Moreover, the flexible tunability of the Rydberg blockade enables the realization of the less explored non-crystalline \cite{Debanjan24,Patrick24}, amorphous \cite{Grushin23,Cassella2023} or non-Archimedean lattice geometries \cite{Moessner20}, and spin liquid models with higher spins \cite{Pohle23,Ma23,Pohle24,Yan25,ikegami2025}, on which it is possible to tune custom flux patterns and tailored magnetic‐field profiles \cite{Willian25}, warranting new directions for studying localization phenomena and anomalous transport in spin liquids defined on these synthetic platforms.

%%%%%%%%%%%%%%%%%%%%%%%%%%%%%%%%%%%%
\section*{Acknowledgement}
%%%%%%%%%%%%%%%%%%%%%%%%%%%%%%%%%%%%
S.F. acknowledges the support of U.S. National Science Foundation's Materials Research Science and Engineering Center under award number DMR-2011876. 
S.F. is also supported by Deutsche Forschungsgemeinschaft (DFG, German Research Foundation) under Germany’s Excellence Strategy--EXC--2111--390814868 as well as the Munich Quantum Valley, which is supported by the Bavarian state government with funds from the Hightech Agenda Bayern Plus.
N.T. acknowledges the support of U.S. National Science Foundation through award DMR-2138905.
Authors are grateful to Penghao Zhu, David Ronquillo, Kang Wang, Kevin Zhang, Peng Rao, Xu Yang, Ryan Buechele, Sasank Budaraju, Tao Xiang, Mohit Randeria, Natalia Perkins, Yuan-Ming Lu, Adhip Agarwala, Eun-Ah Kim, Subhro Bhattacharjee, Qiang Luo, Willian Natori, Peru d'Ornellas, Frank Pollmann, Johannes Knolle, and Michael Knap for related collaborations, comments and discussions.

\bibliography{biblio.bib}
\end{document}